\documentclass[pdflatex,sn-mathphys-num,iicol]{sn-jnl}

\DeclareUnicodeCharacter{202F}{\,}
\usepackage{amsmath,amssymb}
\usepackage[version=3]{mhchem}
\usepackage[]{siunitx}
\usepackage{dcolumn}
\usepackage{subfig}
\usepackage{float}
\usepackage[section]{placeins}
\usepackage{afterpage}

\newcommand{\subfigimg}[4][,]{%
  \setbox1=\hbox{\includegraphics[#1]{#3}}
  \leavevmode\rlap{\usebox1}
  \rlap{\hspace*{#4pt}\raisebox{\dimexpr\ht1-2\baselineskip}{#2}}
  \phantom{\usebox1}
}

\newcommand{\um}{\unit{\micro\metre}}

\newcommand\Tstrut{\rule{0pt}{2.6ex}}         
\newcommand\Bstrut{\rule[-1.1ex]{0pt}{0pt}}   

\begin{document}

\title[Proton-Induced Reactions on Lanthanum from 55--200 MeV]{Measurement of Proton-Induced Reactions on Lanthanum from 55--200 MeV by Stacked-Foil Activation}

\author*[1,2]{\fnm{Jonathan T.} \sur{Morrell}}\email{jmorrell@bnl.gov}
\author[2]{\fnm{Ellen M.} \sur{O'Brien}}
\author[1]{\fnm{Michael} \sur{Skulski}}
\author[3]{\fnm{Andrew S.} \sur{Voyles}}
\author[1]{\fnm{Dmitri G.} \sur{Medvedev}}
\author[2]{\fnm{Veronika} \sur{Mocko}}
\author[4,3]{\fnm{Lee A.} \sur{Bernstein}}
\author[2]{\fnm{C. Etienne} \sur{Vermeulen}}

\affil*[1]{\orgname{Brookhaven National Laboratory}, \orgaddress{\city{Upton}, \state{NY}, \postcode{11973}, \country{USA}}}
\affil[2]{\orgname{Los Alamos National Laboratory}, \orgaddress{\city{Los Alamos}, \state{NM}, \postcode{87545}, \country{USA}}}
\affil[3]{\orgname{University of California, Berkeley}, \orgaddress{\city{Berkeley}, \state{CA}, \postcode{94720}, \country{USA}}}
\affil[4]{\orgname{Lawrence Berkeley National Laboratory}, \orgaddress{\city{Berkeley}, \state{CA}, \postcode{94720}, \country{USA}}}

\abstract{
Cerium-134 is an isotope desired for applications as a chemical analogue to the promising therapeutic radionuclide \ce{^{225}Ac}, for use in bio-distribution assays as an \textit{in vivo} generator of the short-lived positron-emitting isotope \ce{^{134}La}. In the 50--100 MeV energy range relevant to the production of \ce{^{134}Ce} by means of high-energy proton bombardment of lanthanum, existing cross section data are discrepant and have gaps at important energies. To address these deficiencies, a series of 17 \ce{^{139}La} foils (99.919\% natural abundance) were irradiated in two stacked-target experiments: one at the Los Alamos National Laboratory's Isotope Production Facility (IPF) with an incident proton energy of 100 MeV, and a second at Brookhaven National Laboratory's Brookhaven Linac Isotope Producer (BLIP) with an incident proton energy of 200 MeV --- a complete energy range spanning approximately 55--200 MeV. Cross sections are reported for 30 products of \ce{^{139}La}(p,x) reactions (representing up to 55\% of the total non-elastic cross section), in addition to 24 residual products measured in the \ce{^{nat}Cu} and \ce{^{nat}Ti} foils that were used as proton flux monitors.  The measured production cross sections for \ce{^{139}La} reactions were compared to literature data as well as default calculations from the nuclear reaction modeling codes TALYS, EMPIRE and ALICE, as well as the TENDL-2023 library.  The default calculations typically exhibited poor predictive capability, due to the complexity of multiple interacting physics models in this energy range, and deficiencies in preequilibrium reaction modeling. Building upon previous efforts to evaluate proton-induced reactions in this energy range, a parameter adjustment procedure was performed upon the optical model and the two-component exciton model using the TALYS-2.0 code. This resulted in an improvement in \ce{^{139}La}(p,x) cross sections for applications including isotope production, over default predictions.
}


\maketitle
\twocolumn

\section{Introduction}
Proton accelerators operating in the approximately 10--200 MeV energy range are advantageous for the production of radioisotopes having characteristics of simultaneous high activity and high \textit{specific} activity, \textit{i.e.}, having a high ratio of the desired radionuclide to ``cold" (stable) impurities. These characteristics are generally advantageous to the field of nuclear medicine for the creation of radiopharmaceuticals used in the diagnosis and treatment of various diseases, such as cancers. For example, one such radionuclide we are interested in producing is \ce{^{134}Ce}, which has applications as a positron-emitting analogue of the therapeutic isotope \ce{^{225}Ac}.

Actinium-225 is an alpha-emitting radionuclide that is currently under study for the treatment of various forms of cancer, such as advanced prostate cancer and acute myeloid leukemia \cite{Kratochwil1941, CRMPC_study, AML_study}. With a half-life of 9.9203 (3) days, \ce{^{225}Ac} quickly decays to \ce{^{209}Bi} through the emission of four 5--8 MeV $\alpha$ particles and two $\beta^-$ particles, with the longest-lived intermediate decay product being the 3.234 (7) h \ce{^{209}Pb} \cite{A225, A221, A217, A213, A209}. Having a characteristic range of 50--100 \um\ in human tissue, these emitted alpha particles have a high likelihood of killing cancerous cells while sparing nearby healthy tissue, provided a sufficiently specific cancer-targeting vector. One example of a promising vector is PSMA-617 (prostate-specific membrane antigen), which has shown efficacy (defined as a decrease in prostate-specific antigen, or PSA, serum concentration of $\geq$50\%) in 66\% of patients afflicted with advanced (metastatic) prostate cancer, according to a recent meta-analysis \cite{actinium_meta}.

An important aspect of treatment planning in targeted radionuclide therapy is the ability to assay the bio-distribution of the injected radiopharmaceutical, typically with clinical positron-emission tomography (PET) scanners \cite{alpha_review}. Unfortunately, \ce{^{225}Ac} lacks positron emissions in any of its decay products, negating the possibility of performing such scans directly. Instead, the therapeutic targeting vector must be radiolabeled with a positron-emitting chemical analogue of actinium. Recent studies have demonstrated the ability of \ce{^{134}Ce} to act as a PET imaging surrogate for drug conjugates incorporating \ce{^{225}Ac} for long-term tumor targeting \cite{Bailey2021, BAILEY202228, Bobba265355}. Cerium-134 decays with a half-life of 3.16 (4) days to \ce{^{134}La} ($T_{1/2}=6.45$ (16) min), which emits a positron in approximately 64\% of decays \cite{A134}. Therefore, \ce{^{134}Ce}, acting as a chemical analogue to actinium, serves as an \emph{in-vivo} generator of the positron-emitter \ce{^{134}La}. 

Cerium-134 could be produced by proton accelerators using natural targets of lanthanum, cerium, praseodymium, neodymium or even samarium, however the (p,6n) reaction on \ce{^{139}La} (99.9119\% natural abundance) is generally predicted to have the highest \ce{^{134}Ce} yield of these options. The \ce{^{140}Ce}(p,7n)\ce{^{134}Pr} production route is theoretically interesting because it could produce \ce{^{134}Ce} with a higher radiopurity than using a lanthanum target; however, the $\approx$11 min half-life of \ce{^{134}Pr} is too short for practical radiochemical separations, which may take several hours to days \cite{A134}. By this time, the product \ce{^{134}Ce} would be chemically indistinguishable from the target material, and therefore the separation would be impossible.

Currently available cross section data for the \ce{^{139}La}(p,6n)\ce{^{134}Ce} reaction extends from approximately 50--90 MeV \cite{Tarkanyi2017, Morrell_La, BECKER202081}. However, there are only two data points above 70 MeV, where the peak of the cross section is expected, and the three available data sets are generally discrepant beyond their reported error margins, leading to calls for improvements in these data \cite{Qaim_data}. The primary goal of this study was to address these discrepancies with additional measurements in the 55--100 MeV energy range, as well as to extend the data set with measurements up to approximately 200 MeV, which would characterize the preequilibrium tail for the \ce{^{134}Ce} product. This was done using the stacked-foil activation technique, in which a set of thin foils are simultaneously irradiated with a proton beam, such that the beam loses energy as it passes through each foil in the stack \cite{Niobium_voyles, GRAVES201644, Niobium_voyles}. By measuring the induced radioactivity for each foil in the stack, the radionuclide production cross sections can be inferred as a function of proton energy. One benefit of this technique is that while the experiment was designed around the measurement of the (p,6n) cross section, many more radionuclide production cross sections could be simultaneously determined. In this experiment, 30 distinct reaction channels were observed in the lanthanum targets, comprising an estimated 55\% of the total non-elastic cross section (based on the TENDL-2023 non-elastic cross section), with an additional 24 products measured in the \ce{^{nat}Cu} and \ce{^{nat}Ti} foils used as proton flux monitors. This represents the most extensive measurement of proton-induced reactions on lanthanum to date, and the first measurement above 100 MeV. The resulting data set provides an exquisite sensitivity to the nuclear reaction processes relevant to the formation of the associated radionuclides, such as preequilibrium particle emission.

These data were used to qualitatively assess the predictions of various nuclear reaction modeling codes, as well as to determine how well the fitting procedure developed in Fox \textit{et al.} for the existing lanthanum data set (previously extending only up to 90 MeV) extrapolated to the new 100--200 MeV measurements \cite{Fox}. While the extrapolation was generally better than the default predictions, there was considerable room for improvement, which we performed with an optimization of selected preequilibrium and optical model parameters using the TALYS-2.0 code \cite{TALYS}. As part of this optimization procedure we compared various optimization algorithms, figure-of-merit weighting procedures, and sensitivities for various TALYS parameters. These results are vital to improving the quality of nuclear data evaluations for these high-energy charged particle reaction pathways.

\section{Experimental Methods and Materials}

In this work, a number of proton-induced nuclear reaction cross sections were measured using the stacked-foil activation method. These measurements were performed as part of a Tri-lab collaboration between Los Alamos National Laboratory (LANL), Brookhaven National Laboratory (BNL) and Lawrence Berkeley National Laboratory (LBNL). Two sets of foils were irradiated: one at the LANL Isotope Production Facility (IPF) with an incident proton energy of 100 MeV, and a second at the BNL Brookhaven Linac Isotope Producer (BLIP) with 200 MeV protons. We will be referring to these irradiations as the ``LANL stack" and the ``BNL stack" respectively.

\subsection{Stack Design and Irradiations}

In the stacked-foil (sometimes called stacked-target) activation method, a set of thin foils are arranged in a ``stack" and irradiated with charged particles: protons in this case. As the proton beam traverses each successive foil in the stack, its average energy is degraded via stopping interactions within the foils, such that each foil is activated at a slightly lower proton energy than the foil preceding it. Each foil was paired with multiple ``monitor" foils, which served as a measure of the beam current and energy at each point in the stack. In our experiment, additional ``degraders" made of thick pieces of aluminum or copper were used to further decrease the energy between each set of foils. Following a stacked-foil irradiation, the proton-induced reaction products created in each foil are assayed via gamma spectrometry. The assayed production rates for each isotope, $R_i$, are related to the foil's areal density, $\rho r$, the average proton current, $I_p$, and the cross section for the production of each isotope, $\sigma_i$ according to

\begin{equation}
R_i = (\rho r) I_p \sigma_i
\label{eq:activation}
\end{equation}

where the standard units for $R_i$, $\rho r$, $I_p$ and $\sigma_i$ are $s^{-1}$, $atoms\cdot cm^{-2}$, $s^{-1}$ and $cm^2$, respectively. Using this equation, the reaction cross section can be calculated so long as the beam current and areal densities are well-known. This is often referred to as the ``thin-target" activation equation \cite{GRAVES201644, Niobium_voyles}. It is based on the assumption that both the proton current and the cross section are approximately constant within the foil --- which can only be the case for foils thin enough such that their energy degradation is much less than the incident proton energy. An additional feature of these experiments is in the determination of proton beam currents, $I_p$, using ``monitor" foils, rather than external diagnostics such as a Faraday cup --- the use of which is precluded by the physical layouts of both the IPF and BLIP irradiation stations. This monitor foil method relies upon the activation of thin foils having well-characterized ``monitor reaction" cross sections, and using Eq. \ref{eq:activation} to calculate $I_p$ using a known value of $\sigma_i$.

\begin{figure}[!htbp]
\includegraphics[width=\columnwidth]{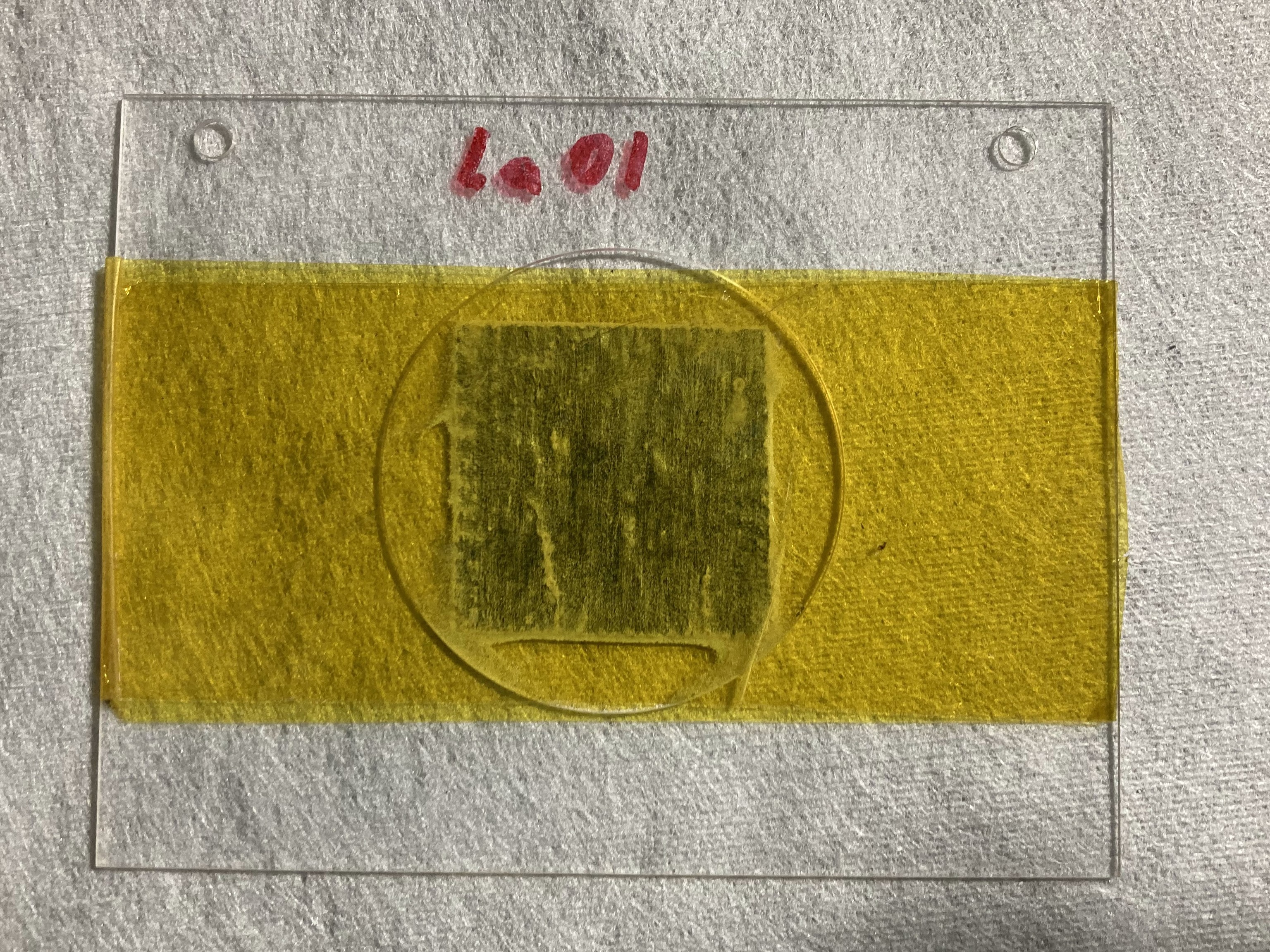}
\caption{Photo of an individual lanthanum foil sealed in Kapton polyimide tape, and secured to the acrylic frame before the irradiation. 
}
\label{fig:lanthanum_foil}
\end{figure}

Our experiment consisted of two separate irradiations: one stack of ten lanthanum foils irradiated at the LANL IPF with 100 MeV protons, and a stack of seven lanthanum foils irradiated at the BNL BLIP with 200 MeV protons. The lanthanum foils used in this experiment were identical for both irradiations, all of at least 99\% purity by metals basis, purchased from Goodfellow Corporation (Coraopolis, PA 15108, USA). These consisted of natural (\textit{i.e.,} non-enriched) lanthanum metal, which is constituted of 99.9119\% \ce{^{139}La} and the remainder of \ce{^{138}La} \cite{Huang_2021}.  Each foil had a nominal thickness of 25 \um\ and was approximately 25.4 mm square. Oxidation of the lanthanum metal was minimized by handling the foils inside of a glove box with an inert (dry argon) cover gas. Figure \ref{fig:lanthanum_foil} shows a typical preparation of one of the lanthanum foils used in the experiment. Slight oxidation can be seen along the lower edge of the foil, however, oxidation in the beam-strike area (center) was minimal prior to irradiation.

Prior to irradiation, the dimensions and masses of each of these foils were measured in the argon-filled glovebox, with a 1--2\% accuracy in the areal density. The foils were then suspended in the center of a plastic mounting frame, sealed between two pieces of 3M 5413-Series Kapton polyimide film tape, which consist of a 25 \um\ polyimide backing material and a $\approx 43$ \um\ layer of acrylic adhesive. The adhesive was specifically chosen to be free of silicon, which under proton bombardment has been shown to lead to the production of \ce{^{24}Na}: a contaminant that would interfere with the \ce{^{27}Al}(p,x)\ce{^{24}Na} monitor reaction channel, and also produce an unwanted $\gamma$-ray spectrometry background in the form of 1.368 and 2.754 MeV photo-peaks and associated Compton scattering events \cite{A24, Niobium_voyles}. The mounting frames were also machined out of acrylic, for this same reason.

The LANL stack made use of natural titanium and copper as monitor foils, observing the \ce{^{46}Sc} and \ce{^{48}V} monitor reaction products in titanium and \ce{^{62}Zn}, \ce{^{65}Zn}, \ce{^{56}Co} and \ce{^{58}Co} in copper. In the BNL stack, aluminum was used instead of titanium, where the \ce{^{22}Na} and \ce{^{24}Na} monitor channels were observed. All of these foils were cut from sheets of 25 \um\ nominal thickness into approximately 25.4 mm squares, which were also measured, weighed and mounted to their frames in Kapton tape in a similar manner as described for the lanthanum foils. For both stacks, one monitor foil of each material was co-irradiated with each lanthanum foil, such that there were 20 monitor foils irradiated in the LANL stack and 14 in the BNL stack. The respective metals-basis purities for the titanium, copper and aluminum monitor foils used in this experiment were 99.95\%, 99.99\% and 99.999\%, respectively.

The degraders used in the LANL stack were made from 110-Copper alloy sheets for degraders 1--4 and 6065-Aluminum alloy sheets for degraders 5--9, with thicknesses ranging from approximately 0.6--1.5 mm.  The BNL stack made use of 110-Copper degraders, machined from plates ranging between 3.7--5.1 mm in thickness. Each degrader was also carefully measured and weighed to determine its areal density. Finally, a stainless steel ``profile monitor" foil was placed in front of and behind each stack, with a nominal thickness of 100 \um. The purpose of these foils was to be used as a post-irradiation check of the beamspot entering and exiting the stack, by mapping the induced activation profile with radiochromic film (Gafchromic EBT3). For both irradiations, the profile monitor foils confirmed that the beam was well-contained within the approximately 25 mm by 25 mm area of the foils, in the center of the stack. The complete specifications of both BNL and LANL stacks can be found in Appendix \ref{stack_appendix}.

\begin{figure}[!htbp]
\includegraphics[width=\columnwidth]{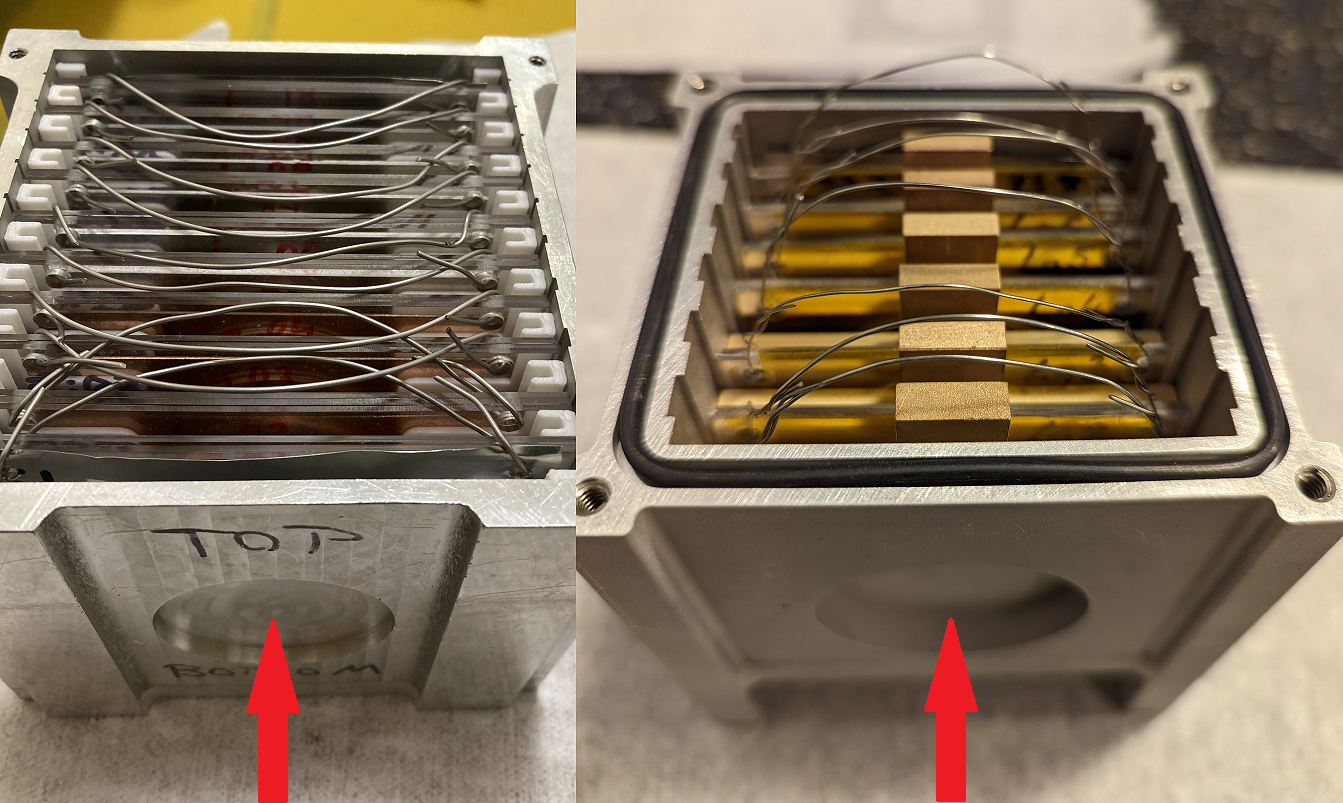}
\caption{Photos of the experimental target boxes used in the LANL (left) and BNL (right) irradiations. Individual foil packets can be seen bundled together with wire, for ease of manipulation inside of the hot cells. The red arrows indicate the orientation of the incident proton beam.
}
\label{fig:target_boxes}
\end{figure}

Since both IPF and BLIP are high-power production facilities with water-cooled irradiation stations that are only accessible remotely, and require that all materials be introduced into a hot cell and handled with telemanipulators, a specially designed sample irradiation box was required for both irraditions. It was critical that this sample irradiation box was leak tight, as the lanthanum foils would rapidly oxidize if exposed to the cooling water. This had the added benefit of significantly minimizing surface contamination on the foils. Photos of the target boxes used in the LANL and BNL irradiations can be seen in Fig. \ref{fig:target_boxes}, with red arrows indicating the direction of the incident beam. Both target boxes were relatively similar in design. They were both machined from aluminum, with a cutout in the front to minimize the thickness of the entrance window, and both had a water-tight seal provided by an aluminum lid (not pictured) that bolts to the top of the box. In both cases, the experimental foils in each compartment between degraders --- La, Cu and Ti for LANL and La, Cu and Al for BNL --- were bundled together into packets using baling wire, to aid in sample retrieval with hot cell telemanipulators after the irradiation.

Both stacks were irradiated with a nominal beam current of 200 nA for a duration of 1 hour, 6 seconds at LANL and 1 hour, 1 minute at BNL. The LANL irradiation took place on 13 September, 2022 and the BNL irradiation took place on 17 March, 2023. Both facilities utilized an inductive pickup current monitor, with current readings recorded at a frequency of approximately 20 times per minute. The resolution of each current monitor was 1 nA, and the beam current was stable over the duration of both irradiations with an RMS fluctuation less than 3 nA. The measurements of the inductive pickup current monitors agreed quite well with the current determined using the monitor foils in the first (highest-energy) compartment. However, scattering and absorption of protons within the stacks themselves caused the beam current to drop as the beam traversed the stack.

\subsection{Gamma Spectrometry}
\label{gamma_spec}

Following each irradiation the target boxes were retrieved from each respective hot cell, and the foils were separated and placed into individual plastic bags made of 50 \um\ thick polyethylene in order to prevent the spread of radioactive contamination during the gamma-ray assay. The time between end-of-bombardment and the first count of the lanthanum foils was approximately 2 hours at LANL and 1 hour at BNL, which meant the product with the shortest half-life that was independently measured was \ce{^{132m}La}, with a half-life of $24.3(5)$ min \cite{A132}. The gamma counting process had differences between each site, but in general, short, initial counts were taken at the respective production facilities to capture the short-lived products, after which the foils were transferred to a secondary on-site counting laboratory where the gamma-ray background was lower. All gamma-ray assays were performed using mechanically cooled High-Purity Germanium (HPGe) detectors.

At LANL, the first counts of the lanthanum foils were performed using an ORTEC GEM p-type coaxial HPGe detector (model GEM20P-PLUS), at a distance ranging from 35--44 cm from the entrance window of the detector. These counts ranged in duration from 2--4 minutes, with the five highest-energy foils being counted twice, and the five lowest-energy foils being counted once. Following this, all the foils were transferred to a secondary on-site counting laboratory --- internally referred to as ``TarDIS" --- which houses two ORTEC GEM p-type coaxial HPGe detectors (model GEM20P4-70-PL). All of the three detectors described utilize transistor-reset preamplifiers, which allowed for count rates of approximately 3 kHz at a dead-time of approximately 5\%, while still maintaining good energy resolution (approximately 1.85 keV FWHM at 1.33 MeV). At the TarDIS counting lab, the count plan was to cycle all of the foils between both detectors over a 45 day period, with the count length increasing from 10 minutes at the beginning of the period to 24 hours at the end. However, one of the two detectors suffered an electronic malfunction after approximately 36 hours, and was replaced with the GEM20P-PLUS that had been used previously at the IPF location. Initially the foils were counted 45 cm from the detector face, and were gradually brought closer in as they decayed, to as close as 10 cm. The same Eckert \& Ziegler source set was used to calibrate all three detectors, which consisted of \ce{^{152}Eu}, \ce{^{133}Ba}, \ce{^{137}Cs} and \ce{^{60}Co} sealed point sources, of known activity (provided by the manufacturer) with a listed 95\% confidence interval of $\pm$3\%.

At BNL, two ORTEC GEM p-type coaxial HPGe detectors were used for the initial lanthanum foil counts at the BLIP facility --- one having a transistor-reset preamplifier and another having a standard charge-sensitive preamplifier. Each of the 7 lanthanum foils were counted once on each detector, for approximately 10 minutes per count, at a distance of 62 cm on the charge-sensitive detector and 45 cm on the transistor-reset detector. Lead shielding was used to minimize the environmental background in the detector, however the samples were sufficiently far from the lead shielding so as to avoid substantial Compton-scattered background and lead x-rays. After the BLIP counts, the foils were transferred to a separate on-site counting laboratory, where three p-type HGPe detectors were utilized for gamma counting: an ORTEC IDM-200-V, an ORTEC Trans-SPEC-DX-100T, and an ORTEC GEM20P4-70-PL. The IDM and Trans-SPEC utilize charge-sensitive preamplifiers, while the GEM20P4 makes use of a transistor-reset preamplifier. The lanthanum foils were preferentially counted on the transistor-reset detector for the sake of improved statistics, however each lanthanum, copper and aluminum foil was counted on every detector to minimize the potential of systematic effects arising from any single detector. The foil-to-detector counting distances varied from 10--59 cm, depending on the detector, the foil being counted and the decay time. The BLIP and on-site detectors were all calibrated using the same source set: \ce{^{152}Eu}, \ce{^{133}Ba}, \ce{^{137}Cs}, \ce{^{60}Co}, and \ce{^{241}Am} point sources procured from Eckert \& Ziegler, with a listed 95\% confidence interval of $\pm$3\%. After 7 days of counting at the BNL on-site counting laboratory, the foils were shipped to the TarDIS laboratory used in the LANL irradiation, where they were counted on the two ORTEC GEM20P4-70-PL detectors for an additional 35 day period, at a distance of 10 cm. The calibration source set was the same as previously described for that location.

\section{Data Analysis}

The residual nuclide production cross sections were extracted from the measured gamma-ray spectra in a consistent manner for both irradiations. The open-source CURIE python library was used for isotope identification and to extract the number of measured counts from each photopeak in the measured spectra, as well as to generate the response functions (energy, resolution and efficiency) for each detector \cite{Curie}. CURIE was also used to calculate production rates for each product isotope based on the peak fits in each foil. For the monitor foils, these production rates were used to calculate the beam current witnessed by each foil, and CURIE was used to determine the proton energy distributions within each foil. These calculated proton energies were then refined using a variance minimization procedure. Finally, the beam currents from the monitor foils were used to calculate the residual nuclide production cross sections from the production rates measured in the lanthanum foils, according to Eq. \ref{eq:activation}. The uncertainties in the reported cross sections are attributable to: statistical and systematic uncertainties from the peak fits, uncertainties in isotope half-lives and decay gamma branching ratios, uncertainties in the measured foil areal density, systematic uncertainties from the production rate determination, uncertainty in the evaluated monitor cross sections, and uncertainty in the detector efficiency.

\subsection{Calibration and Peak Fitting}

The analysis of the measured gamma-ray spectra was performed using the CURIE python library. The detector response functions consisted of energy, resolution and efficiency calibrations which were determined from a set of calibration standards of known activity, as described in Section \ref{gamma_spec}. This procedure was performed for each detector, at every distance that was used to count the experimental foils. Because the BNL foils were counted at BNL and LANL, with two different sets of calibration sources used between them, a cross-calibration was performed to ensure the detector efficiencies were consistent between the two sites. The calibration functions for the energy and resolution calibration were quadratic and linear, respectively, whereas the efficiency function used by CURIE consists of a semi-empirical ``physical" efficiency model, based on the work of Vidmar \textit{et al.} \cite{VIDMAR2001533}.

%
%
%
%

\begin{figure}[!htbp]
\includegraphics[width=\columnwidth]{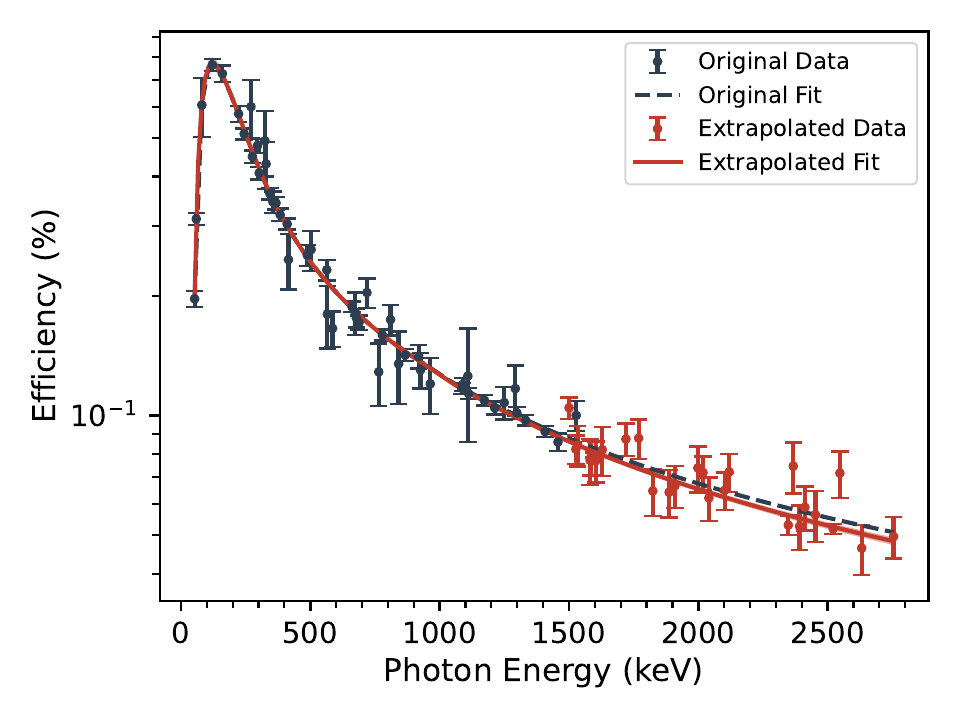}
\caption{Comparison of ``physical" efficiency model fit to original efficiency data taken from the calibration sources (black, dashed line), versus the fit to efficiency data that was extrapolated from the experimental foil measurements (red, solid line).
}
\label{fig:extrapolated_efficiency}
\end{figure}

The main advantage of using such a model for the efficiency is in its ability to extrapolate well to energies above which calibration data are available, as demonstrated in Fig. \ref{fig:extrapolated_efficiency}. The black, dashed line shows the original fit to the efficiency data which were taken from the calibration sources, for which the photopeak energy ranged from 53.1622 keV (a minor line in \ce{^{133}Ba}) to a maximum energy of 1528.1 keV, from \ce{^{152}Eu} \cite{A133, A152}. However, the gamma spectra collected from the experimental foils extended as high as the 3009.645 keV gamma line in \ce{^{56}Co}, well above the highest calibration energy \cite{A56}.

To correctly estimate the detector efficiencies above approximately 1500 keV, an extrapolation procedure was performed from the spectra collected during the experiment, using isotopes having decay gammas both below 1500 keV (known efficiency) \textit{and} above 1500 keV (unknown efficiency). The high-energy efficiency was estimated by multiplying the known low-energy efficiency by the appropriate ratios of $I_{\gamma}$ and measured photopeak counts. Examples of isotopes with gamma emissions both above and below 1500 keV that were used in this extrapolation are \ce{^{133m}Ce}, \ce{^{129m}Ba}, \ce{^{132}La} and \ce{^{56}Co}. This extrapolation is also shown in Fig. \ref{fig:extrapolated_efficiency} as red points, with the solid red line indicating a fit to both the calibration data (black points) and the extrapolated data (red points). As can be seen, the original calibration (dashed black) extrapolated very well to higher energy. This would likely \textit{not} have been the case for a log-polynomial fit to the same data.

Minor correction factors were applied to the efficiency, with a geometry correction factor accounting for the fact that the experimental ``source" (activated foil) was not a perfect point source, but rather a distributed source over the beam-strike area. An additional correction factor accounted for attenuation through the Kapton tape and polyethylene encapsulation, as well as within the foils themselves (self-attenuation). The attenuation correction factors were derived from the XCOM library of photon attenuation coefficients \cite{berger2011xcom, hubbell1995tables}. These corrections were generally quite small, with the largest geometry correction being 0.35\% and the largest attenuation correction being 7.91\%.

%
%

\begin{figure*}[!htbp]
\includegraphics[width=0.9\textwidth]{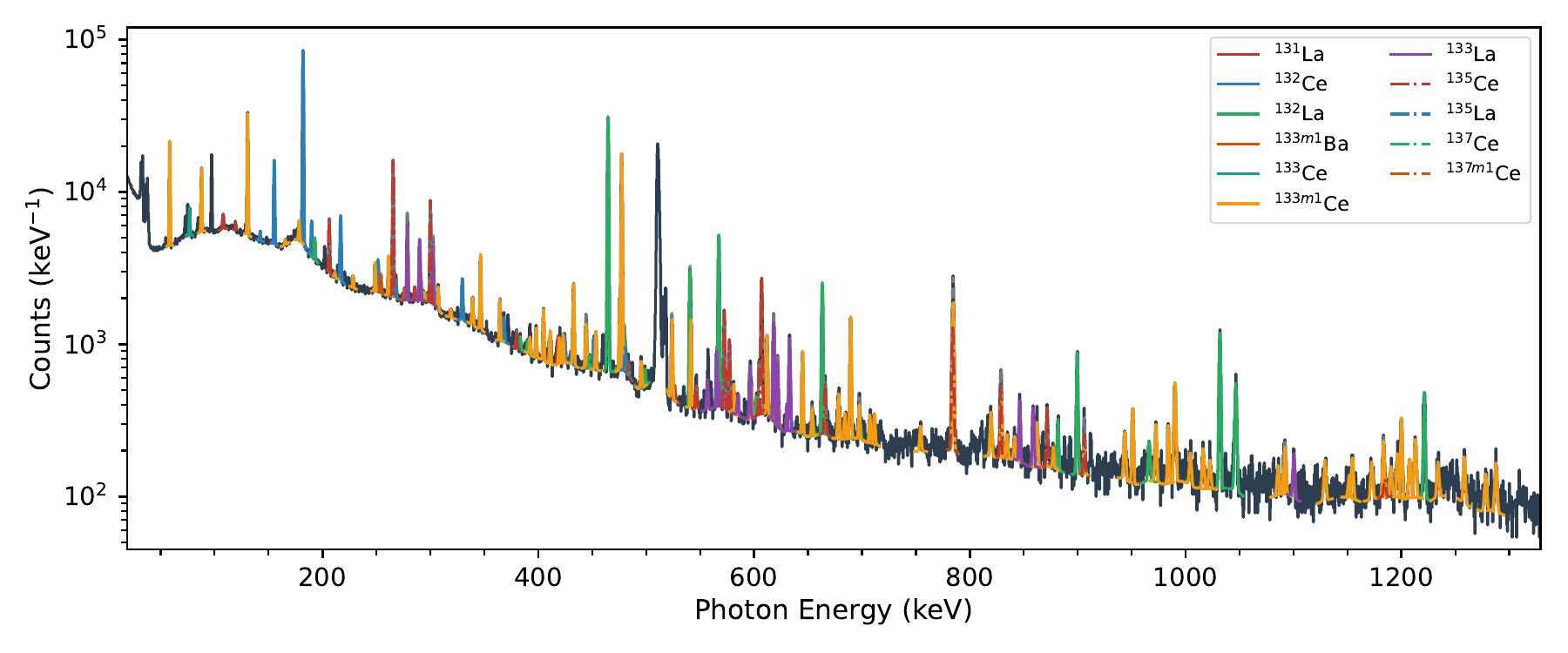}
\caption{Example of a gamma-ray spectrum with peaks identified and fit using CURIE -- from the first lanthanum foil in the LANL stack, irradiated with approximately 91.3 MeV protons. This spectrum was collected approximately 4 hours after the end-of-bombardment.
}
\label{fig:spectrum}
\end{figure*}

An example of a collected $\gamma$-ray spectrum with peaks fit using the CURIE code can be seen in Fig. \ref{fig:spectrum}. In general, the resolution of the HPGe detectors was quite good, such that there were very few peaks which could not be resolved due to issues such as interference from neighboring lines. However, certain isotopes share decay gammas of the \textit{exact} same energy, for example \ce{^{132m}La} and \ce{^{132g}La}, which both decay into \ce{^{132}Ba} and therefore have a number of identical lines. In these cases, the overlapping peaks were excluded in favor of isolated decay gammas from those isotopes. Fig. \ref{fig:spectrum} also illustrates the advantage of using a semi-automated peak fitting code like CURIE for the analysis, as the spectrum is clearly very complex and contains too many individual gamma lines for identifying and/or fitting individual peaks to be practical. In total, the entire data set (comprised of spectra from both irradiations) contained approximately 75,000 individual peak fits from 644 gamma spectra.

\subsection{Production Rate Determination}
\label{prod_rates}

The production rates were determined for each individual isotope by first using the number of counts, $N_c$, from each observed photopeak, and calculating the corresponding number of decays according to:

\begin{equation}
N_D = \frac{N_c}{I_{\gamma}\epsilon(E_{\gamma})} \frac{t_{real}}{t_{live}}
\end{equation}

where $I_{\gamma}$ is the gamma branching ratio, $\epsilon(E_{\gamma})$ is the detector efficiency at energy $E_{\gamma}$, and $t_{real}$ and $t_{live}$ are the real and live times of the data acquisition system, respectively. These decays are associated with a given start and stop time, $t_{start}$ and $t_{stop}$. The CURIE library implements the general solution to the Bateman equations for radioactive production and decay \cite{Bateman}, and solves for the production rates of any number of isotopes in a decay chain (originating from the same parent radionuclide) by taking advantage of the fact that the Bateman solutions are additive, and converting the problem into a linear system of the form:

\begin{equation}
\bar{M} \cdot \vec{R} = \vec{N_D}
\label{eq:bateman}
\end{equation}

where $\vec{N_D}$ is a vector of observed decays for all isotopes in the chain, $\vec{R}$ is a vector of production rates, and $\bar{M}$ is a matrix of decays calculated from a decay chain of unit production rates for a single isotope in $\vec{R}$, during the interval $t_{start}$ to $t_{stop}$ of the associated element of $\vec{N_D}$. While CURIE only implements this approach for decay chains with a single parent isotope, this methodology was extended to work for multiple decay chains feeding the same isotopes, which was required in this work to fit the $A=132$ and $A=133$ decay chains.

\begin{figure}[!htbp]
\includegraphics[width=\columnwidth]{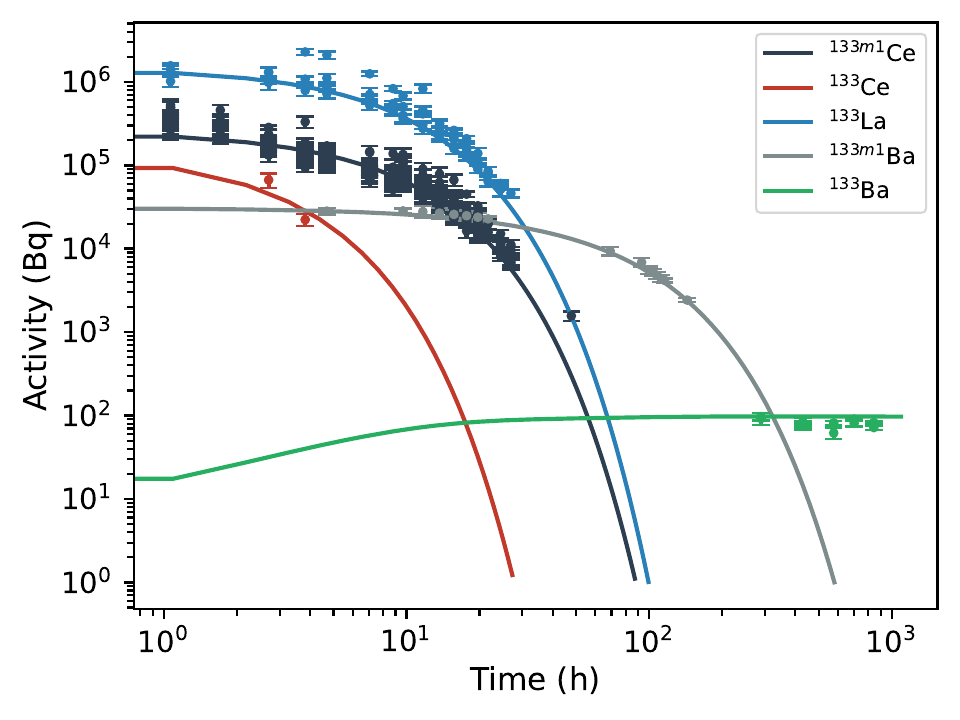}
\caption{Example of a fit to the $A=133$ decay chain in a lanthanum foil irradiated with approximately 160 MeV protons at BNL.
}
\label{fig:decay_chain}
\end{figure}

An example of this fit to the $A=133$ decay chain is shown in Fig. \ref{fig:decay_chain}, which fit the production rates of \ce{^{133m}Ce}, \ce{^{133g}Ce}, \ce{^{133}La} and \ce{^{133m}Ba}. While \ce{^{133g}Ba} is likely also independently produced, it was not quantified with sufficient sensitivity to distinguish its independently produced activity from in-feeding from the other isotopes. Such an approach is very necessary in this case, as both \ce{^{133m}Ce} and \ce{^{133g}Ce} decay into \ce{^{133}La} (\textit{i.e.,} \ce{^{133}La} has two parent isotopes), and all isotopes in this system feed into \ce{^{133g}Ba}. Therefore, there is neither a simple equation for calculating the production rate of a given isotope from measured activities, nor for calculating the measured activities from the number of observed decays.

The half-lives and major gamma branching ratios used for each of the isotopes observed in this study are listed in Tables \ref{table:decay_data_lanthanum} and \ref{table:decay_data_monitor} in Appendix \ref{nudat_appendix}. The uncertainties in the half-lives and gamma branching ratios were used in the determination of the uncertainties in production rates, however in general these were much smaller than the systematic uncertainties associated with the gamma counting process.

One additional feature of this data set was the presence of secondary (mostly thermal) neutrons, resulting from spallation reactions in the stack, and subsequent thermalization in the surrounding cooling water. This could potentially inflate our determination of reaction rates through (n,x) reactions resulting in the same product. Conveniently, we were able to use the \ce{^{140}La} ($T_{1/2}=1.67855(12)$ d) product, resulting from neutron capture on the lanthanum foils, as a neutron flux monitor \cite{A140}. This was used in combination with a FLUKA Monte Carlo simulation to determine the neutron flux profile throughout the stack \cite{FLUKA}, showing that only the \ce{^{64}Cu} production rate in the copper monitor foils had a non-negligible contribution from secondary neutrons (from the capture reaction on \ce{^{63}Cu}). This effect has been corrected for in the reported cross sections.

\subsection{Beam Currents and Energy Determination}
\label{monitors}

The proton beam currents witnessed by each foil in each of the two irradiated stacks were determined using a set of monitor foils, which have well-characterized reaction channels that are included in the list of beam monitor reactions from the IAEA charged-particle cross section database for medical radionuclide production \cite{IAEACPR}. This method is well described in other publications \cite{Niobium_voyles, Fox}, and has some advantages and disadvantages compared to other methods of measuring beam current. The main disadvantage is that the accuracy of the measured cross sections cannot exceed that of the monitor reaction cross sections, and in particular, any unknown systematic errors in the monitor cross sections will be applied (proportionally) to the measured cross sections. However, these effects can be mitigated by using multiple monitor reaction channels in multiple foil materials, and by comparing the monitor foil-derived beam currents to beam currents determined by electronic measurement, such as an inductive-pickup current monitor. The main advantage of this method is that it can correctly account for beam current losses within the stack, which would not be observable with an electronic beam current monitor. Also, issues such as electron suppression or other sources of ``dark current" do not impact the measurement.

The beam current for an individual monitor channel is calculated from a measured production rate, $R_i$, using Eq. \ref{eq:activation} where the cross section used is the flux-averaged cross section:

\begin{equation}
\bar{\sigma} = \frac{\int_0^{\infty}\sigma(E)\psi(E)dE}{\int_0^{\infty}\psi(E)dE}
\label{eq:avg_xs}
\end{equation}

where $\sigma(E)$ is the recommended cross section from the IAEA beam monitor library \cite{IAEACPR}, and $\psi(E)$ is the proton flux calculated using a 1-D Monte Carlo stopping power code implemented in CURIE, which uses the Anderson \& Ziegler formalism for stopping powers \cite{Curie, ZIEGLER20101818}. This treatment accounts for the beam straggling which occurs towards the back of the stack. This was particularly significant for the BNL data set, which was degraded from 200 MeV incident energy down to just under 100 MeV in the rear foil.

\begin{figure}[!htbp]
\includegraphics[width=\columnwidth]{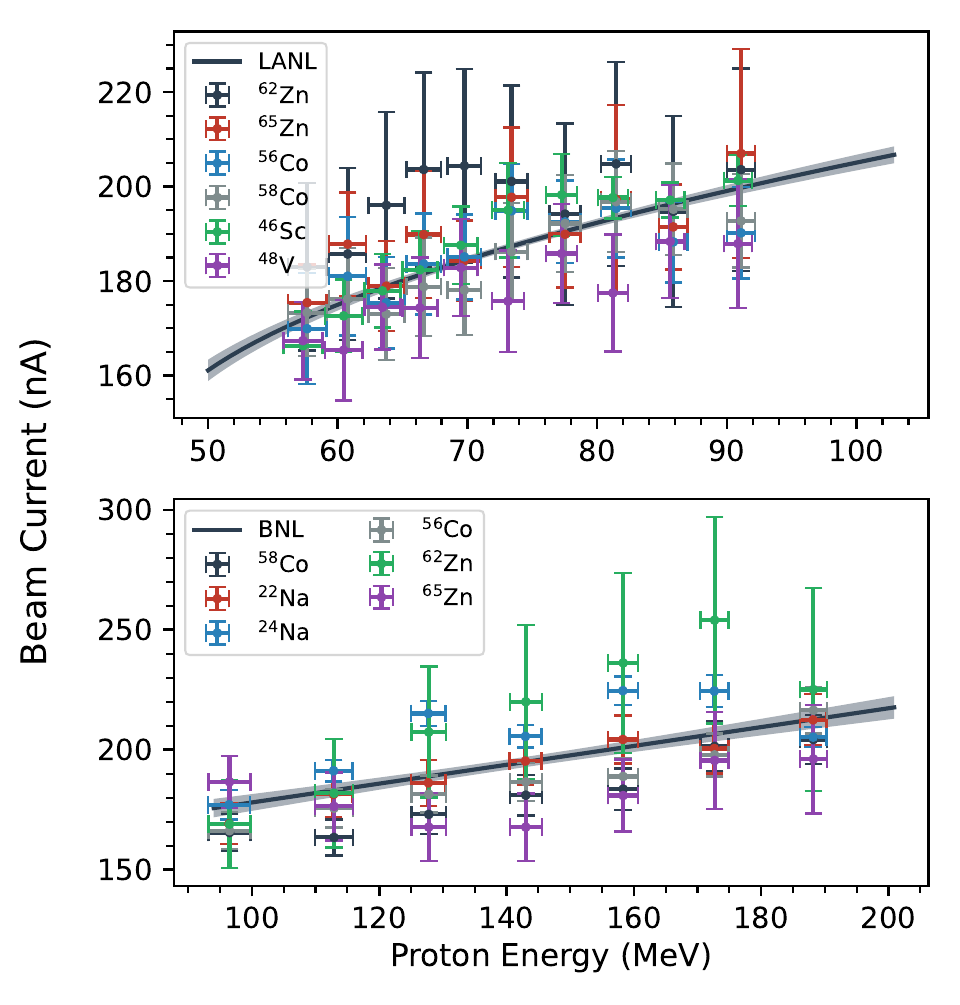}
\caption{Beam currents measured in each of the 6 monitor reaction channels utilized by the LANL stack (top) and BNL stack (bottom). The black line on each plot indicates a fit to the calculated beam currents as a function of energy.
}
\label{fig:beam_current}
\end{figure}

The measured beam currents from each reaction channel as a function of proton energy were compiled for both stacks, and fit with a linear function ($I = I_0 + m\cdot E$) for the BNL stack and a square-root function ($I = I_0 + m\cdot \sqrt{E-45}$) for the LANL stack. The measured per-channel beam currents and attending fits can be seen in Fig. \ref{fig:beam_current}, with the LANL stack shown in the upper panel and the BNL stack shown in the lower panel.

One complication for the BNL stack was that only the \ce{^{27}Al}(p,x)\ce{^{22}Na} and \ce{^{nat}Cu}(p,x)\ce{^{58}Co} reactions have recommended cross sections above 100 MeV, despite many experimental measurements in this energy region for the production of \ce{^{24}Na} (from \ce{^{27}Al}) and \ce{^{56}Co}, \ce{^{62}Zn} and \ce{^{65}Zn} (from \ce{^{nat}Cu}), which were all observed reaction products in the monitor foils, and have IAEA-recommended cross sections below 100 MeV. To improve the quality of the beam current determination for the BNL stack, a down-selected set of experimental data were used to generate monitor reaction cross sections in this energy region. This was done by performing a regression of the following 3-parameter log-polynomial function to the down-selected cross section data:

\begin{multline}
\sigma(E) = \sigma_{100} + \Big(\frac{c_1}{E}\ln(E-60)^2 \\
+ \frac{c_2}{E} \ln(E-60) + \frac{c_3}{E}  \Big) (E-100)
\end{multline}

where $\sigma_{100}$ is the IAEA-recommended cross section for that channel at 100 MeV, and $c_1$, $c_2$ and $c_3$ are adjustable parameters. This functional form was selected to ensure that the fitted cross section agreed with the IAEA evaluation at 100 MeV. The resulting fits to these additional monitor channels can be seen in Appendix \ref{additional_cross_sections}. There were significant discrepancies between our apparent cross sections and the \ce{^{nat}Cu}(p,x)\ce{^{58}Co} IAEA reference cross sections, which may indicate the need for future study of this channel \cite{IAEACPR}.

\begin{figure}[!htbp]
\includegraphics[width=\columnwidth]{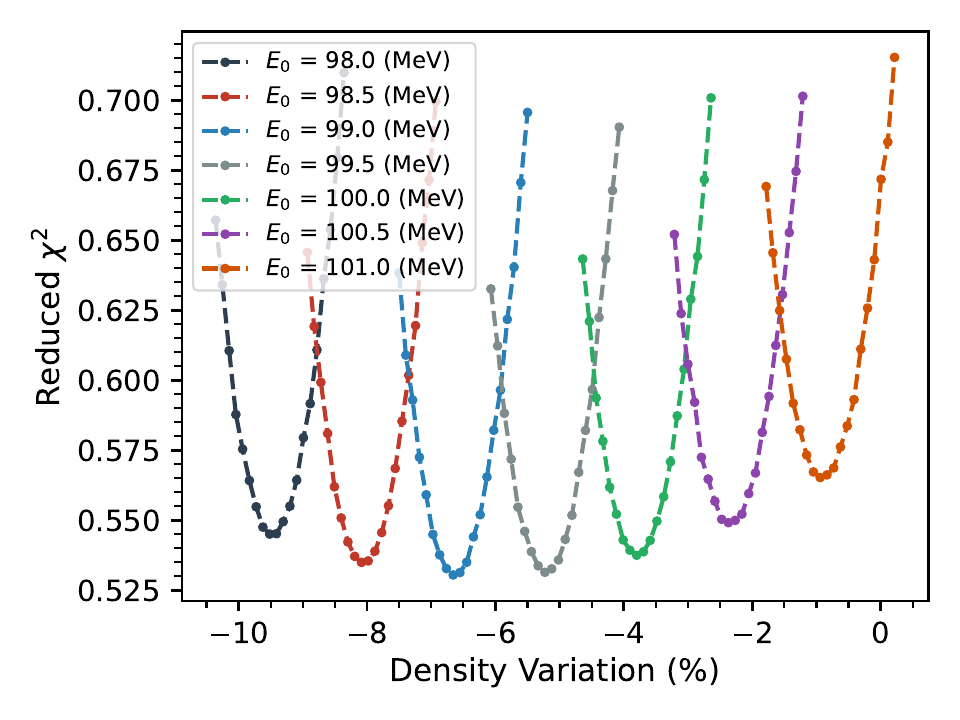}
\caption{Reduced $\chi^2$ of the fit to the LANL monitor foil data, plotted as a function of degrader density variation and incident proton energy.
}
\label{fig:minimization}
\end{figure}

The energy distributions predicted by the Anderson-Ziegler calculation in CURIE were adjusted using a ``variance minimization" methodology, similar to the methods described in Graves \textit{et al.} and Voyles \textit{et al.} \cite{GRAVES201644, Niobium_voyles}. In this approach, a multiplier to the areal densities of all stack elements ($\Delta \rho$), as well as the incident energy ($E_0$) are treated as free parameters, and are varied in order to minimize a goodness-of-fit parameter: in this case the reduced $\chi^2$ (or $\chi^2_{\nu}$). The purpose of this is not to presume that the densities of the stack elements or the incident energy were not well-known, but rather to correct for unaccounted systematics in the calculation of the proton energy spectra. A plot showing the calculation of the $\chi^2_{\nu}$ for variation of $\Delta \rho$ and $E_0$ for the LANL stack can be seen in Fig. \ref{fig:minimization}. For the LANL stack, the optimum $E_0$ and $\Delta \rho$ were found to be 99 MeV and -6.65\%. For the BNL stack, only $\Delta \rho$ was optimized using this method, as there was not sufficient sensitivity of $\chi^2_{\nu}$ to the incident energy. The resulting optimum density change for the BNL stack was found to be $+1.75\%$.

\section{Results and Discussion}
\label{results_section}

The cross sections for the products observed in this work were calculated from the measured production rates, $R_i$, the areal densities $(\rho r)$ and monitor-foil derived beam currents $I_p$, using Eq. \ref{eq:activation}. The uncertainties in the cross sections were calculated as the quadrature sum of the uncertainties from these three sources. The results of these measurements are tabulated in Table \ref{table:139LA_XS} for the \ce{^{139}La} products, Table \ref{table:natCU_XS} for products in the copper foils, and similarly in Table \ref{table:natTI_XS} for titanium products. The results for selected reaction channels that are of particular interest for medical applications or from a nuclear reaction modeling perspective are discussed below.

\begin{table*}[!htbp]
\scriptsize
\setlength{\tabcolsep}{4pt}
\caption{Summary of lanthanum cross sections measured in this work. Subscripts $(c)$ and $(i)$ indicate cumulative and independently measured cross sections, respectively. The subscript $(m+)$ indicates the cross section includes contributions from one or more isomers of the same product isotope, but is otherwise independent of feeding from other products. Uncertainties are given in the least significant digit, \emph{i.e.}, 188.3(21) means $188.3 \pm 2.1$, \emph{etc.}}
\begin{tabular}{lccccccccc}
 & \multicolumn{9}{c}{\ce{^{139}La}(p,x) Production cross sections (mb)} \Bstrut \\
\hline
\Tstrut E$_p$ (MeV) & 188.3(21) & 172.8(22) & 158.5(23) & 143.2(25) & 128.0(27) & 113.1(29) & 96.7(33) & 91.3(10) & 86.1(11) \\
 & 81.7(11) & 77.8(12) & 73.7(12) & 70.0(13) & 66.9(13) & 64.0(14) & 61.1(14) & 58.0(15) & - \Bstrut \\
\hline
\Tstrut \ce{^{139}Ce_{(m+)}} & 3.92(48) & 4.46(53) & 4.94(58) & 5.43(62) & 6.04(71) & 6.48(71) & 7.63(89) & 7.99(88) & 8.14(89) \\
 & 8.92(98) & 8.50(94) & 9.5(10) & 9.6(10) & 11.3(12) & 10.7(12) & 11.5(13) & 11.5(13) & - \\
\ce{^{137m}Ce_{(i)}} & 14.27(78) & 16.59(90) & 17.96(97) & 20.3(11) & 23.5(13) & 26.5(14) & 31.0(17) & 34.3(18) & 37.0(19) \\
 & 40.3(21) & 40.8(21) & 44.9(24) & 48.0(25) & 54.9(29) & 57.1(30) & 60.5(32) & 67.4(36) & - \\
\ce{^{137g}Ce_{(i)}} & - & 4.63(66) & 4.95(82) & - & 7.1(22) & - & 8.0(33) & 13.6(16) & 13.4(14) \\
 & 12.2(24) & 15.3(26) & 17.1(35) & 20.8(25) & 23.0(33) & 24.5(33) & 23.0(32) & 28.0(44) & - \\
\ce{^{136}Cs_{(m+)}} & 0.466(23) & 0.467(28) & 0.428(31) & 0.369(25) & 0.318(29) & 0.249(24) & 0.172(14) & 0.102(14) & 0.088(31) \\
 & 0.104(22) & 0.147(74) & 0.077(27) & - & 0.121(34) & 0.111(24) & 0.093(24) & 0.085(31) & - \\
\ce{^{135}La_{(i)}} & 85(17) & 104(15) & 96(13) & 118(17) & 124(20) & 137(15) & 138(11) & 134.4(94) & 147(10) \\
 & 143.1(99) & 143(11) & 141(11) & 138(12) & 150(13) & 143(13) & 139(13) & 139(12) & - \\
\ce{^{135}Ce_{(m+)}} & 24.0(11) & 27.7(13) & 31.0(14) & 35.4(15) & 44.1(20) & 51.3(23) & 66.7(31) & 77.4(33) & 85.4(35) \\
 & 98.3(43) & 105.5(46) & 130.5(54) & 158.9(76) & 211.4(93) & 258(12) & 322(14) & 412(18) & - \\
\ce{^{135m}Ba_{(i)}} & 17.00(89) & 17.93(97) & 16.76(86) & 16.54(84) & 15.69(86) & 14.90(87) & 12.48(66) & 11.35(64) & 12.4(10) \\
 & 10.4(13) & 10.8(12) & 6.50(66) & 9.4(13) & 6.2(10) & - & - & - & - \\
\ce{^{134}Ce_{(i)}} & 30.1(34) & 36.7(40) & 35.6(32) & 40.1(31) & 48.6(32) & 62.6(37) & 87.4(52) & 97.7(56) & 119.5(68) \\
 & 152.8(86) & 183(10) & 240(13) & 267(15) & 287(16) & 247(14) & 182(10) & 108.6(63) & - \\
\ce{^{133}La_{(i)}} & 99.5(53) & 118.0(56) & 124.7(57) & 143.4(68) & 157.2(74) & 158.1(75) & 165.0(77) & 193.6(85) & 186.4(80) \\
 & 136.8(60) & 96.0(42) & 63.9(29) & 27.9(14) & 15.31(86) & 4.28(27) & 1.413(89) & 1.138(73) & - \\
\ce{^{133m}Ce_{(i)}} & 15.75(84) & 17.69(84) & 20.84(96) & 26.2(12) & 35.3(17) & 44.4(21) & 72.4(34) & 105.7(47) & 120.4(52) \\
 & 129.7(57) & 99.8(44) & 66.8(30) & 30.6(15) & 11.12(62) & 2.62(17) & 0.469(29) & 1.138(73) & - \\
\ce{^{133g}Ce_{(i)}} & 3.17(17) & 3.39(16) & 4.44(20) & 5.09(24) & 6.44(30) & 9.54(45) & 12.58(59) & 17.37(77) & 22.90(98) \\
 & 24.2(11) & 21.34(94) & 15.10(69) & 7.24(36) & 2.65(15) & 0.630(40) & - & - & - \\
\ce{^{133m}Ba_{(i)}} & 20.6(11) & 20.69(98) & 19.87(91) & 19.49(92) & 18.42(86) & 15.25(72) & 10.39(49) & 10.66(47) & 8.54(37) \\
 & 6.82(30) & 5.68(25) & 5.29(24) & 5.38(26) & 6.10(34) & 6.62(42) & 7.37(46) & 9.35(60) & - \\
\ce{^{132m}La_{(i)}} & 35.2(23) & 33.0(21) & 39.8(25) & 44.6(28) & 58.7(37) & 56.9(36) & 26.2(17) & 23.0(14) & 14.13(91) \\
 & 9.23(63) & 3.80(34) & 1.61(15) & - & - & - & - & - & - \\
\ce{^{132}Cs_{(i)}} & 5.45(36) & 5.11(33) & 4.58(29) & 4.18(26) & 3.61(23) & 3.13(20) & 2.73(18) & 2.52(16) & 2.52(16) \\
 & 2.43(15) & 2.18(14) & 1.97(13) & 1.72(11) & 1.477(99) & 1.106(74) & 0.694(53) & 0.381(33) & - \\
\ce{^{132}Ce_{(m+)}} & 16.5(11) & 16.7(11) & 21.0(13) & 24.7(16) & 33.7(21) & 46.6(30) & 53.5(34) & 66.4(42) & 36.3(23) \\
 & 14.7(10) & 3.46(31) & 0.394(38) & - & - & - & - & - & - \\
\ce{^{131}La_{(c)}} & 67.0(29) & 71.9(30) & 78.6(33) & 89.5(37) & 97.2(41) & 87.6(39) & 16.86(99) & 7.48(67) & 2.09(42) \\[5pt]
\ce{^{131}Ba_{(m+)}} & 18.8(15) & 18.9(15) & 13.8(16) & 9.5(16) & 7.2(19) & - & 7.66(77) & 11.89(77) & 15.21(72) \\
 & - & 20.70(78) & 21.37(81) & 16.33(64) & 12.48(49) & 6.88(29) & 2.81(14) & 0.674(54) & - \\
\ce{^{130}Ce_{(i)}} & 7.25(76) & 6.98(63) & 8.99(95) & 11.39(90) & 11.41(93) & 3.04(53) & - & - & - \\[5pt]
\ce{^{129}Cs_{(m+)}} & 14.82(61) & 11.98(50) & 7.20(28) & 7.96(37) & 8.70(41) & 6.00(30) & 1.96(10) & - & 0.745(49) \\[5pt]
\ce{^{129m}Ba_{(c)}} & 22.67(93) & 20.66(86) & 18.91(73) & 15.75(73) & 11.47(53) & 12.78(65) & 8.76(47) & 2.44(16) & 1.399(92) \\[5pt]
\ce{^{129g}Ba_{(c)}} & 33.6(14) & 33.4(14) & 34.9(14) & 25.8(12) & 11.10(52) & 6.07(31) & 3.50(19) & 2.75(18) & 1.276(84) \\[5pt]
\ce{^{128}Ba_{(c)}} & 34.9(22) & 31.2(19) & 23.8(15) & 15.17(93) & 13.40(82) & 10.10(63) & 0.694(60) & - & - \\[5pt]
\ce{^{127}Xe_{(c)}} & 32.9(18) & 25.8(14) & 22.9(12) & 19.6(10) & 11.18(66) & 2.18(16) & 0.692(63) & - & - \\[5pt]
\ce{^{127}Cs_{(c)}} & 34.7(21) & 28.2(16) & 23.4(12) & 19.3(14) & 12.8(12) & - & - & - & - \\[5pt]
\ce{^{126}I_{(i)}} & 1.55(20) & 1.60(38) & 0.86(14) & 0.57(13) & 0.65(26) & 0.52(13) & - & - & - \\[5pt]
\ce{^{126}Ba_{(c)}} & 6.62(83) & 3.83(50) & 3.40(42) & 2.59(41) & 1.57(48) & - & - & - & - \\[5pt]
\ce{^{125}Xe_{(c)}} & 16.01(72) & 12.46(55) & 7.88(35) & 2.83(13) & 1.368(77) & 0.576(52) & - & - & - \\[5pt]
\ce{^{125}Cs_{(c)}} & 17.8(26) & 12.7(27) & 9.5(31) & - & - & - & - & - & - \\[5pt]
\ce{^{123}Xe_{(c)}} & 4.32(27) & 2.06(12) & 1.259(92) & 0.590(75) & - & - & - & - & - \\[5pt]
\ce{^{123}I_{(i)}} & 1.30(24) & 1.02(10) & 0.707(97) & 0.383(87) & - & - & - & - & - \\[5pt]
\end{tabular}
\label{table:139LA_XS}
\end{table*}

\begin{table*}[!htbp]
\scriptsize
\setlength{\tabcolsep}{4pt}
\caption{Summary of copper cross sections measured in this work. The uncertainty format and subscript designations $(c)$, $(i)$ and $(m+)$ remain the same as in Table \ref{table:139LA_XS}. Products which were used as beam current monitors (and are therefore somewhat circular) are indicated with a $*$, \textit{e.g.} \ce{^{65}Zn^{*}_{(i)}}.}
\begin{tabular}{lccccccccc}
 & \multicolumn{9}{c}{\ce{^{nat}Cu}(p,x) Production cross sections (mb)} \Bstrut \\
\hline
\Tstrut E$_p$ (MeV) & 188.2(21) & 172.7(22) & 158.3(23) & 143.0(25) & 127.8(27) & 112.9(29) & 96.5(33) & 91.1(11) & 85.9(11) \\
 & 81.5(11) & 77.5(12) & 73.4(12) & 69.8(13) & 66.6(13) & 63.7(14) & 60.8(14) & 57.6(15) & - \Bstrut \\
\hline
\Tstrut \ce{^{65}Zn^{*}_{(i)}} & 1.217(97) & 1.45(12) & 1.530(96) & 1.58(11) & 1.72(12) & 1.96(14) & 2.43(13) & 2.53(26) & 2.543(95) \\
 & 2.82(27) & 2.92(15) & 3.29(23) & 3.30(12) & 3.63(23) & 3.66(16) & 4.12(21) & 4.18(15) & - \\
\ce{^{64}Cu_{(i)}} & 24.5(20) & 28.8(29) & 31.8(34) & 30.3(25) & 37.3(25) & 35.4(34) & 41.7(29) & 37.4(34) & 40.8(38) \\
 & 44.3(39) & 39.8(53) & 42.2(40) & 48.6(46) & 48.6(46) & 48.8(42) & 50.1(45) & 50.9(46) & - \\
\ce{^{62}Zn^{*}_{(i)}} & 2.12(20) & 2.70(24) & 2.85(25) & 3.10(28) & 3.50(31) & 3.74(36) & 4.47(41) & 5.01(45) & 5.20(45) \\
 & 5.90(53) & 6.02(52) & 6.80(60) & 7.55(66) & 8.24(73) & 8.74(77) & 9.34(81) & 10.64(93) & - \\
\ce{^{61}Cu_{(c)}} & 29.3(49) & 36.5(60) & 37.4(61) & 38.2(62) & 43.8(72) & 45.5(75) & 51.8(85) & 55.1(88) & 59.8(95) \\
 & 64(10) & 67(11) & 73(12) & 79(13) & 82(13) & 84(13) & 84(13) & 89(14) & - \\
\ce{^{60}Co_{(m+)}} & 10.6(10) & 11.49(73) & 10.39(79) & 11.19(64) & 11.17(74) & 10.95(66) & 11.36(68) & 28(12) & 12.1(19) \\
 & 14.9(69) & 10.7(11) & - & 10.08(81) & - & 10.47(84) & 15.3(39) & 11.03(65) & - \\
\ce{^{59}Fe_{(i)}} & 1.188(58) & 1.163(54) & 1.112(47) & 1.065(48) & 1.031(61) & 0.953(47) & 0.849(41) & 0.797(92) & 0.792(33) \\
 & 0.797(86) & 0.777(32) & 0.74(12) & 0.715(32) & 0.701(76) & 0.634(28) & 0.584(71) & 0.498(20) & - \\
\ce{^{58}Co^{*}_{(m+)}} & 42.7(20) & 44.4(21) & 42.8(18) & 45.1(17) & 46.2(18) & 46.3(18) & 49.1(20) & 50.3(21) & 50.8(19) \\
 & 50.5(22) & 47.9(19) & 44.1(18) & 39.5(15) & 36.9(15) & 33.2(12) & 31.7(14) & 30.6(11) & - \\
\ce{^{57}Ni_{(c)}} & 1.84(10) & 1.939(99) & 1.87(10) & 1.91(10) & 1.95(10) & 1.85(10) & 1.569(90) & 1.46(11) & 1.305(68) \\
 & 1.239(85) & 1.071(67) & 1.268(86) & 1.393(80) & 1.57(10) & 1.822(85) & 2.07(13) & 2.33(11) & - \\
\ce{^{57}Co_{(i)}} & 40.4(24) & 41.3(22) & 40.6(23) & 41.9(22) & 42.3(22) & 41.6(23) & 40.2(23) & 34.7(44) & 38.1(17) \\
 & 35.7(30) & 35.2(17) & 36.2(37) & 36.3(17) & 39.7(35) & 43.4(18) & 41.1(46) & 54.2(24) & - \\
\ce{^{56}Ni_{(c)}} & 0.125(29) & 0.155(29) & 0.177(22) & 0.148(12) & 0.121(12) & 0.0768(76) & 0.0690(79) & - & 0.097(10) \\
 & - & 0.109(12) & - & 0.133(23) & - & - & - & - & - \\
\ce{^{56}Mn_{(i)}} & 2.28(12) & 2.44(14) & 2.31(18) & 1.93(11) & 1.99(12) & 1.519(92) & 1.105(86) & 1.173(66) & 1.152(63) \\
 & 1.117(72) & 1.053(61) & 1.003(64) & 0.793(66) & 0.659(52) & 0.511(49) & 0.428(51) & 0.346(43) & - \\
\ce{^{56}Co^{*}_{(c)}} & 12.45(53) & 11.99(52) & 11.85(46) & 11.98(47) & 11.65(46) & 11.04(47) & 10.22(45) & 10.42(45) & 10.82(39) \\
 & 12.22(55) & 12.89(46) & 13.83(57) & 12.78(46) & 11.19(50) & 8.15(30) & 5.50(28) & 2.71(10) & - \\
\ce{^{55}Co_{(c)}} & 1.99(13) & 1.99(12) & 1.94(13) & 1.84(12) & 1.74(12) & 1.57(11) & 1.61(12) & 1.71(13) & 1.54(11) \\
 & 1.251(93) & 0.904(77) & 0.464(47) & 0.219(37) & 0.105(25) & 0.056(23) & 0.046(16) & 0.061(23) & - \\
\ce{^{54}Mn_{(i)}} & 16.39(70) & 15.43(63) & 14.36(95) & 13.46(59) & 11.93(53) & 10.43(46) & 6.96(59) & 6.23(30) & 4.73(17) \\
 & 3.96(31) & 3.59(20) & 3.70(28) & 4.02(17) & 4.57(37) & - & - & - & - \\
\ce{^{52}Mn_{(c)}} & 5.18(22) & 4.78(20) & 4.09(18) & 3.53(15) & 2.66(12) & 1.933(88) & 1.769(88) & 1.76(10) & 1.262(51) \\
 & 0.866(77) & 0.400(22) & - & 0.048(20) & - & - & - & - & - \\
\ce{^{51}Cr_{(c)}} & 11.32(49) & 8.9(11) & 7.92(33) & 6.27(27) & 5.07(24) & 3.88(20) & 1.46(11) & - & - \\[5pt]
\ce{^{48}V^{*}_{(i)}} & 1.922(78) & 1.455(65) & 1.048(41) & 0.688(29) & 0.450(21) & 0.300(18) & 0.0555(65) & - & - \\[5pt]
\end{tabular}
\label{table:natCU_XS}
\end{table*}

\begin{table*}[!htbp]
\scriptsize
\setlength{\tabcolsep}{4pt}
\caption{Summary of titanium cross sections measured in this work. The uncertainty format and subscript designations $(c)$, $(i)$ and $(m+)$ remain the same as in Table \ref{table:139LA_XS}. Products which were used as beam current monitors (and are therefore somewhat circular) are indicated with a $*$, \textit{e.g.} \ce{^{48}V^{*}_{(i)}}.}
\begin{tabular}{lccccccccc}
 & \multicolumn{9}{c}{\ce{^{nat}Ti}(p,x) Production cross sections (mb)} \Bstrut \\
\hline
\Tstrut E$_p$ (MeV) & 90.9(11) & 85.6(11) & 81.2(11) & 77.3(12) & 73.1(12) & 69.5(13) & 66.3(13) & 63.4(14) & 60.4(14) \Bstrut \\
\hline
\Tstrut \ce{^{48}V^{*}_{(i)}} & 7.21(34) & 7.89(32) & 8.06(41) & 9.11(35) & 9.39(42) & 10.60(45) & 10.91(55) & 11.76(45) & 12.12(68) \\
\ce{^{48}Sc_{(i)}} & 1.97(20) & 2.05(21) & 1.93(18) & 1.98(18) & 1.89(21) & 1.89(17) & 1.90(19) & 1.87(18) & 1.81(17) \\
\ce{^{47}Sc_{(i)}} & 20.0(10) & 19.7(10) & 20.5(12) & 20.6(12) & 20.7(11) & 20.93(95) & 20.6(20) & 21.13(96) & 21.29(95) \\
\ce{^{46}Sc^{*}_{(m+)}} & 39.9(17) & 40.8(15) & 42.5(16) & 44.1(16) & 45.2(20) & 45.5(16) & 46.5(17) & 47.8(17) & 49.4(18) \\
\ce{^{44m}Sc_{(i)}} & 18.46(79) & 19.04(87) & 20.94(90) & 21.34(89) & 21.38(97) & 20.25(84) & 18.8(12) & 17.19(80) & 14.80(67) \\
\ce{^{43}K_{(c)}} & 1.725(94) & 1.610(99) & 1.503(74) & 1.379(71) & 1.414(83) & 1.367(73) & 1.318(54) & 1.380(62) & 1.446(74) \\
\ce{^{42}K_{(i)}} & 5.88(42) & 6.00(32) & 5.89(33) & 6.22(36) & 6.26(38) & 6.55(31) & 6.51(34) & 6.19(35) & 5.27(27) \\
\end{tabular}
\label{table:natTI_XS}
\end{table*}

The results are shown in comparison to previous measurements from T{\'a}rk{\'a}nyi \textit{et al.}, Morrell \textit{et al.} and Becker \textit{et al.} \cite{Tarkanyi2017, Morrell_La, BECKER202081}. In general, the results show good agreement with previous measurements. Interestingly, the \ce{^{134}Ce} production cross section above 100 MeV was much larger than expected, which could have implications for medical isotope production applications. This work represents the first measurement of proton-induced reactions on lanthanum above 100 MeV, as well as the first measurement of many of these reaction products. Notably, the observation of \ce{^{130}Ce} represents the first measurement of an exclusive (p,10n) excitation function (\textit{i.e.,} a cross section curve measured at multiple energies, including the threshold) in this energy range.

In addition, we compare our results to the TENDL-2023 evaluation, as well as default predictions from TALYS-2.0, EMPIRE-3.2.3, and ALICE-20 \cite{rochman2017tendl, TALYS, HERMAN20072655, ALICE}. A complete description of the default models used by each of these codes that are relevant to intermediate-energy proton-induced reactions can be found in Fox \textit{et al.} \cite{Fox}. The relevant differences are in the optical models (Koning-Delaroche in TALYS and EMPIRE versus Becchetti-Greenlees in ALICE), the level density models (CT Fermi gas in TALYS, enhanced GSM in EMPIRE, and shell-dependent Kataria-Ramamurthy in ALICE) as well as the default preequilibrium models. TALYS implements a two-component exciton model, which has been parameterized with a global fit to emission spectra for $A \geq 24$ and incident energies between 7 and 200 MeV \cite{KONING200415}. The default preequilibrium model in EMPIRE is \texttt{PCROSS}, a more simplistic one-component exciton model, while ALICE makes use of the Hybrid Monte-Carlo Simulation (HMS) precompound decay model. Additionally, the treatment of angular momentum in ALICE is generally poorer than the other two codes, which tends to lead to inaccurate isomeric ratio predictions. Overall, all three codes struggled to predict many of the multi-particle-out channels, which would seem to indicate a deficiency in the preequilibrium models used by these codes. A selected parameter adjustment using the TALYS-2.0 code was performed in this work, which is described in Section \ref{reaction_modeling}, and the results are shown here for comparison.

\subsection{\ce{^{139}La}(p,6n)\ce{^{134}Ce} Cross Section}

Cerium-134 is a challenging isotope to quantify, due to its relatively weak gamma-emission probability, with the 0.209(15)\%, 130.4 keV gamma being the strongest emission \cite{A134}. Because the \ce{^{134}Ce} production cross section is generally quite high in the energy region explored in this work, it is not the counting statistics that are a challenge. Rather, it is the fact that in this photon energy range there is a significant Compton background for most types of HPGe detectors, including those used in this measurement, arising from higher energy gamma-rays scattering out of the detector. Additionally, above proton energies of approximately 60 MeV, the ($T_{1/2}=11.50$ day) product \ce{^{131}Ba} is produced, which has a 2.18(3)\% gamma emission at 133.617 keV, which may interfere with the 130.4 keV \ce{^{134}Ce} emission depending on the resolution of the detector.

Instead, the preferred signature for \ce{^{134}Ce} is the 5.04(20)\%, 604.721 keV gamma emitted by its decay product \ce{^{134}La}. Lanthanum-134 has a half-life of only 6.45 minutes, much shorter than the 3.16 day half-life of \ce{^{134}Ce}, which means it reaches secular equilibrium with \ce{^{134}Ce} after only a few hours. However, even this approach is not foolproof, as the ($T_{1/2}=17.7$ h) \ce{^{135}Ce} isotope produced via the (p,5n) channel emits multiple gamma-rays which overlap at this energy, meaning it must have time to sufficiently decay ($\approx 7$ days) before the \ce{^{134}La} decay can be observed without contamination. 

\begin{figure}[!htbp]
\includegraphics[width=\columnwidth]{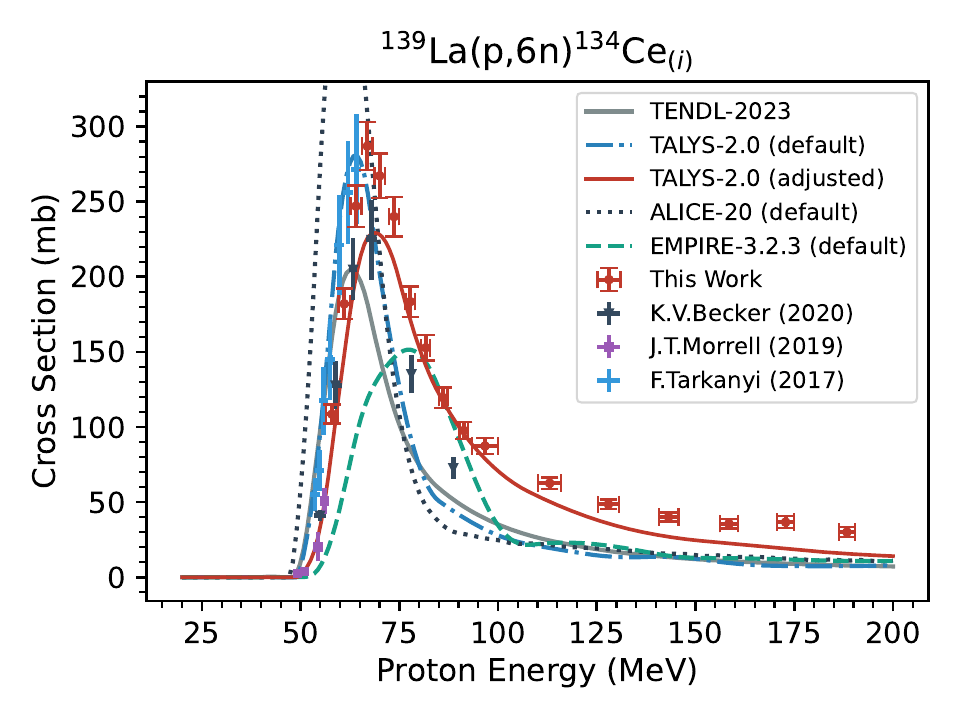}
\caption{Measured cross sections for the \ce{^{139}La}(p,6n)\ce{^{134}Ce} reaction.
}
\label{fig:134CE}
\end{figure}

The measured cross sections for \ce{^{134}Ce} production can be seen in Fig. \ref{fig:134CE} in comparison to other measurements and theoretical predictions. The most notable result is that the measured cross sections were much higher than theoretical predictions in the 100--200 MeV energy region. This may have implications for bulk production of this isotope for medical applications, as it would imply a higher production rate at facilities utilizing these higher energy beams than theoretical estimates had previously suggested. 

In general these results agree quite well with previous measurements. However, there is a significant difference in the two highest energy data points from Becker \textit{et al.}, which is not explained by the reported error margins. One potential source of discrepancy could be the choice of gamma line used for quantifying \ce{^{134}Ce}, as the Becker \textit{et al.} measurement relied solely upon the 0.209\%, 130.4 keV gamma branch discussed previously. While that gamma decay was observed in this work, it was supplemented with the \ce{^{134}La} decay gamma at 604.721 keV, which typically had lower peak fitting uncertainties than the 130.4 keV gamma.

In addition to under-predicting the cross section at high energy, the default models generally struggled to predict the centroid energy of the ``compound peak", though not as significantly as has been reported in previous versions of these codes \cite{Morrell_La}. We expect this is due to differences in the predicted neutron emission spectra arising from the different level densities and preequilibrium reaction models used by each code, which would affect how much energy is ``carried away" by these neutrons prior to the nucleus reaching a compound state. In the adjusted TALYS-2.0 model, this effect has been improved, though not completely corrected, with modifications to the default preequilbrium parameters, to be discussed in more detail below.

\subsection{\ce{^{139}La}(p,n)\ce{^{139}Ce} Cross Section}

\begin{figure}[!htbp]
\includegraphics[width=\columnwidth]{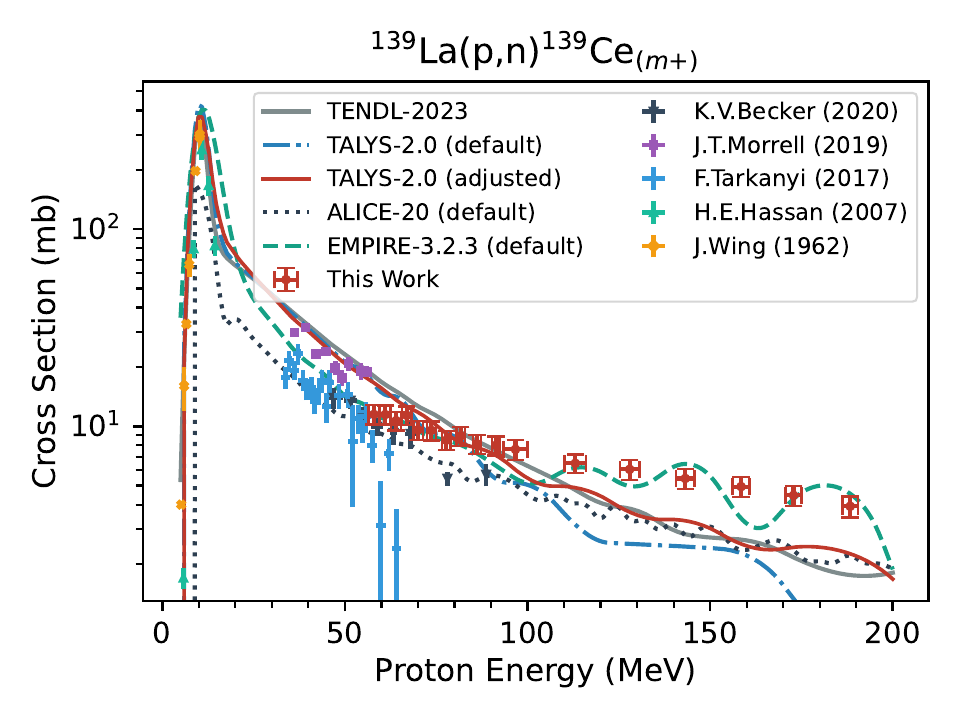}
\caption{Measured cross sections for the \ce{^{139}La}(p,n)\ce{^{139}Ce} reaction.
}
\label{fig:139CE}
\end{figure}

The product \ce{^{139}Ce} was measured via its 79.90(5)\% intensity, 165.86 keV gamma emission, which was clearly observable several days after the end-of-bombardment due to this isotope's relatively long half-life of 137.641(20) days. This also included feeding from the \ce{^{139m}Ce} isomer, which has a half-life of only 57.58(32) s, and decays via a 100\% isomeric transition to the ground state. The measured cross sections for the \ce{^{139}La}(p,n)\ce{^{139}Ce_{(m+)}} channel can be seen in Fig. \ref{fig:139CE}.

The results agree quite well with most of the existing measurements, although there is a large amount of scatter in these data. It is unclear why this might be the case, as the 165.86 keV gamma line is well-isolated from contaminants and quite strong, but could be attributed to poor counting statistics in these measurements if the lanthanum foil were not counted a sufficient length of time --- in our experiment the 165.86 keV line typically had a count rate below 1 s$^{-1}$. 

In general the reaction modeling is quite good, with ALICE generally performing best at low energy and EMPIRE being the best code at predicting the high-energy behavior. Similar to the (p,6n) channel, the cross section in the 100--200 MeV region was higher than most of the codes predicted, although by a far smaller percentage. This may also have significance for medical isotope production applications, as \ce{^{139}Ce} is the primary contaminant of concern when producing \ce{^{134}Ce}, due to its long half-life and inability to be chemically separated from \ce{^{134}Ce} \cite{Bailey2021, Bobba265355}. Therefore, this cross section sets a fundamental lower limit to the possible radiopurity with which \ce{^{134}Ce} can be produced.

\subsection{\ce{^{139}La}(p,3n)\ce{^{137m,g}Ce} Cross Sections}

\begin{figure}[!htbp]
    \sloppy
    \centering
    \subfloat{
        \centering
        \subfigimg[width=\columnwidth]{}{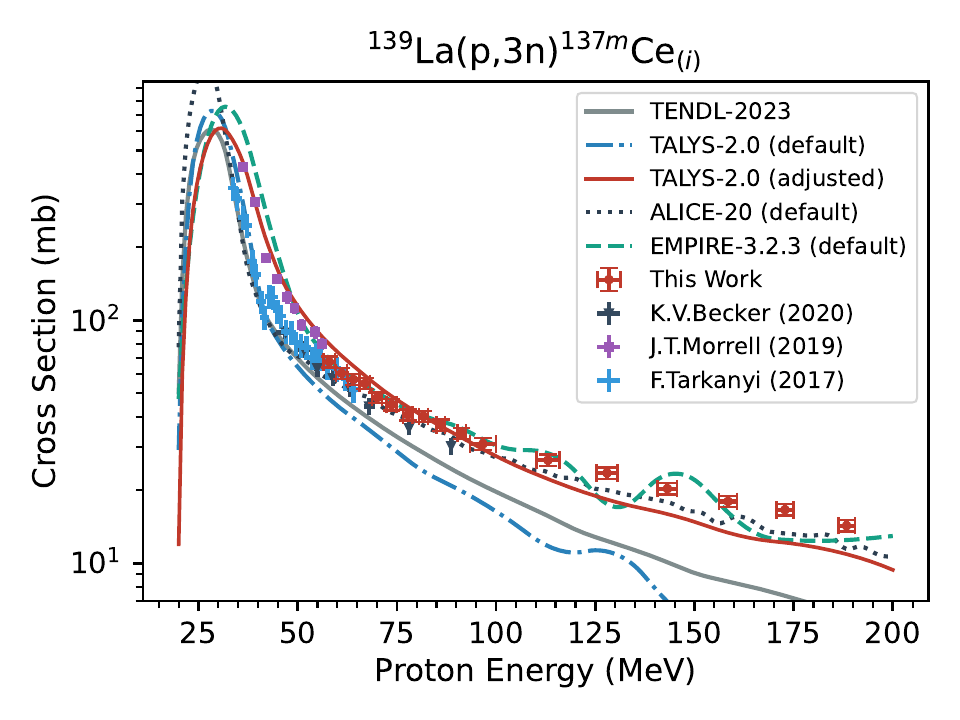}{100}
   \hspace{-10pt}}
    \\
    \subfloat{
        \centering
        \subfigimg[width=\columnwidth]{}{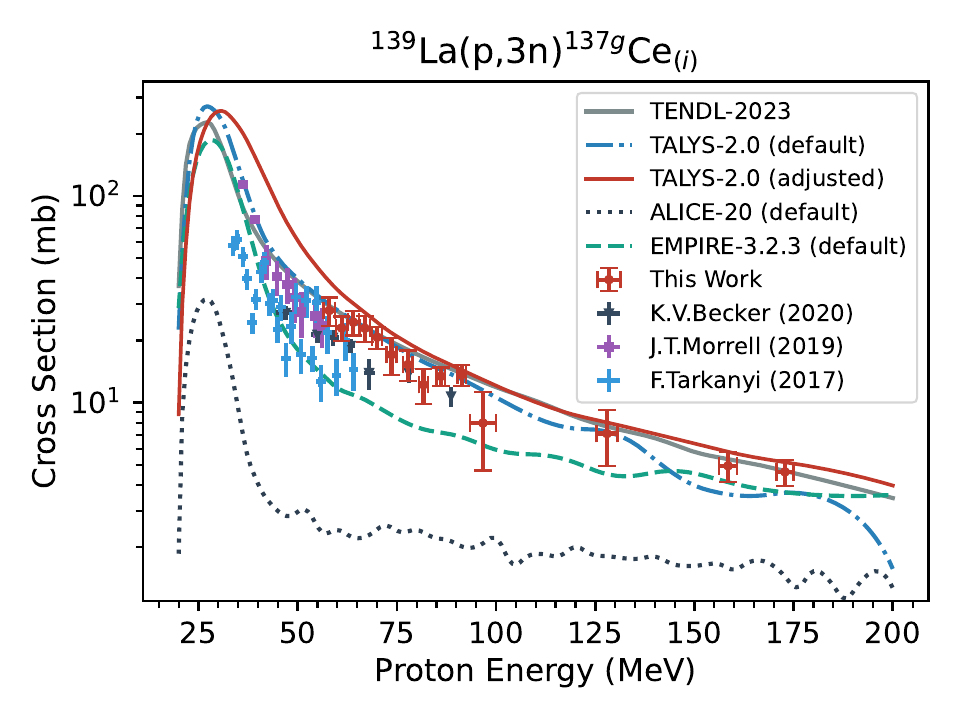}{100}
   \hspace{-10pt}}
    \caption{Measured cross sections for the \ce{^{139}La}(p,3n)\ce{^{137m}Ce} (top) and \ce{^{139}La}(p,3n)\ce{^{137g}Ce} (bottom) reactions.
    }
    \label{fig:137CE}
\end{figure}

Both the isomer ($J_{\pi}=11/2^-$) and ground state ($J_{\pi}=3/2^+$) of \ce{^{137}Ce} were independently observed in this work, formed through the (p,3n) reaction channel. The measured cross sections for the formation of these products are shown in Fig. \ref{fig:137CE}.

The \ce{^{137m}Ce} isomer decays with a 1.433(13) day half-life, mostly via an isomeric transition to the ground state, accompanied by a 254.29 keV gamma emission with an 11.1(4)\% branching ratio. This allowed the production of the isomer to be quantified relatively accurately. Unfortunately, the fact that the isomer feeds into the shorter-lived ($T_{1/2}=9.0(3)$ h) ground state makes the quantification of the ground state population somewhat difficult. This is compounded by the relatively weak gamma-emission probability of the ground state, with a 1.680(84)\% intense 447.15 keV gamma being the strongest emission. Distinguishing the independent \ce{^{137}Ce} ground state production from isomeric feeding required fitting spectra shortly (a few hours) after the end-of-bombardment, where the signal-to-noise ratio of the 477.15 keV gamma line was quite low due to the Compton continuum arising from many higher energy gamma decays. Therefore the uncertainty on this cross section was generally very large, and a few of the BNL data points were omitted as the uncertainty was greater than 50\%.


In both the isomer and ground state, there is good agreement with existing measurements. While there is a significant amount of fluctuation in the ground state data, this is generally within the large error bounds that arise from the difficulty described above. Most of the reaction modeling codes produced reasonable predictions of the \ce{^{137m}Ce} (isomer) population, with ALICE-20 giving the best result across the span of energies. However, ALICE greatly under-predicted the ground state population, and due to the rather large variability in the measured data for \ce{^{137g}Ce}, it is unclear which of the other models performed best. Because most of the (p,3n) cross section populates the isomer, the total cross section for this channel is relatively well-predicted by the models, with only ALICE greatly over-predicting the m/g (isomer-to-ground state) ratio. This is in accordance with a general observation of relatively poor angular momentum modeling in ALICE \cite{Fox, Arsenic_Fox}.

\subsection{\ce{^{139}La}(p,5n)\ce{^{135}Ce} Cross Section}

\begin{figure}[!htbp]
\includegraphics[width=\columnwidth]{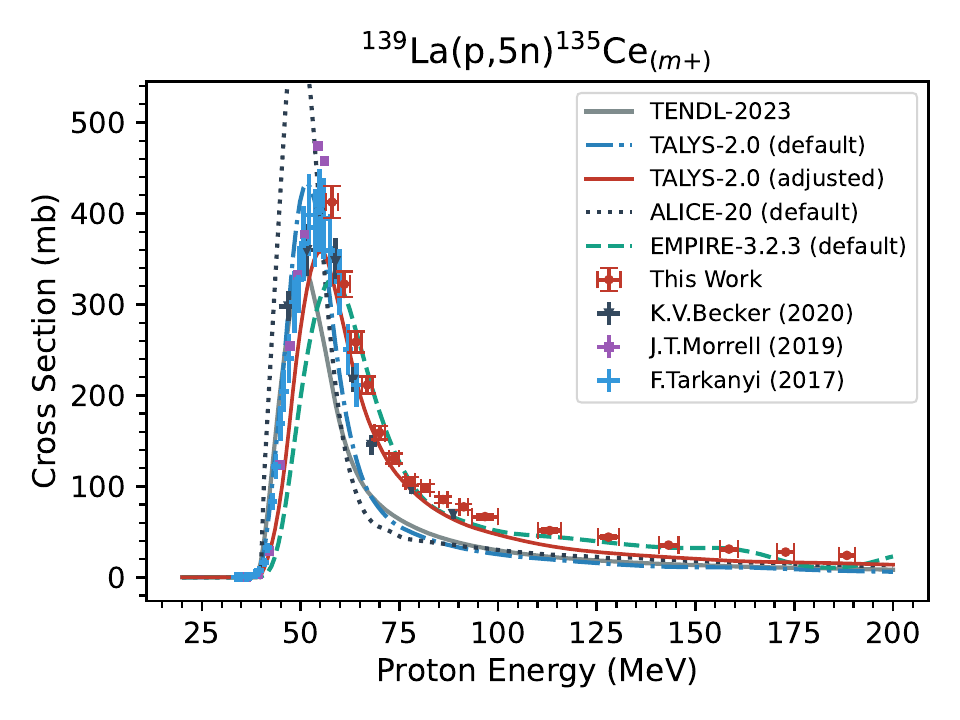}
\caption{Measured cross sections for the \ce{^{139}La}(p,5n)\ce{^{135}Ce} reaction.
}
\label{fig:135CE}
\end{figure}

The population of \ce{^{135}Ce} via the (p,5n) reaction channel was relatively well-quantified, due to a large number of intense gamma emissions ranging from 119.52 keV ($I_{\gamma}=1.30(9)$\%) to 1184.09 keV ($I_{\gamma}=1.09(5)$\%), six of which had a gamma branching ratio greater than 10\%. This cross section also includes the population of the \ce{^{135m}Ce} isomer, as its 20(1) second half-life was too short to be independently measured before it decayed with a 100\% isomeric transition to the ground state. The measured cross sections for this channel are plotted in Fig. \ref{fig:135CE}.

The results of this work show excellent agreement with previous measurements, likely due to the large production cross section and high accuracy with which this isotope's activity can be quantified. TALYS and ALICE show good agreement in the prediction of the compound peak, though ALICE somewhat over-estimates the maximum cross section. EMPIRE struggled to predict both the centroid energy and magnitude of the compound peak, although it has the best prediction above $\approx$70 MeV. This could be a result of better preequilibrium modeling parameters used in the EMPIRE exciton model. However, it could also be a numerical artifact arising from EMPIRE's limitations on the number of tracked ejectile particles, essentially forcing the cross section to be artificially higher at high energies.

For medical isotope production applications, \ce{^{135}Ce} ($T_{1/2}=17.7(3)$ h) is the longest-lived contaminant to \ce{^{134}Ce} production, aside from the much longer-lived \ce{^{139}Ce}, and cannot be chemically removed. However, because the \ce{^{135}Ce} half-life is shorter than \ce{^{134}Ce}, the radiopurity of \ce{^{134}Ce} can be improved by waiting for \ce{^{135}Ce} to decay away --- although the decay time cannot be so long that the \ce{^{134}Ce}/\ce{^{139}Ce} ratio is significantly impacted. Because of the trade-off between \ce{^{135}Ce} and \ce{^{139}Ce} impurity, it is essential to measure the magnitude and energy dependence of each of these cross sections if one is to optimize the \ce{^{134}Ce} production scenario.

\subsection{\ce{^{139}La}(p,7n)\ce{^{133m,g}Ce} Cross Sections}

\begin{figure}[!htbp]
    \sloppy
    \centering
    \subfloat{
        \centering
        \subfigimg[width=\columnwidth]{}{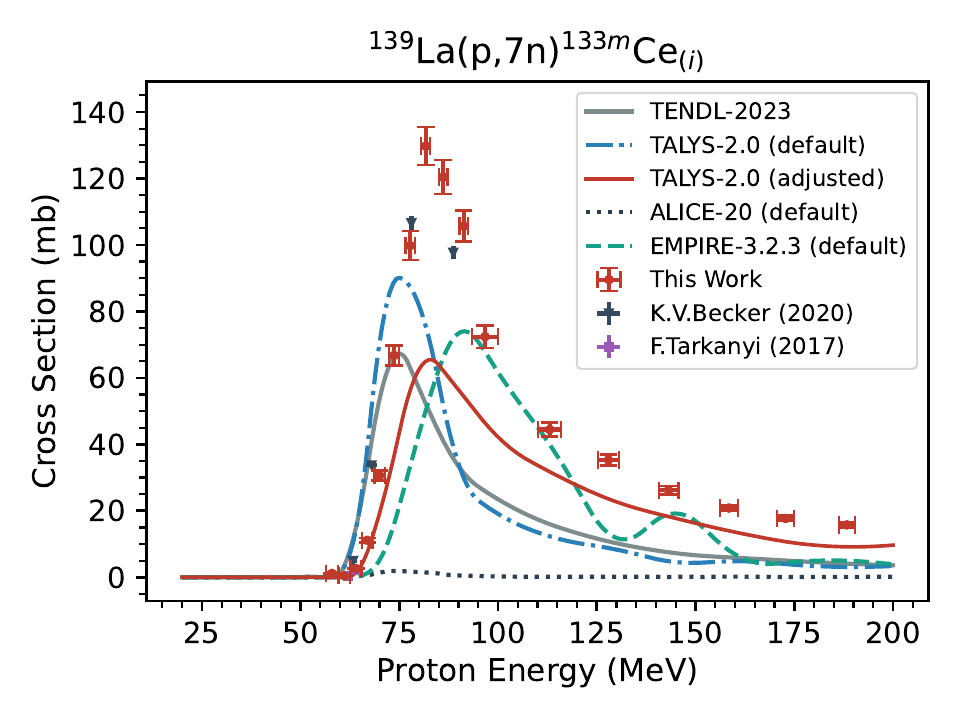}{100}
   \hspace{-10pt}}
    \\
    \subfloat{
        \centering
        \subfigimg[width=\columnwidth]{}{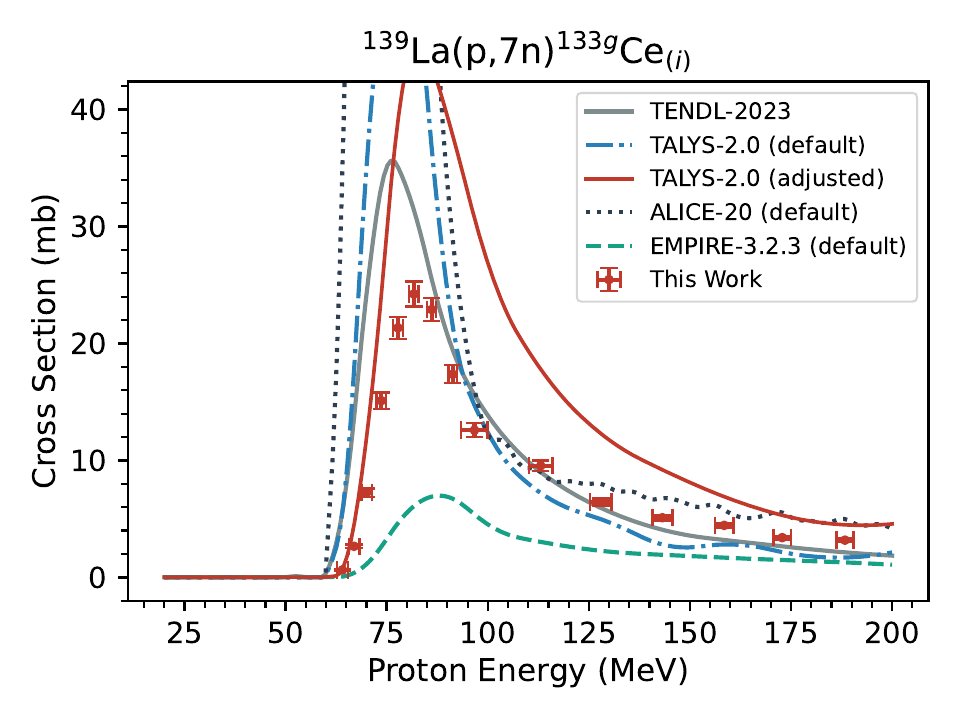}{100}
   \hspace{-10pt}}
    \caption{Measured cross sections for the \ce{^{139}La}(p,7n)\ce{^{133m}Ce} (top) and \ce{^{139}La}(p,7n)\ce{^{133g}Ce} (bottom) reactions.
    }
    \label{fig:133CE}
\end{figure}

The isomer ($J_{\pi}=9/2^-$) and ground state ($J_{\pi}=1/2^+$) of \ce{^{133}Ce} were accurately quantified via a large number of gamma decays in the ($T_{1/2}=5.1(3)$ h) isomer, and the 15.9(23)\% intense 76.9 keV gamma emission from the ($T_{1/2}=97(4)$ min) ground state. Unlike the (p,3n) reaction populating \ce{^{137}Ce}, the ground state of \ce{^{133}Ce} was relatively straightforward to independently quantify due to the isomer exclusively decaying to \ce{^{133}La}, with no population of the ground state of \ce{^{133}Ce}. The main challenge associated with these (p,7n) channels was to ensure consistency with the other products in the overlapping $A=133$ decay chains: \ce{^{133}La}, \ce{^{133m}Ba} and \ce{^{133g}Ba}. However, this did not have any impact on the accuracy of the \ce{^{133m,g}Ce} products themselves. The measured cross sections can be seen in Fig. \ref{fig:133CE}.

The experimental data on these channels are quite sparse, with only four measurements of \ce{^{133m}Ce} from Becker \textit{et al.} and two from T{\'a}rk{\'a}nyi \textit{et al.} \cite{Tarkanyi2017, BECKER202081}. However, where they overlap, the agreement is quite good. The consistency of the measurements in this and other channels would seem to indicate that the beam current determination and energy assignments of the various measurements agree quite well, and that any systematic differences are likely attributable to specifics of the activity determination (counting, peak-fitting, decay-corrections, \textit{etc.}) associated with individual isotopes, rather than some global difference.

For both \ce{^{133m}Ce} and \ce{^{133g}Ce}, no particular reaction model gives a satisfactory prediction of the cross section. Most of the models over-predict the ground state population and under-predict the isomer, except for EMPIRE which under-predicts them both. Once again, ALICE gives very poor predictions of the m/g ratio, with almost no population of the isomer. Also, the default codes give generally poor predictions of the centroid energy of the compound peak and the overall shape, which we likely attribute to the preequilibrium modeling.

\subsection{\ce{^{139}La}(p,10n)\ce{^{130}Ce} Cross Section}

While many other reaction channels were observed, one final product isotope that is particularly interesting is \ce{^{130}Ce}, resulting from the exclusive (p,10n) reaction on \ce{^{139}La}. We believe that this represents the first measurement of an independent, exclusive (p,10n) excitation function in this energy range in the literature. While there are two measurements of the \ce{^{124}Sn}(p,10n)\ce{^{115}Sb} reaction on tin targets enriched in \ce{^{124}Sn} from Alexandryan \textit{et al.} and Balabekyan \textit{et al.}, they were each only measured at one energy point --- 660 MeV and 3.65 GeV, respectively --- and therefore did not capture either the threshold of the reaction or the shape of its excitation function \cite{tin660, tin365}. Also, while there are several measurements of residual products resulting from (p,10n) reactions, they are either cumulative cross sections or include contributions from other target isotopes.

\begin{figure}[!htbp]
\includegraphics[width=\columnwidth]{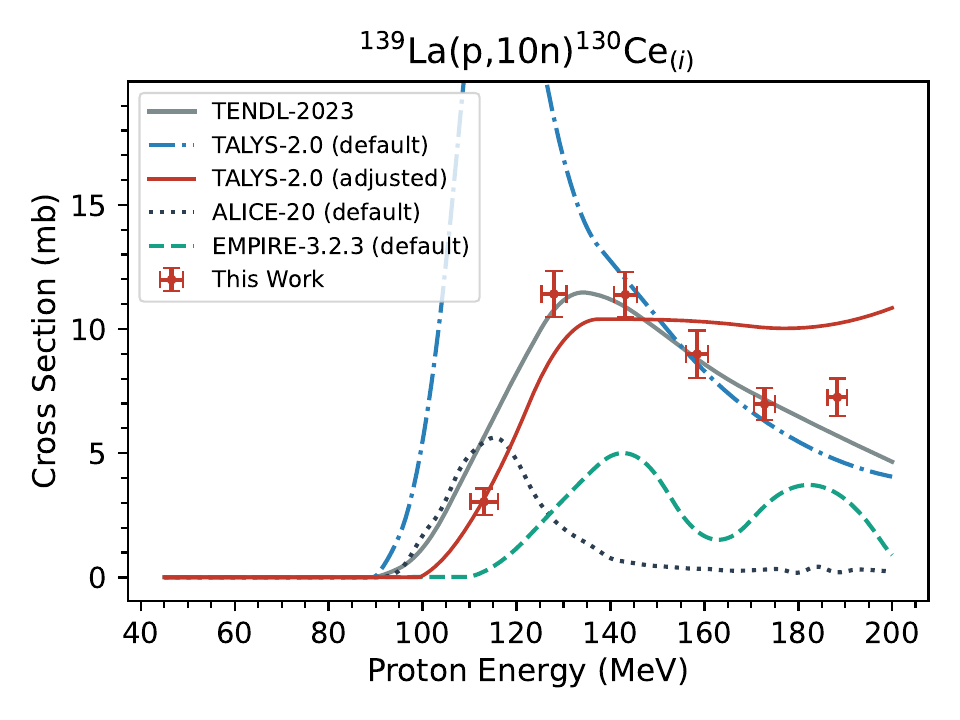}
\caption{Measured cross sections for the \ce{^{139}La}(p,10n)\ce{^{130}Ce} reaction.
}
\label{fig:130CE}
\end{figure}

While \ce{^{130}Ce} does have uniquely identifying gamma emissions, the initial counts after end-of-bombardment had contaminating gamma lines in the \ce{^{130}Ce} peaks, and therefore the 357.4 keV ($I_{\gamma}=81(4)$\%) gamma from \ce{^{130}La} was used to quantify the \ce{^{130}Ce} production rate instead. Because of the 8.7(1) min half-life of \ce{^{130}La}, after approximately 90 minutes of decay the entire \ce{^{130}La} population was attributable to \ce{^{130}Ce} decay, which was confirmed by fitting the decay of the 357.4 keV line with the 22.9(5) minute half-life of \ce{^{130}Ce}. The results of this measurement are plotted in Fig. \ref{fig:130CE}.

The results are significantly discrepant with the modeled cross sections, indicating that a significant parameter adjustment is necessary. Interestingly, the measurement showed almost no compound peak. Although it is possible there are not enough energy points to resolve the compound peak, the data seem to suggest that the compound mechanism is significantly damped in this channel, and presumably for higher (p,xn) channels as well.

\section{Charged Particle Reaction Modeling}
\label{reaction_modeling}

The comparison of the cross sections measured in this work to the three nuclear reaction modeling codes TALYS, EMPIRE and ALICE, as well as TENDL-2023, showed a clear need for a parameter adjustment from the default models. Additionally, while the adjusted parameter set from Fox \textit{et al.} fit the available data well, it did not extrapolate to our new measurements above 100 MeV. In this work, we apply a modified version of the TALYS-based residual product fitting procedure that was introduced in Fox \textit{et al.} to the new set of measured cross sections \cite{Fox}. In conjunction with literature data, these new measurements represent up to approximately 55\% of the entire non-elastic cross section of proton reactions on lanthanum, for a combined 30 reaction channels, ranging in energy from 4.5--188.4 MeV. This provided a unique opportunity to explore the nuclear reaction physics of this system. The motivation for using TALYS over other reaction modeling codes has been well described in Fox \textit{et al.}, but we will reiterate the benefits to TALYS being the accessibility of this code (and its underlying models), as well as the perception that the two-component exciton model employed by TALYS is the most physically justified out of the default models in the most prominently used codes (\emph{e.g.} EMPIRE, CoH, ALICE) \cite{Fox, TALYS, HERMAN20072655, CoH, ALICE}.

\subsection{Fitting Procedure}

The parameter adjustment procedure used to improve the predicted cross sections was inspired by the recent work of Fox \textit{et al.}, which proposed a particular sequence of parameters to be optimized, with the specific goal of fitting high-energy proton-induced reaction data \cite{Fox}. Our general approach can be outlined as follows:

\begin{enumerate}
\item Down-select data to strongest, independently measured reaction channels. We will refer to this as ``training" data.
\item Choose a goodness-of-fit metric to be optimized.
\item Optimize base-model selection parameters, \textit{e.g.,} \texttt{ldmodel}, \texttt{strength}, \texttt{deuteronomp} and \texttt{alphaomp}.
\item Adjust major optical model parameters, primarily \texttt{rvadjust} and \texttt{avadjust} for both neutrons and protons.
\item Adjust major preequilibrium exciton model parameters, primarily \texttt{M2constant}, \texttt{M2limit} and \texttt{M2shift}.
\item Adjust minor parameters, \textit{e.g.,} \texttt{Rgamma}, \texttt{Cstrip a}, \textit{etc.}
\item Iterate through steps 4--6 until convergence in the goodness-of-fit metric.
\item Validate using the unused (cumulative) channels excluded from step 1.
\end{enumerate}

There are three primary differences between the proposed approach and the work of Fox \textit{et al}. First, we adjusted the optical model parameters before other parameters, rather than after the preequilibrium optimization. This was motivated by the fact that optical model parameters determine the magnitude of the non-elastic cross section, which should be fixed before modifying parameters which affect channel-to-channel competition. Second, we greatly increased the number of parameters considered in the optimization, particularly the neutron optical model parameters and additional preequilibrium model parameters: specifically the \texttt{Rnunu}, \texttt{Rnupi}, \texttt{Rpinu} and \texttt{Rpipi} modifiers to the matrix element, as well as \texttt{Rgamma}. Finally, we did not perform any isotope-specific level density adjustments using the \texttt{ptable} parameter, however this would certainly be useful if one wished to provide the best interpolation of the data.

While a number of parameter optimization algorithms were considered for this work, the method with the best performance (best improvement over default TALYS) was actually the simplest method: sequentially optimizing one parameter at a time. Conventional regression and simulated annealing methods were also attempted, but failed to produce a satisfactory result. This method of sequentially optimizing the parameters for specific models within TALYS has a number of advantages over less physically-motivated evaluation methods. One of the most obvious advantages is that TALYS contains over 400 adjustable parameters, some of which are numerical, some are yes/no, some require a choice from a number of options, and some even require a file input. This means that one could not optimize \textit{all} parameters in TALYS using conventional methods such as linear regression, and even if one were to use a more global optimization approach such as a machine learning model, there are likely not enough experimental data available to train such a model on the full parameter space. Instead, a physically-motivated evaluation approach allows one to down-select to a much smaller subset of parameters, based on which parameters are relevant to the data being fit, as well as eliminating parameters which have been well-constrained by other experiments. Even then, there may still be dozens of relevant parameters to a data set like the stacked-foil measurement performed here, which may overwhelm a conventional regression algorithm, particularly when the effects of one parameter are highly correlated with another. We hypothesize that the strong sensitivities and correlations of our problem to these parameters are the reason that sequential parameter optimization provided the best results. 

Following the procedure outlined above, we selected our ``training" data set from the 18 largest cross sections in Table \ref{table:139LA_XS} which were independently measured, including products with contributions from an isomer but no feeding from other isotopes, \textit{i.e.,} cross sections subscripted (\textit{i}) or (\textit{m+}). The 12 remaining cross section channels were reserved for validation. In Fox \textit{et al.} two goodness-of-fit metrics were used: reduced-$\chi^2$ ($\chi^2_{\nu}$) weighted by the maximum cross section in a channel, and $\chi^2_{\nu}$ weighted by the integrated cross section in a channel. The reasoning behind using a weighted $\chi^2_{\nu}$ is that the larger reaction channels (either by maximum or integrated cross section) contain more information relevant to the models being optimized, and thus should be weighted heavier than the minor channels. Because we preferred to optimize on a single figure-of-merit, we opted to use the average of these two (max and integral) weights. Also, because \ce{^{134}Ce} is a reaction channel with particular value for applications, the weight for the \ce{^{134}Ce} channel was selectively doubled.

\begin{table}[!htbp]
\small
\caption{Summary of TALYS-2.0 base-model selection parameters.}
\setlength{\tabcolsep}{3pt}
\begin{tabular}{ccccc}
Parameter & Best & Sensitivity & Default & Options \Tstrut \Bstrut \\
\hline
\Tstrut \texttt{ldmodel} & 5 & 84.2 & 1 & 1--6  \\
\texttt{strength} & 10 & 46.1 & 9 & 1--10  \\
\texttt{deuteronomp} & 2 & 21.6 & 1 & 1--5 \\
\texttt{alphaomp} & 4 & 8.48 & 6 & 1--8 \\
\texttt{preeqspin} & 3 & 2.08 & 1 & 1--3 \\
\end{tabular}
\label{table:talys_choice_params}
\end{table}

The first set of parameters to be optimized were related to selection of the base-models, such as level density and gamma strength. The sensitivity to each base-model parameter was first calculated, in order to decide if the parameter should be included, as well as to inform the optimization order. Because these are discrete parameters, the ``sensitivity" was calculated as the maximum percentage change (not necessarily improvement) in weighted-$\chi^2_{\nu}$, out of the whole set of options. The base-model parameters were then optimized in order of highest sensitivity. The resulting ``best" model selection, range of options, default options and sensitivities are given for each of the 5 considered parameters in Table \ref{table:talys_choice_params}. It was found that \texttt{ldmodel 5} and \texttt{strength 10} were the best options, corresponding to the Gogny Hartree-Fock-Bogoluybov (HFB) level densities and Skyrme (HFB) + QRPA corrected strength tables, respectively \cite{talys_level_density, talys_strength}. Also, note that each model was run with the option \texttt{bins 0}, which scales the number of excitation energy bins as the incident energy increases.

\begin{table}[!htbp]
\small
\caption{Summary of optical model parameters explored in TALYS-2.0 modeling. Parameters at the limits of the constraints are indicated with a $*$. Defaults for all parameters were 1.}
\begin{tabular}{cccc}
Parameter & Best & Sensitivity & Constraints \Tstrut \Bstrut \\
\hline
\Tstrut \texttt{rvadjust n} & 0.977 & 4523 & 0.95--1.05  \\
\texttt{avadjust n} & 1.076 & 291 & 0.75--1.25  \\
\texttt{rvadjust p} & 0.9$^*$ & 59.1 & 0.9--1.1 \\
\texttt{avadjust p} & 1.0118 & 30.8 & 0.7--1.3 \\
\texttt{w1adjust n} & 4$^*$ & 5.79 & 0.25--4 \\
\texttt{w2adjust n} & 1.51 & 8.92 & 0.25--4 \\
\texttt{w1adjust p} & 4$^*$ & 18.5 & 0.25--4 \\
\texttt{w2adjust p} & 1.83 & 19.3 & 0.25--4 \\
\end{tabular}
\label{table:talys_optical_params}
\end{table}

Following the base-model selections, the continuous-variable parameters were optimized iteratively, beginning with the optical model parameters. For the optical model and preequilibrium parameters, the sensitivity was calculated as $d\chi^2_{\nu}/dp$, or the ratio of percent change in weighted-$\chi^2_{\nu}$ to the percent change in the parameter --- evaluated at the default value for that parameter. This was used to inform the order in which the parameters were optimized, as well as the range of allowed values. Parameters with very high sensitivities, such as \texttt{rvadjust n}, were constrained to be much closer to the default values. Additionally, because \texttt{rvadjust} and \texttt{avadjust} correspond to the width and diffuseness of the optical-model well potential, which have fairly well-constrained systematics, the allowed range of values was more limited than for other parameters \cite{KONING2003231}. The resulting optimized optical model parameters, as well as the associated sensitivities and parameter constraints, are given in Table \ref{table:talys_optical_params}. In most cases the imposed constraints were more strict than those required by TALYS. Because of the relatively large sensitivity of these calculations to optical model parameters, the resulting ``best" values were very near the default values, with the exception of \texttt{w1adjust} for both protons and neutrons, which relate to the magnitude of the imaginary component of the optical model potential, primarily impacting the elastic to non-elastic cross section ratio.

\begin{figure*}[!htbp]
    \sloppy
    \centering
    \subfloat{
        \centering
        \subfigimg[width=0.496\textwidth]{}{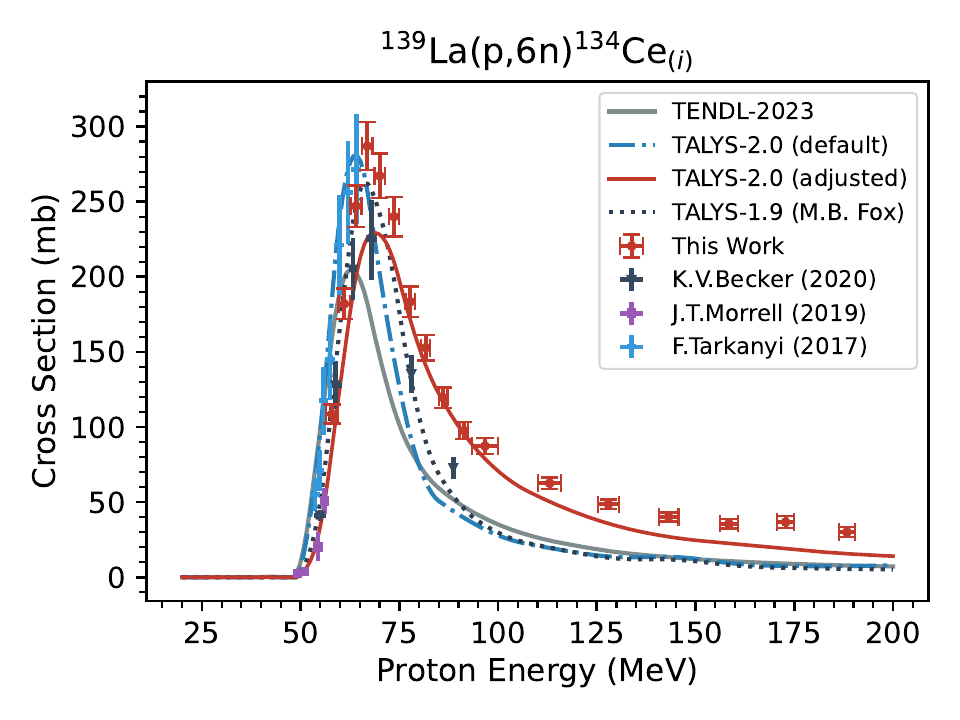}{50}
        \subfigimg[width=0.496\textwidth]{}{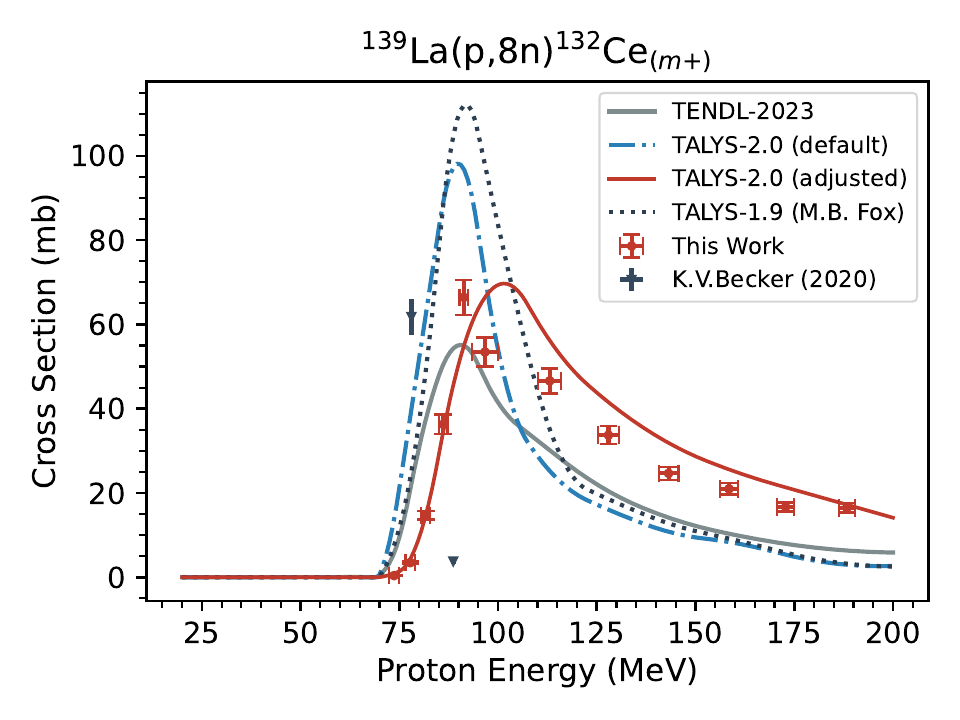}{50}
   \hspace{-10pt}}
    \caption{Comparison of predicted and observed cross sections for the \ce{^{134}Ce} (left) and \ce{^{132}Ce} (right) products, which were used as training data for the parameter adjustments.
    }
    \label{fig:training_plots}
\end{figure*}

\begin{table}[!htbp]
\small
\caption{Summary of preequilibrium parameters explored in TALYS-2.0 modeling. Parameters at the limits of the constraints are indicated with a $*$. Defaults for \texttt{Rnunu} and \texttt{Rgamma} were 1.5 and 2.0, respectively, all others were 1.}
\begin{tabular}{cccc}
Parameter & Best & Sensitivity & Constraints \Tstrut \Bstrut \\
\hline
\Tstrut \texttt{M2constant} & 1.055 & 134 & 0.2--5  \\
\texttt{M2shift} & 1.196 & 109 & 0.2--5  \\
\texttt{M2limit} & 1.926 & 37.1 & 0.2--5 \\
\texttt{Rnunu} & 1.66 & 33.3 & 0.1--10 \\
\texttt{Rnupi} & 0.16 & 35.7 & 0.1--10 \\
\texttt{Rpinu} & 2.21 & 38.0 & 0.1--10 \\
\texttt{Rpipi} & 0.1$^*$ & 6.65 & 0.1--10 \\
\texttt{Rgamma} & 0.1$^*$ & 0.1 & 0.1--10 \\
\texttt{Cstrip a} & 0.1$^*$ & 4.2 & 0.1--10 \\
\end{tabular}
\label{table:talys_preeq_params}
\end{table}

The same approach was taken for optimizing the preequilibrium parameters, with the results shown in Table \ref{table:talys_preeq_params}. The most significant change (from defaults) was to \texttt{M2limit}, which affects the asymptotic behavior of the exciton transition rates. This seemed to be necessary to account for the significant enhancement (over default predictions) of the preequilibrium tail seen in most of the (p,xn) channels, which occurred above 100 MeV. The effects of this are quite apparent in the (p,6n) and (p,8n) cross sections plotted in Fig. \ref{fig:training_plots}, where almost every other prediction is a factor of 2--5 low in this energy region. Other significant changes were to \texttt{Rgamma}, which is a competing mechanism for de-excitation to neutron emission, as well as to \texttt{Cstrip a}, which affects the magnitude of the preequilibrium contributions to (p,$\alpha$) channels.

The completed parameter adjustment resulted in significant improvements to cross section predictions in the ``training" data set, over default predictions. The resulting improvement in the weighted-$\chi^2_{\nu}$ figure-of-merit can be seen in Table \ref{table:chi2_comparison}, in comparison to TENDL-2023, default predictions from TALYS-2.0, EMPIRE-3.2.3, ALICE-20 as well as the parameter adjustment resulting from the work of M.B. Fox \textit{et al.}, which was calculated with TALYS-1.9 as this was the TALYS version used in that work \cite{Fox}. While the Fox adjustment performed well on the data that was available at the time, that data did not include any energies above 100 MeV, and unfortunately as a result did not extrapolate well to the new 100--200 MeV measurements collected in this work. However, the Fox parameters still represent an improvement over default predictions. 

\begin{table}[!htbp]
\small
\caption{Summary of TALYS-2.0 modeling results.}
\begin{tabular}{ccc}
Source & Training $\chi^2_{\nu}$ & Validation $\chi^2_{\nu}$  \Tstrut \Bstrut \\
\hline
\Tstrut TENDL-2023 & 213 & 200 \\
TALYS-2.0 default & 1980 & 2302 \\
TALYS-2.0 adjusted & 48.1 & 1663 \\
TALYS-1.9 (M.B. Fox) & 267 & 2758 \\
EMPIRE-3.2.3 default & 433 & 1189 \\
ALICE-20 default & 3406 & $2.64\times 10^4$ \\
\end{tabular}
\label{table:chi2_comparison}
\end{table}

\begin{figure*}[!htbp]
    \sloppy
    \centering
    \subfloat{
        \centering
        \subfigimg[width=0.496\textwidth]{}{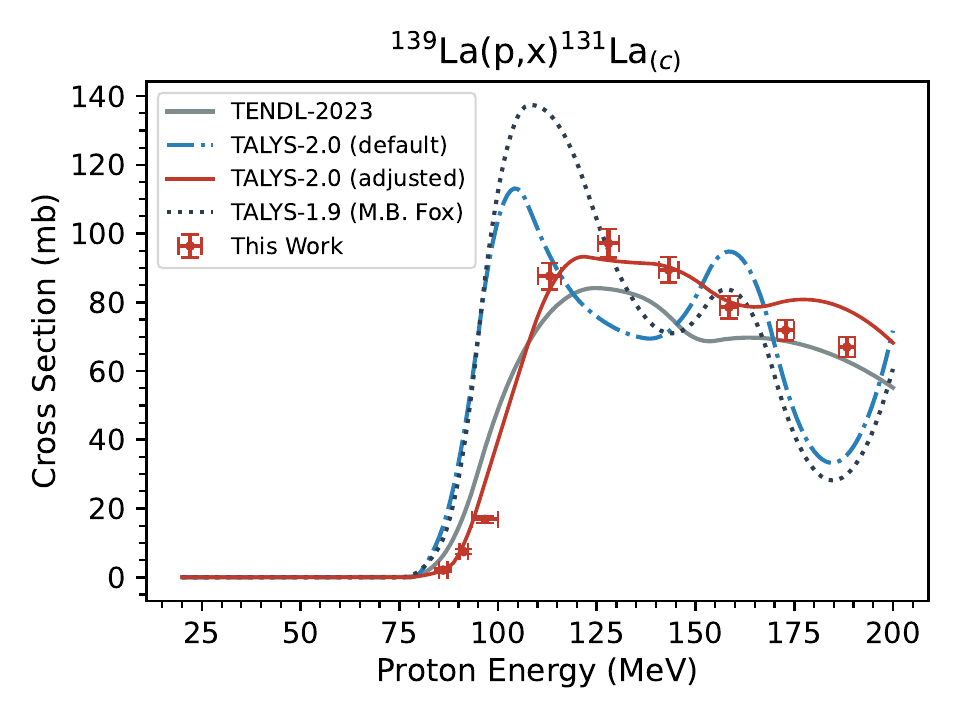}{50}
        \subfigimg[width=0.496\textwidth]{}{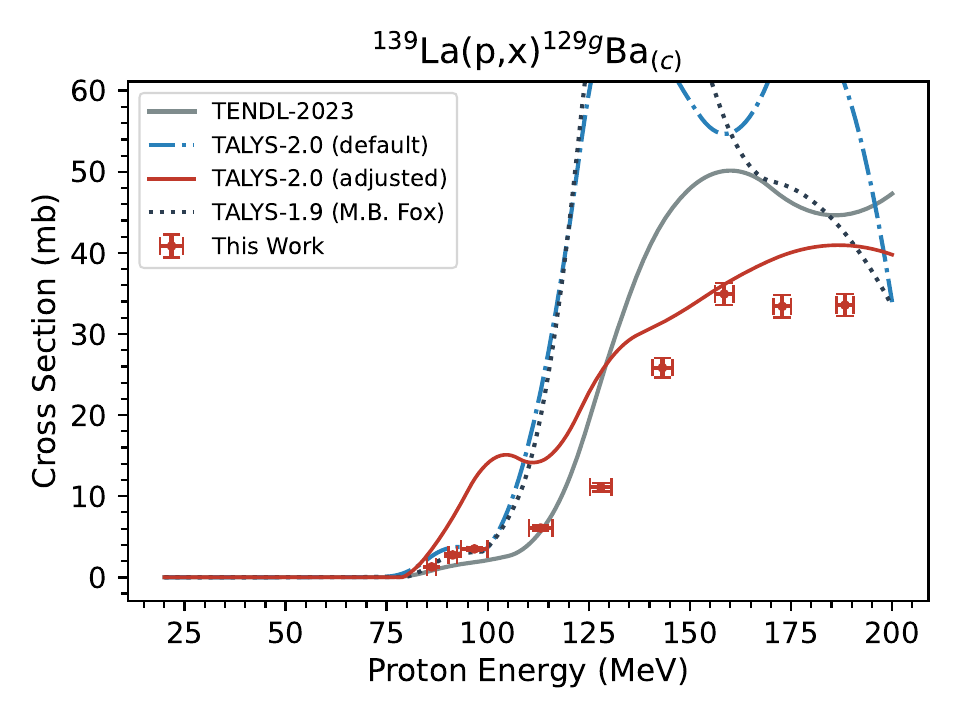}{50}
   \hspace{-10pt}}
    \caption{Comparison of predicted and observed cross sections for the \ce{^{131}La} (left) and \ce{^{129g}Ba} (right) products, which were used to validate the parameter adjustments.
    }
    \label{fig:validation_plots}
\end{figure*}

\begin{figure*}[!htbp]
    \sloppy
    \centering
    \subfloat{
        \centering
        \subfigimg[width=0.329\textwidth]{}{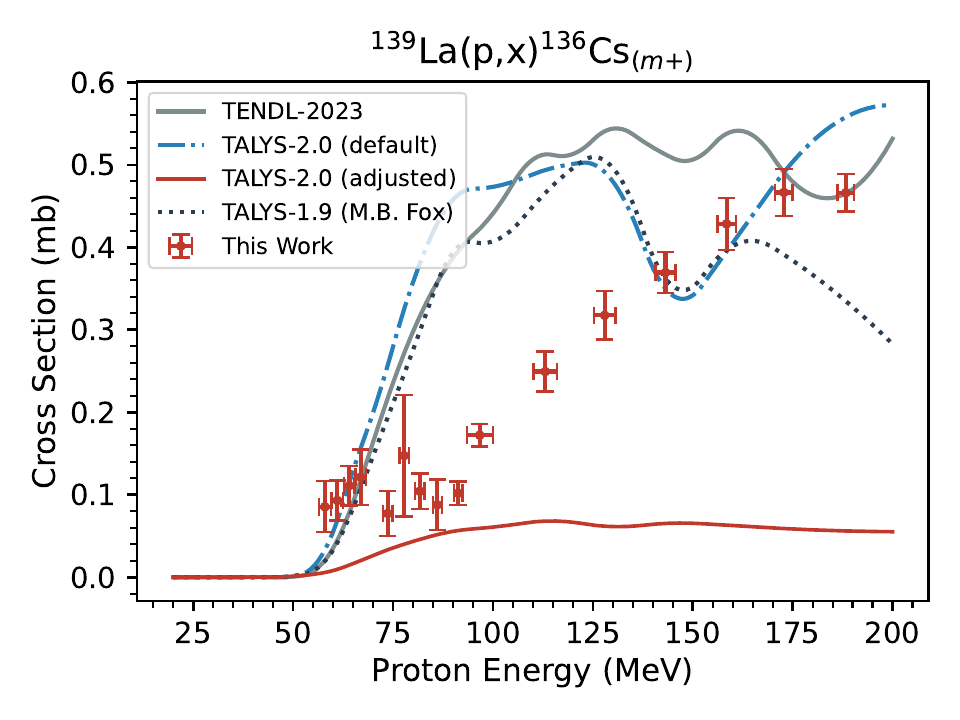}{33}
        \subfigimg[width=0.329\textwidth]{}{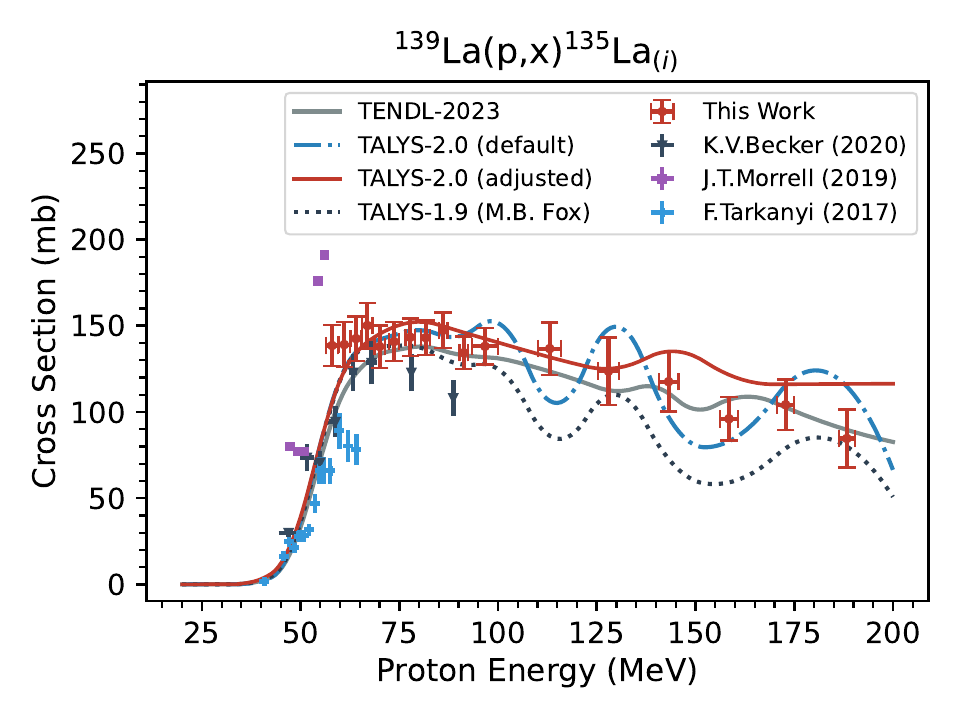}{33}
        \subfigimg[width=0.329\textwidth]{}{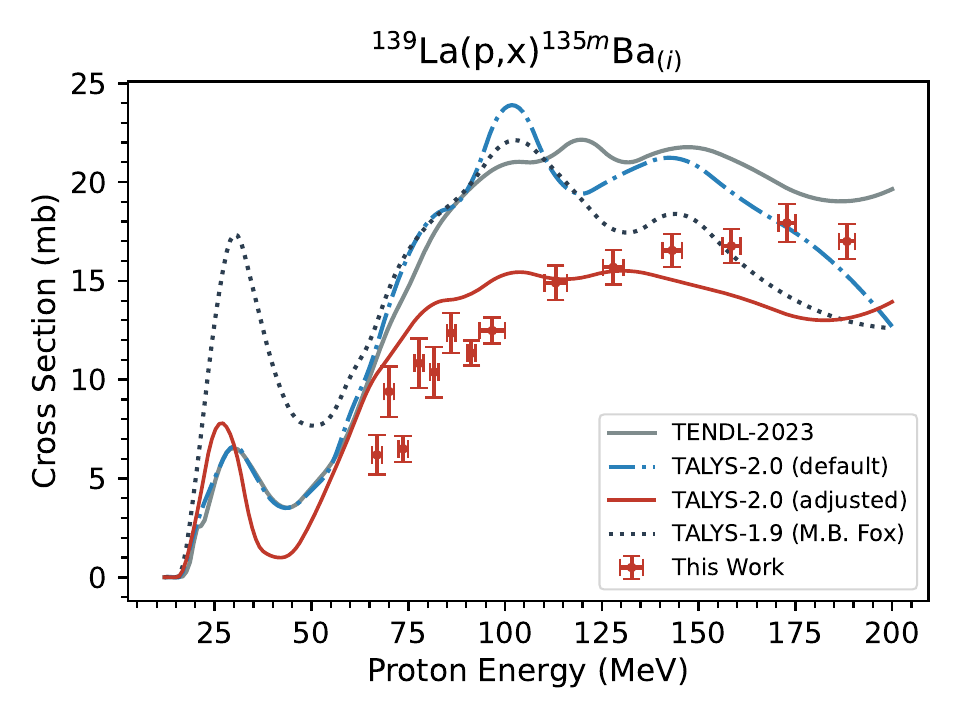}{33}
   \hspace{-10pt}}
    \\
    \subfloat{
        \centering
        \subfigimg[width=0.329\textwidth]{}{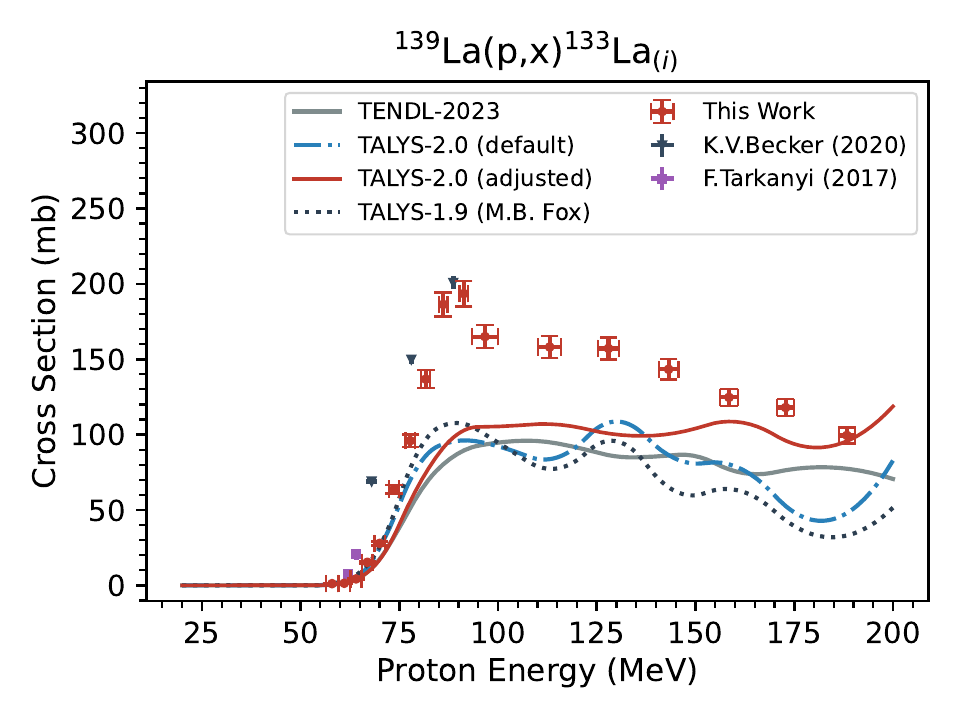}{33}
        \subfigimg[width=0.329\textwidth]{}{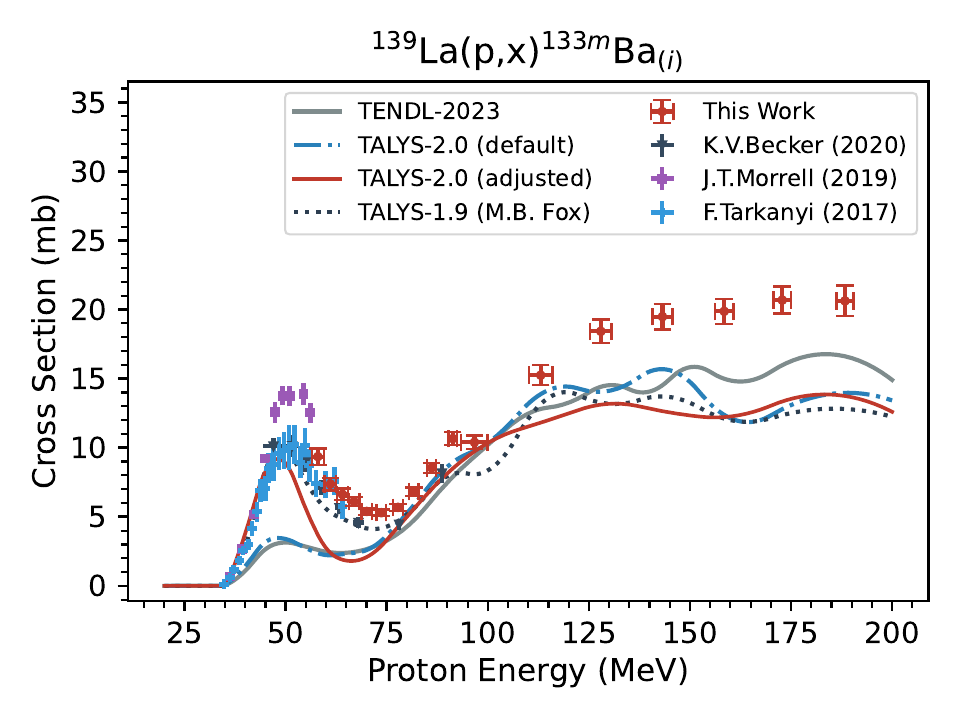}{33}
        \subfigimg[width=0.329\textwidth]{}{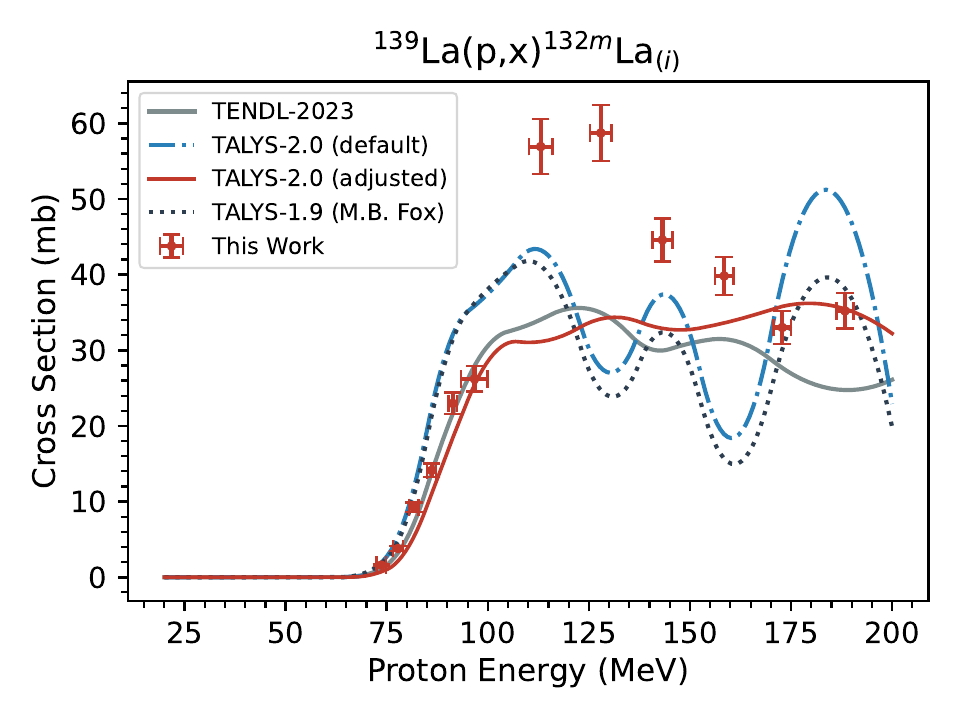}{33}
   \hspace{-10pt}}
    \\
    \subfloat{
        \centering
        \subfigimg[width=0.329\textwidth]{}{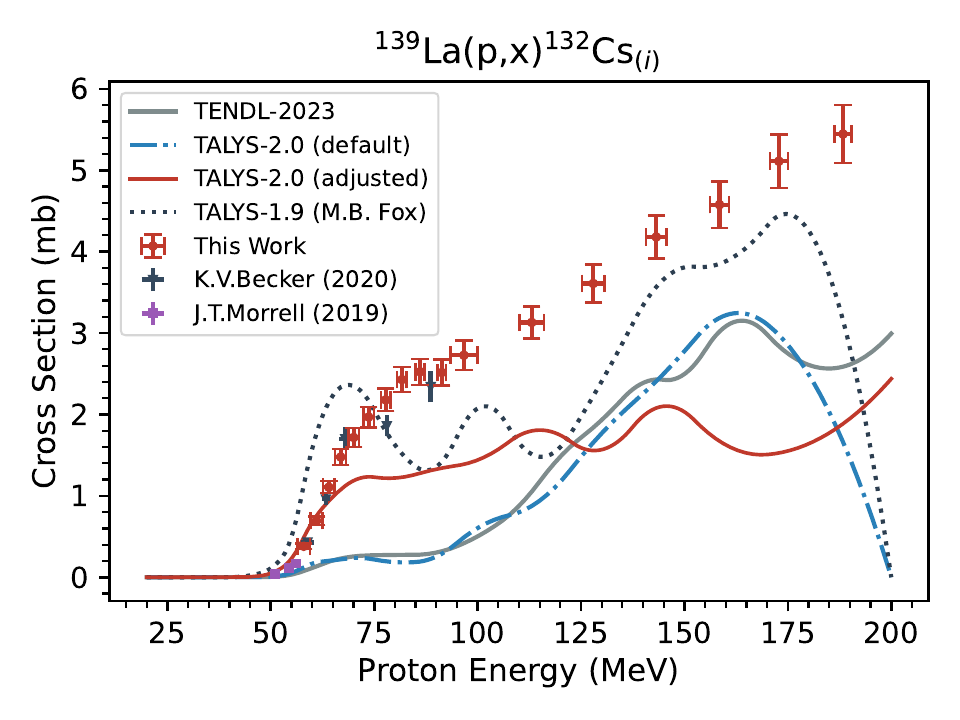}{33}
        \subfigimg[width=0.329\textwidth]{}{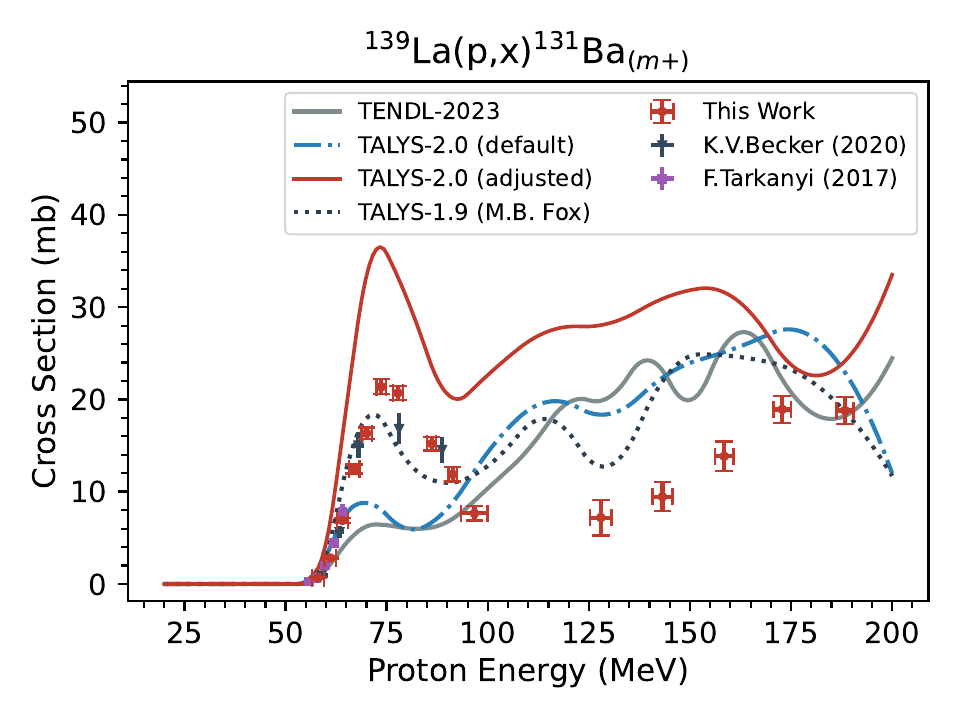}{33}
        \subfigimg[width=0.329\textwidth]{}{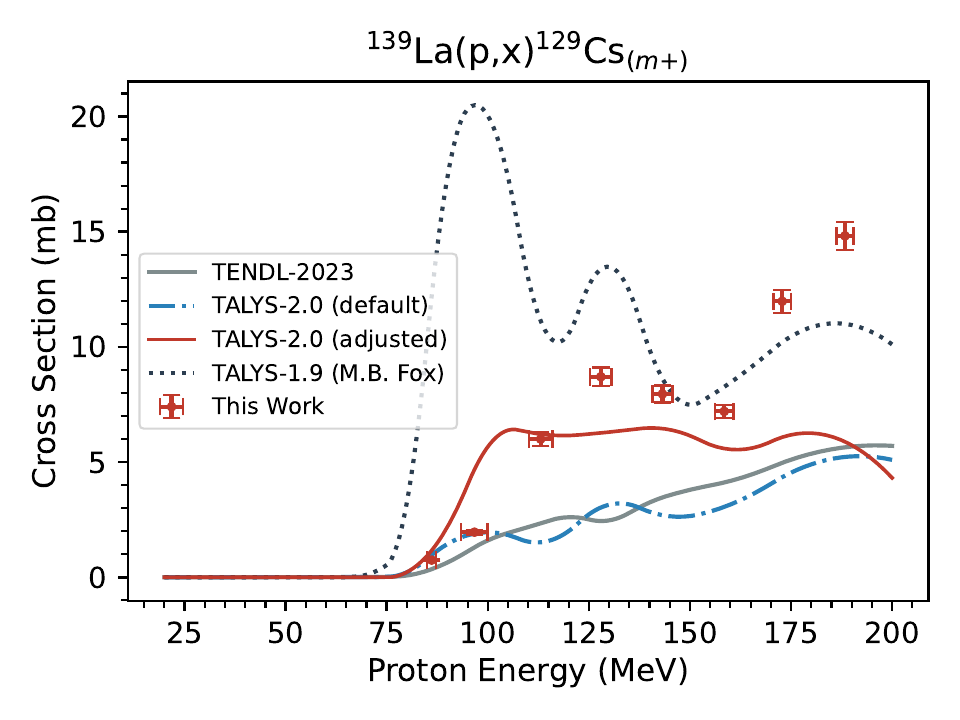}{33}
   \hspace{-10pt}}
    \caption{Comparison of measured cross sections to TENDL-2023 and TALYS-2.0 predictions using default parameters, the parameter adjustments of Fox \textit{et al.}, and the adjustments performed in this work \cite{Fox}. 
    }
    \label{fig:other_xs}
\end{figure*}

We also performed a validation of the parameter adjustments against the various other predictions, using the unused ``validation" data set, mostly comprising the cumulative channels. The weighted-$\chi^2_{\nu}$ for the validation data are also given in Table \ref{table:chi2_comparison}, where the weights were again determined as the average of the two weighting methods (max and integral) from the Fox \textit{et al.} work \cite{Fox}. Two of the 12 validation channels are also plotted in Fig. \ref{fig:validation_plots}, with additional results plotted in Figs. \ref{fig:other_xs} and \ref{fig:additional_1}. In general, the fit to the validation data set was much worse than the training set, as one might expect. However, the fit still represents an improvement over all of the modeling codes run with default values, as well as over the Fox adjustment (for reasons previously discussed). Our adjustment performed worse than the TENDL-2023 evaluation on the validation data, which seems to be attributed to a significant over-prediction of many barium and iodine channels. This could be due to over-fitting, \textit{i.e.,} using too many parameters in our adjustment, or could be due to making adjustments that were too large and are therefore non-physical. While this does highlight the challenges of performing reaction evaluations on incomplete data sets, and the need for more work on this topic, the improvement over default predictions for both training and validation data sets implies some amount of physical consistency in our adjustments.

\section{Summary and Conclusions}

In this work we conducted two stacked-foil irradiations --- at the LANL Isotope Production Facility and the BNL Brookhaven Linac Isotope Producer --- resulting in 30 measurements of \ce{^{139}La}(p,x) cross section channels in the 55--200 MeV proton energy range. Many of these cross sections represent the first observation of certain reaction products, and for all of the observed products this represents the first set of measurements above 100 MeV. Additionally, the fidelity of certain channels between 70--100 MeV was improved with measurements at additional proton energies, notably \ce{^{134}Ce}. The \ce{^{134}Ce} product represents a significant interest to the medical isotope production community. The reported finding that the cross section was higher than predicted above 100 MeV is likely to be of interest for that application. Also, we believe the \ce{^{139}La}(p,10n)\ce{^{130}Ce} reaction measured in this work to be the first measurement of an exclusive (p,10n) excitation function reported in the literature. In general, the significant discrepancies between the observed cross sections and default model predictions suggests a need to use caution when relying upon such predictions for applications where little or no experimental data exist.

The cross sections measured in this work represent a significant proportion of the non-elastic cross section, which we estimate to be as high as 55\% of the entire non-elastic cross section. The extensive nature of this data set enabled a comprehensive parameter adjustment for the p+\ce{^{139}La} system. Using the TALYS-2.0 nuclear reaction modeling code, parameter adjustments were made to the optical model and to the two-component exciton model for preequilibrium. The adjustment procedure resulted in an improvement in the overall fit to the measured data, which we believe to be physically consistent based on an improvement in cumulative reaction channels which were not used as part of the data set for the parameter optimization. The results corroborate previous findings suggesting the need to incorporate residual product excitation functions into exciton model parameterization studies, particularly at high energy \cite{Fox}.

\backmatter

\bmhead{Acknowledgements}

We wish to acknowledge our thanks to the operators of the LANL IPF for their assistance and support during the LANL irradiation: Anthony Koppi, Ross Capon and Nathan Kollarik. Additionally, we thank the BNL BLIP operators Henryk Chelminski and David O'Rourke. We are grateful to Ben Stein for the use of the LANL glovebox for the preparation and storage of the lanthanum foils. We acknowledge Deepak Raparia, head of the Pre-Injector Systems group at BNL-CAD for supporting the BLIP beam tuning. We would also like to thank the LANSCE Accelerator Operations support staff, and acknowledge the support from the LANSCE radiological control technicians. Additionally, we are grateful to Patrick Sullivan and Vicki Litton for their support with radiological controls at BNL.

\section*{Declarations}

\begin{itemize}
\item \textbf{Funding:} This research is supported by the U.S. Department of Energy Isotope Program, managed by the Office of Science for Isotope R\&D and Production. This work was carried out under Lawrence Berkeley National Laboratory (Contract No. DE-AC02-05CH11231), Los Alamos National Laboratory (Contract No. 89233218CNA000001) and Brookhaven National Laboratory (Contract No. DEAC02-98CH10886).
\item \textbf{Conflict of interest:} The authors have no competing interests to declare that are relevant to the content of this article.
\item \textbf{Data availability:} The analysis code used to convert the spectra into cross sections, as well as the gamma-ray spectra and all other raw data generated during this work, have been stored in a LANL archive, and are available upon request to the corresponding author. Upon publication, the cross section data will be incorporated into EXFOR \cite{EXFOR}.
\end{itemize}

This material is approved for public release; distribution unlimited (LA-UR-24-21320).

\begin{appendices}
\section{Cross Sections}
\label{additional_cross_sections}

\begin{figure*}[!htbp]
    \sloppy
    \centering
    \subfloat{
        \centering
        \subfigimg[width=0.496\textwidth]{}{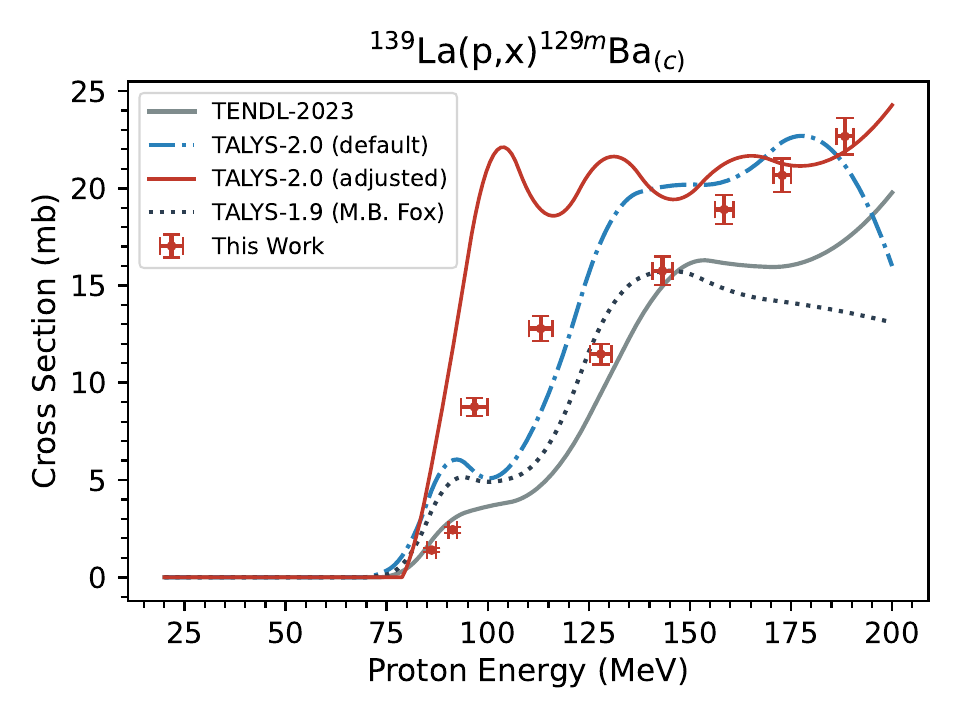}{50}
        \subfigimg[width=0.496\textwidth]{}{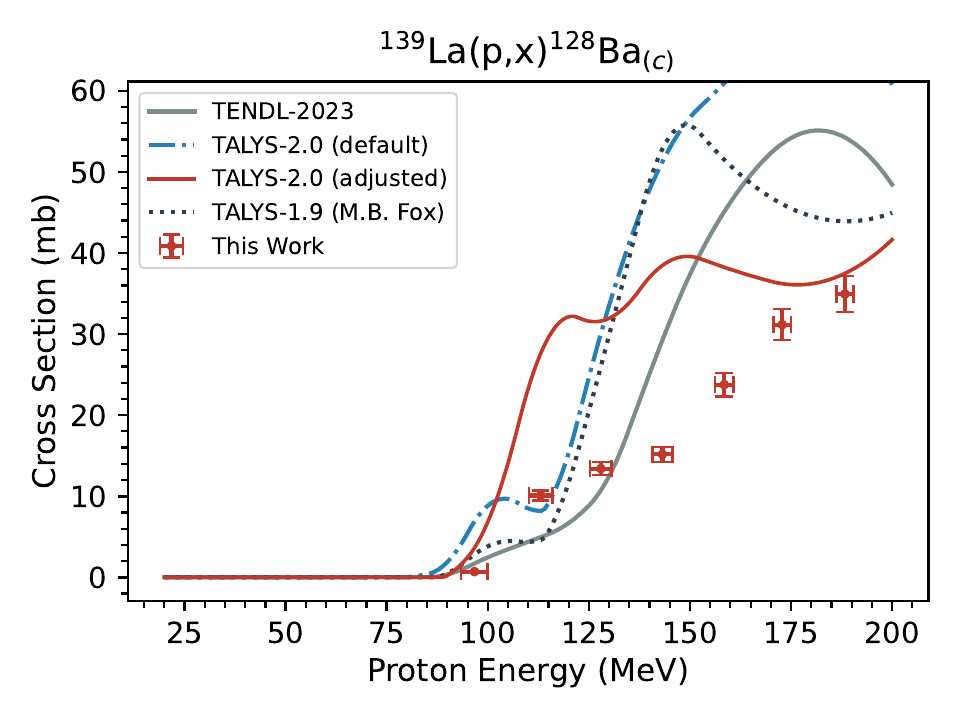}{50}
   \hspace{-10pt}}
    \\
   \subfloat{
        \centering
        \subfigimg[width=0.496\textwidth]{}{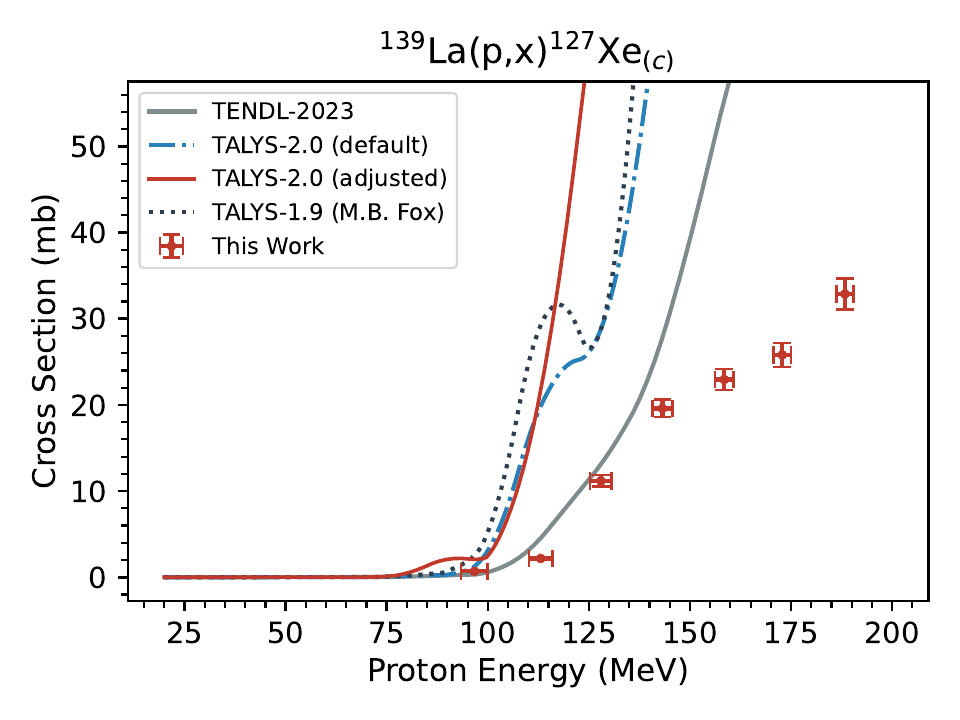}{50}
        \subfigimg[width=0.496\textwidth]{}{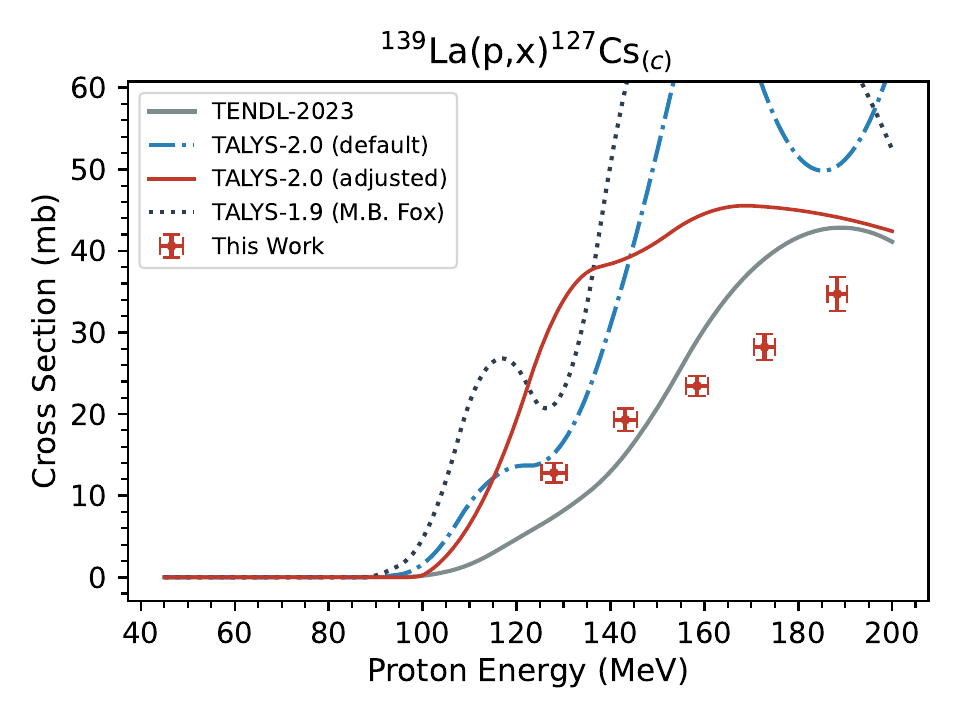}{50}
   \hspace{-10pt}}
    \\
    \subfloat{
        \centering
        \subfigimg[width=0.496\textwidth]{}{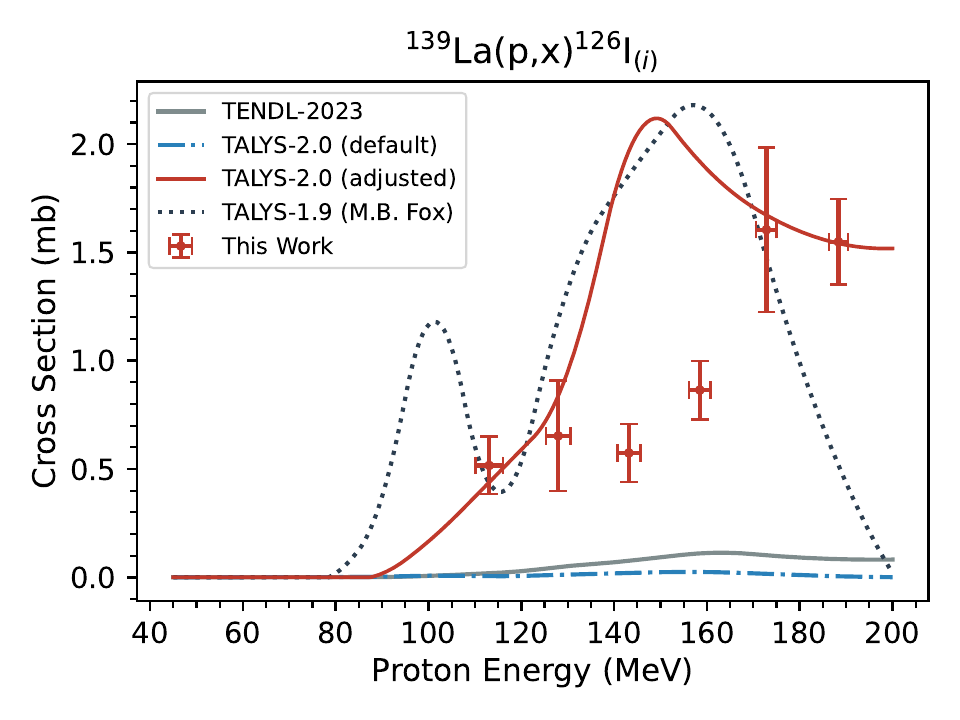}{50}
        \subfigimg[width=0.496\textwidth]{}{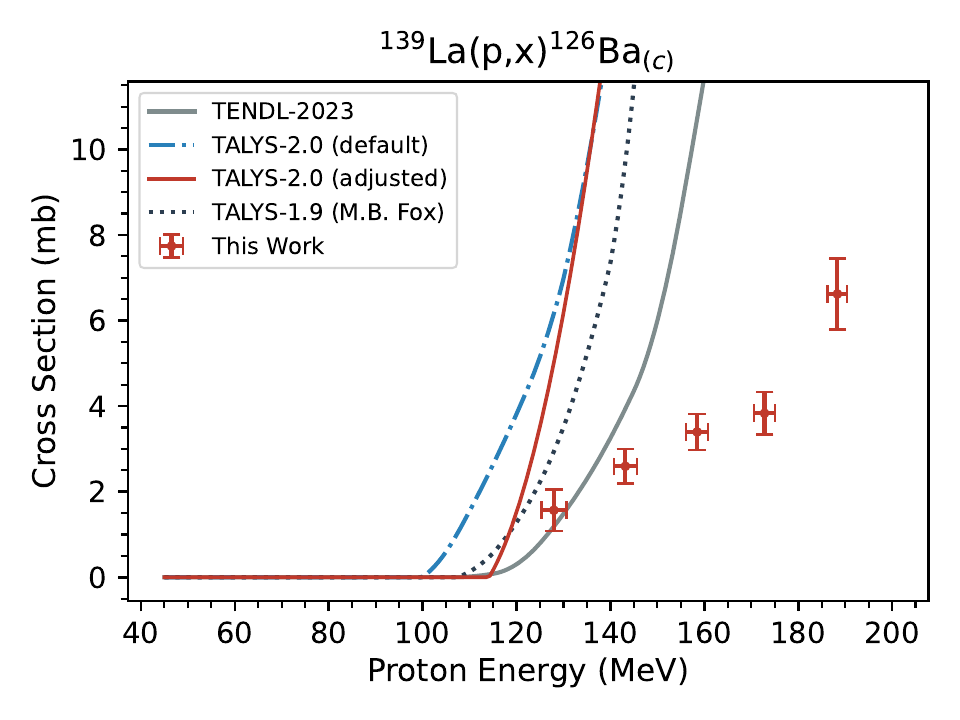}{50}
   \hspace{-10pt}}
    \caption{Comparison of measured cross sections to TENDL-2023 and TALYS-2.0 predictions using default parameters, the parameter adjustments of Fox \textit{et al.}, and the adjustments performed in this work \cite{Fox}. 
    }
    \label{fig:additional_1}
\end{figure*}

\begin{figure*}[!htbp]
    \sloppy
    \centering
   \subfloat{
        \centering
        \subfigimg[width=0.496\textwidth]{}{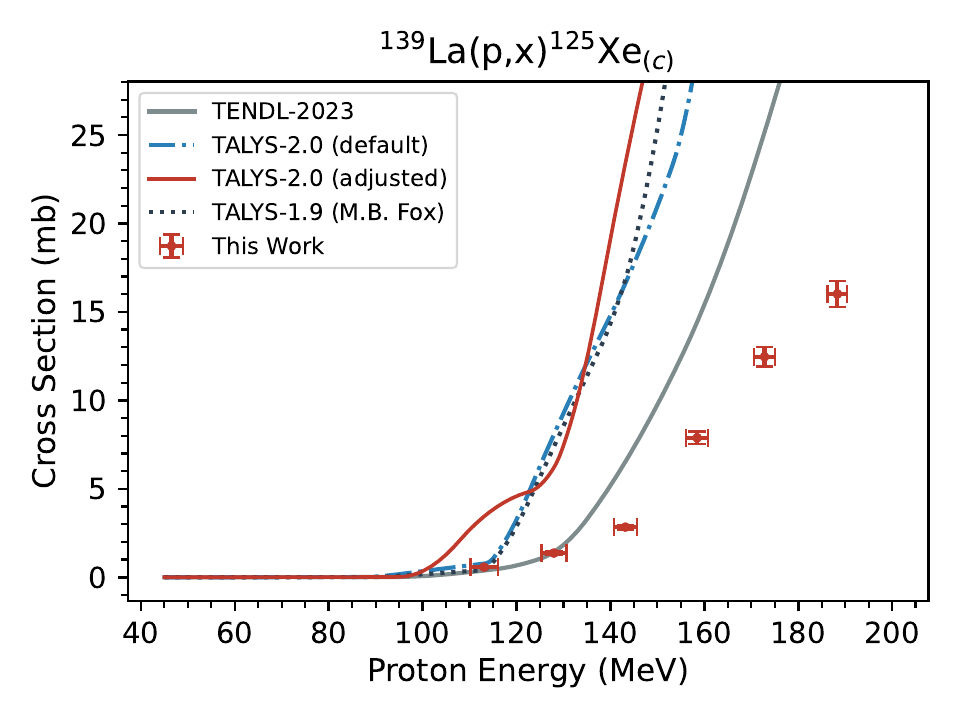}{50}
        \subfigimg[width=0.496\textwidth]{}{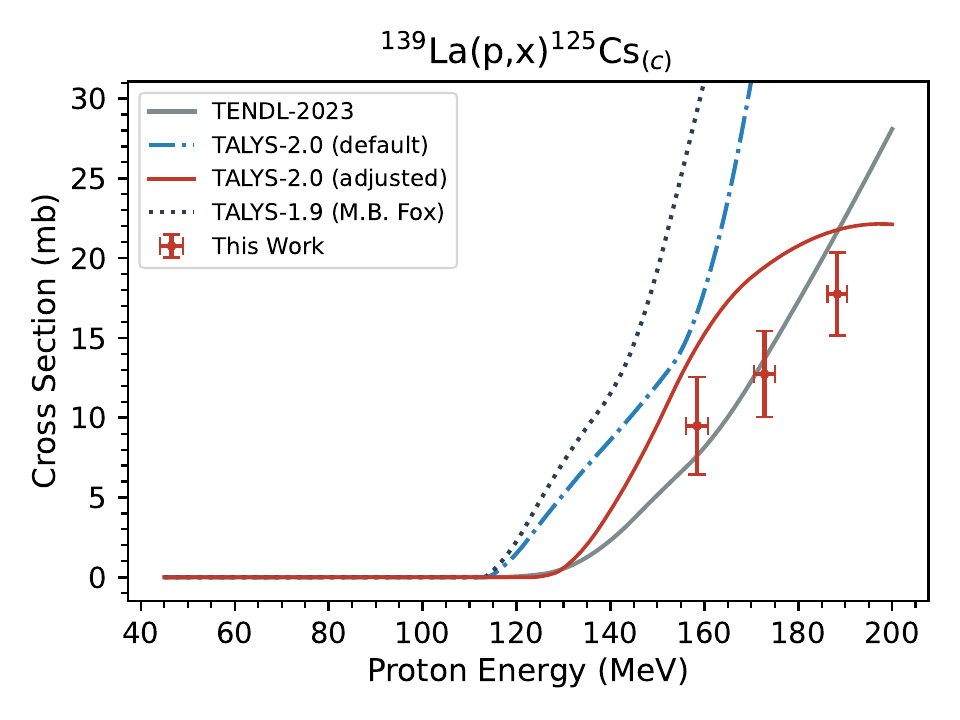}{50}
   \hspace{-10pt}}
    \\
    \subfloat{
        \centering
        \subfigimg[width=0.496\textwidth]{}{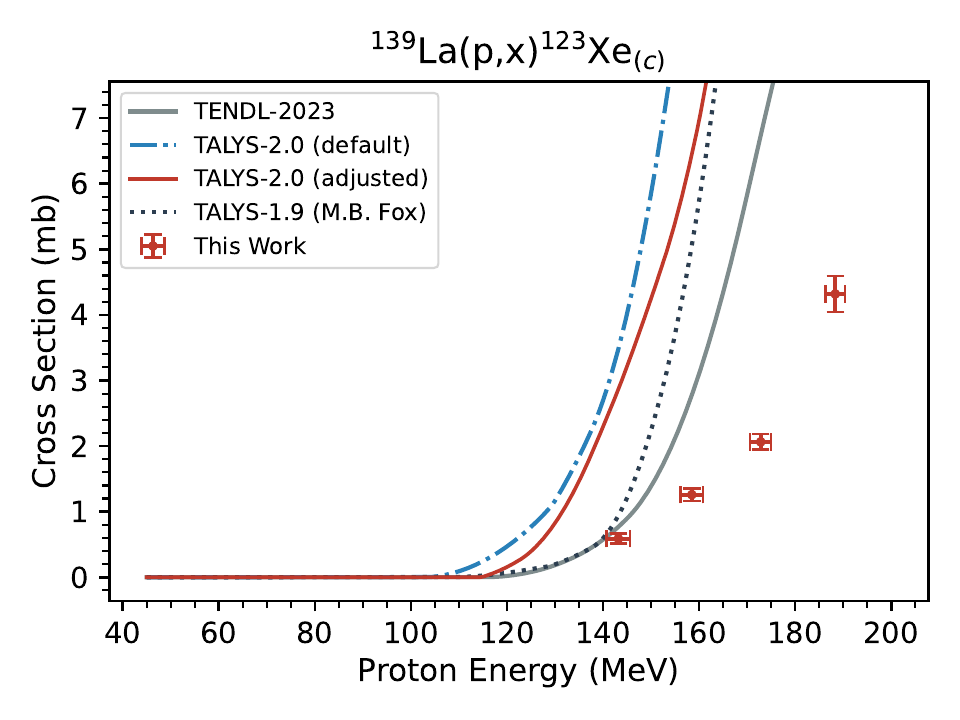}{50}
        \subfigimg[width=0.496\textwidth]{}{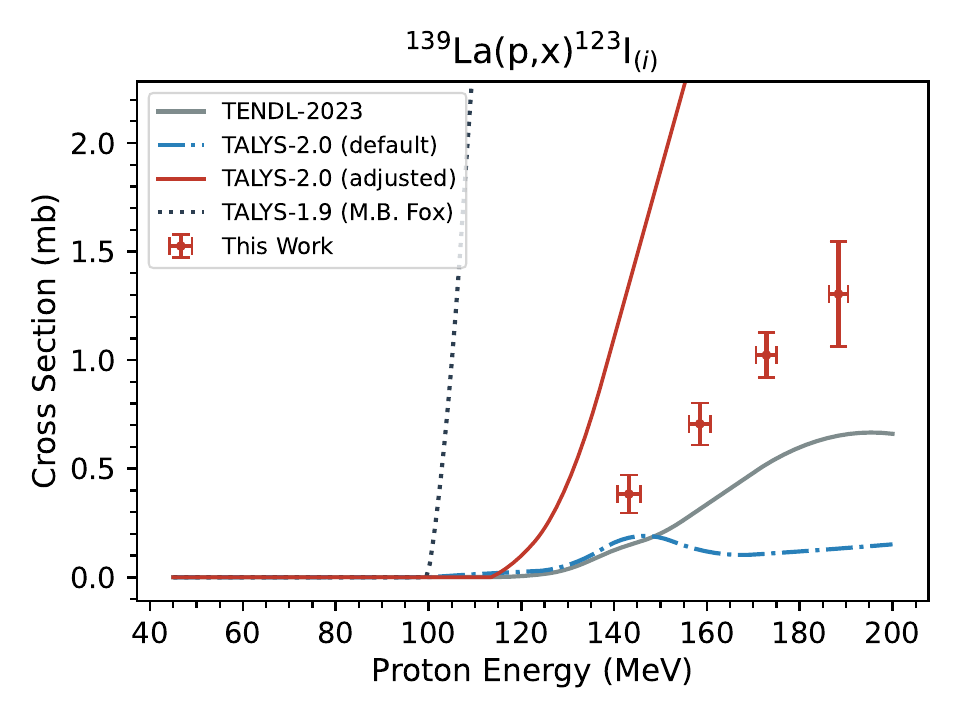}{50}
   \hspace{-10pt}}
    \caption{Comparison of measured cross sections to TENDL-2023 and TALYS-2.0 predictions using default parameters, the parameter adjustments of Fox \textit{et al.}, and the adjustments performed in this work \cite{Fox}. 
    }
    \label{fig:additional_2}
\end{figure*}

Plots of additional cross sections measured in this work that were not discussed in the main text are shown in Figs. \ref{fig:additional_1} and \ref{fig:additional_2}.

\begin{figure*}[!htbp]
    \sloppy
    \centering
    \subfloat{
        \centering
        \subfigimg[width=0.496\textwidth]{}{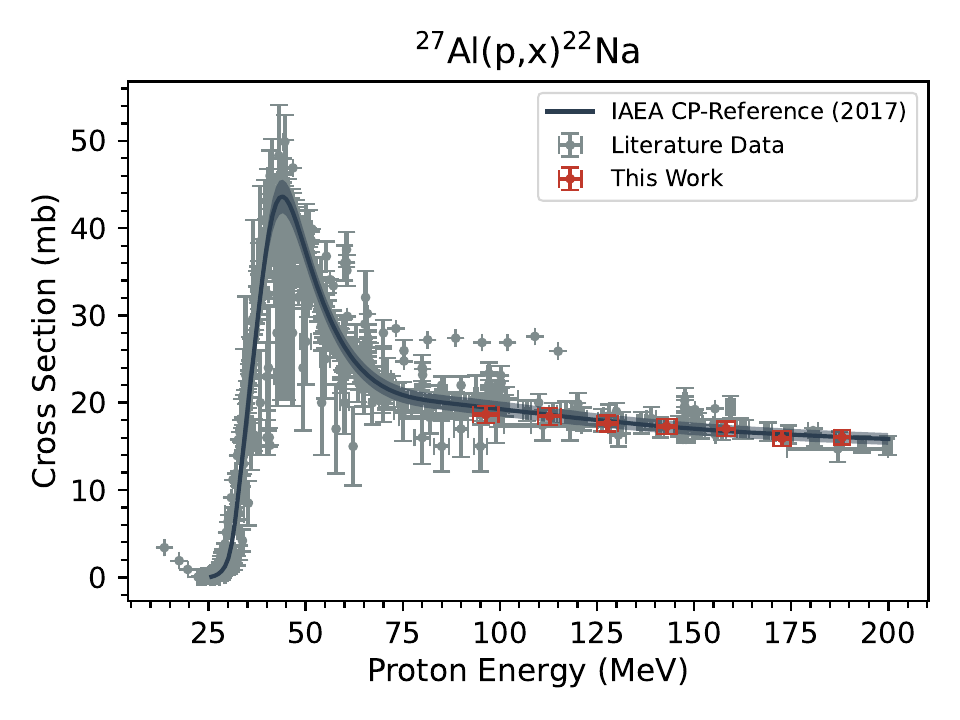}{50}
        \subfigimg[width=0.496\textwidth]{}{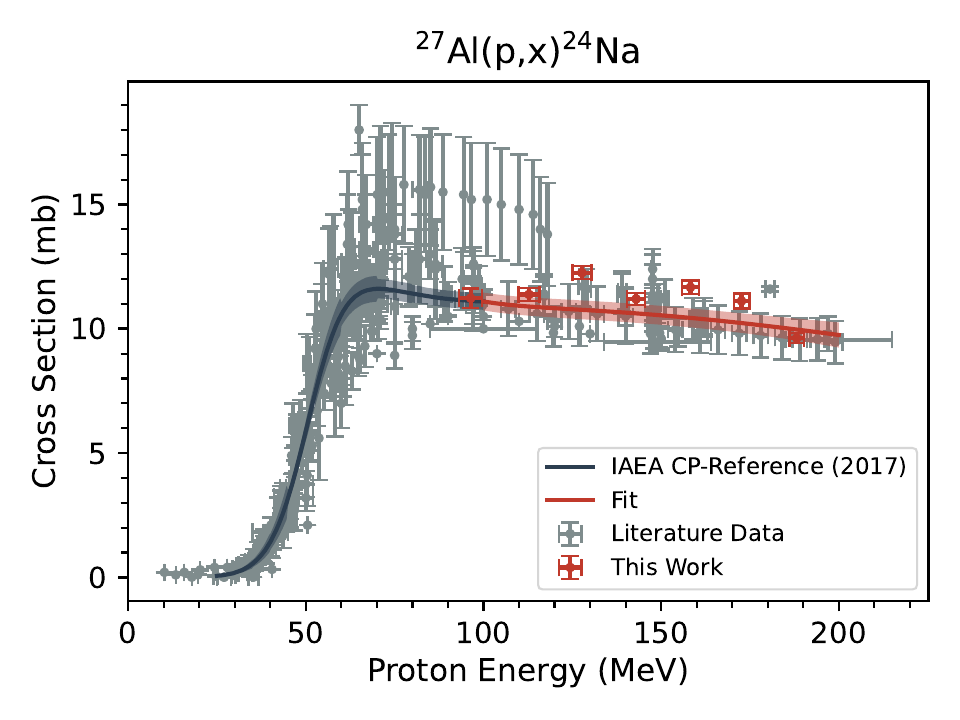}{50}
   \hspace{-10pt}}
    \\
    \subfloat{
        \centering
        \subfigimg[width=0.496\textwidth]{}{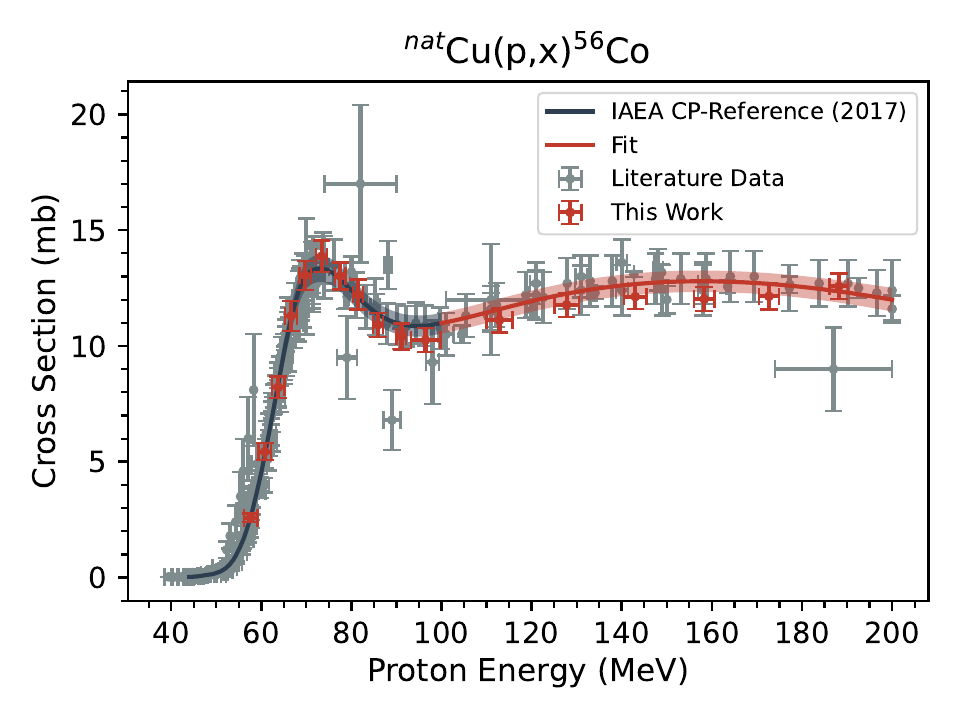}{50}
        \subfigimg[width=0.496\textwidth]{}{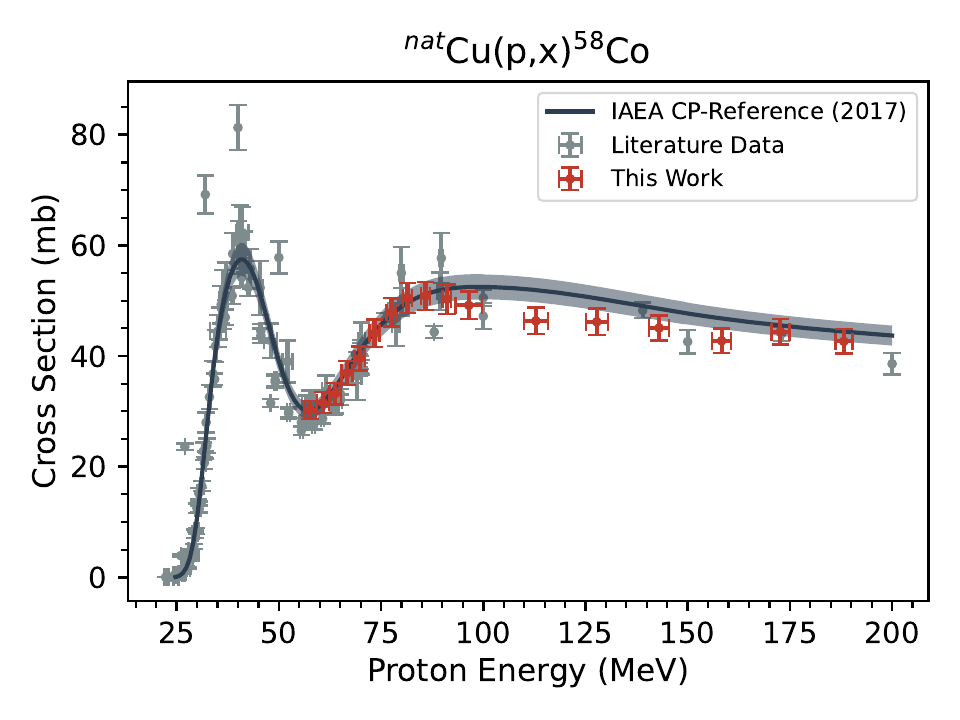}{50}
   \hspace{-10pt}}
    \\
    \subfloat{
        \centering
        \subfigimg[width=0.496\textwidth]{}{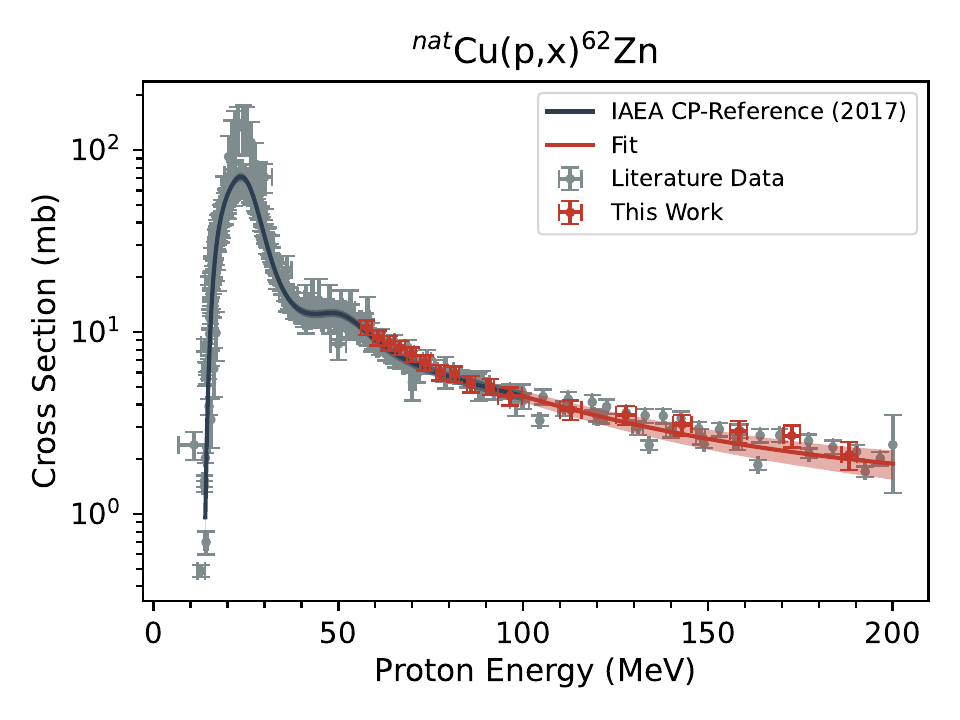}{50}
        \subfigimg[width=0.496\textwidth]{}{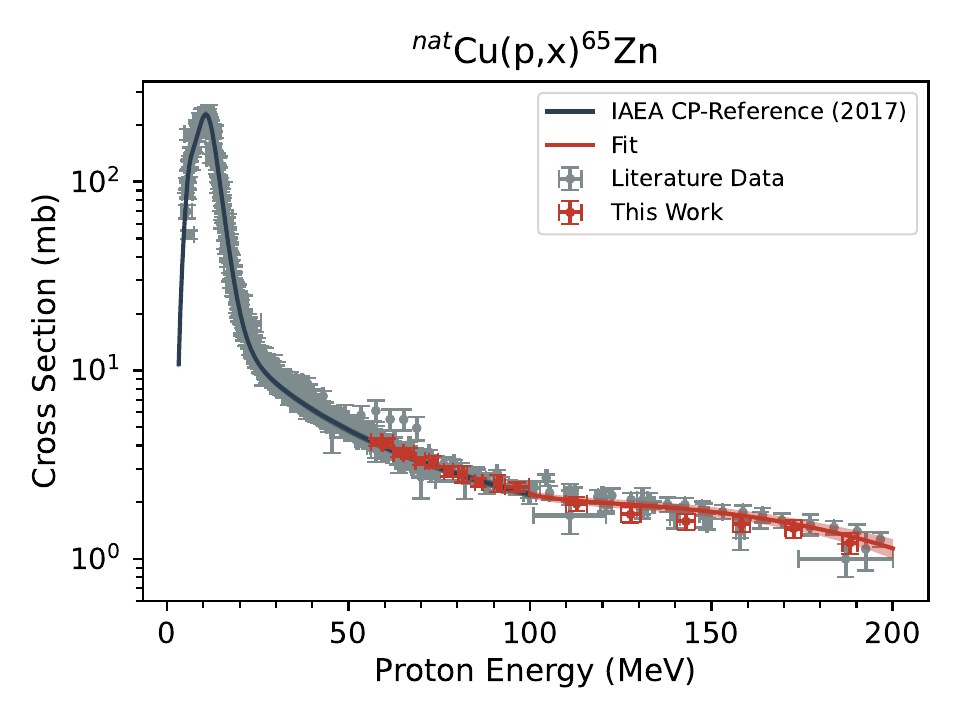}{50}
   \hspace{-10pt}}
    \caption{Experimental and evaluated monitor cross sections for the production of \ce{^{22}Na} (top-left), \ce{^{24}Na} (top-right), \ce{^{56}Co} (middle-left), \ce{^{58}Co} (middle-right), \ce{^{62}Zn} (bottom-left) and \ce{^{65}Zn} (bottom-right) \cite{IAEACPR}.
    }
    \label{fig:monitor_extrapolations}
\end{figure*}

Plots of the monitor reaction cross sections that were extrapolated to the 100--200 MeV energy range are shown in Fig. \ref{fig:monitor_extrapolations}, in comparison with the ``effective" cross sections that are constituted by the measured beam currents.

\section{Stack Design}
\label{stack_appendix}
A list of every component of the experimental stacks used in both irradiations can be found in Table \ref{table:stack_lanl} for the LANL irradiation and Table \ref{table:stack_bnl} for the BNL irradiation.

\begin{table}[!htbp]
\small
\caption{Details of the stack used in the LANL irradiation. Uncertainties are listed in the least significant digit, that is, 15.51(28) means 15.51 $\pm$ 0.28.}
\begin{tabular}{cccc}
Foil Id & Compound & $\Delta x$ (mm) & $\rho \Delta x$ (mg/cm$^2$) \Bstrut \\
\hline
\Tstrut La01 & La & 0.0251 & 15.51(28) \\
Cu01 & Cu & 0.0213 & 19.05(26) \\
Ti01 & Ti & 0.0344 & 15.56(19) \\
D1 & Cu & 1.02 & 909 \\
La02 & La & 0.0288 & 17.81(32) \\
Cu02 & Cu & 0.0214 & 19.15(27) \\
Ti02 & Ti & 0.0346 & 15.65(19) \\
D2 & Cu & 0.794 & 710 \\
La03 & La & 0.0286 & 17.67(32) \\
Cu03 & Cu & 0.0208 & 18.61(26) \\
Ti03 & Ti & 0.034 & 15.37(18) \\
D3 & Cu & 0.663 & 593 \\
La04 & La & 0.025 & 15.45(28) \\
Cu04 & Cu & 0.0214 & 19.11(27) \\
Ti04 & Ti & 0.034 & 15.37(18) \\
D4 & Cu & 0.67 & 599 \\
La05 & La & 0.0251 & 15.48(28) \\
Cu05 & Cu & 0.0215 & 19.20(27) \\
Ti05 & Ti & 0.0346 & 15.64(19) \\
D5 & Al & 1.53 & 414 \\
La06 & La & 0.0243 & 14.98(27) \\
Cu06 & Cu & 0.0213 & 19.05(26) \\
Ti06 & Ti & 0.0339 & 15.30(18) \\
D6 & Al & 1.2 & 324 \\
La07 & La & 0.0283 & 17.47(32) \\
Cu07 & Cu & 0.0215 & 19.24(27) \\
Ti07 & Ti & 0.0343 & 15.49(19) \\
D7 & Al & 1.02 & 275 \\
La08 & La & 0.0246 & 15.17(27) \\
Cu08 & Cu & 0.0213 & 19.10(27) \\
Ti08 & Ti & 0.0348 & 15.72(19) \\
D8 & Al & 1.02 & 275 \\
La09 & La & 0.0241 & 14.85(27) \\
Cu09 & Cu & 0.0208 & 18.65(26) \\
Ti09 & Ti & 0.0343 & 15.52(19) \\
D9 & Al & 1.02 & 275 \\
La10 & La & 0.0305 & 18.81(34) \\
Cu10 & Cu & 0.0208 & 18.63(26) \\
Ti10 & Ti & 0.0334 & 15.11(18) \Bstrut \\
\end{tabular}
\label{table:stack_lanl}
\end{table}

\begin{table}[!htbp]
\small
\caption{Details of the stack used in the BNL irradiation. Uncertainties are listed in the least significant digit, that is, 17.32(31) means 17.32 $\pm$ 0.31.}
\begin{tabular}{cccc}
Foil Id & Compound & $\Delta x$ (mm) & $\rho \Delta x$ (mg/cm$^2$) \Bstrut \\
\hline
\Tstrut La01 & La & 0.028 & 17.32(31) \\
Cu01 & Cu & 0.021 & 18.81(26) \\
Al01 & Al & 0.0267 & 7.19(13) \\
D1 & Cu & 5.11 & 4570 \\
La02 & La & 0.0272 & 16.77(30) \\
Cu02 & Cu & 0.0206 & 18.39(26) \\
Al02 & Al & 0.0274 & 7.40(14) \\
D2 & Cu & 4.46 & 3990 \\
La03 & La & 0.0276 & 17.02(31) \\
Cu03 & Cu & 0.0212 & 18.98(26) \\
Al03 & Al & 0.0265 & 7.15(13) \\
D3 & Cu & 4.46 & 3990 \\
La04 & La & 0.0279 & 17.24(31) \\
Cu04 & Cu & 0.0211 & 18.86(26) \\
Al04 & Al & 0.0269 & 7.26(13) \\
D4 & Cu & 4.12 & 3690 \\
La05 & La & 0.0274 & 16.92(31) \\
Cu05 & Cu & 0.0215 & 19.21(27) \\
Al05 & Al & 0.0269 & 7.26(13) \\
D5 & Cu & 3.7 & 3310 \\
La06 & La & 0.0277 & 17.09(31) \\
Cu06 & Cu & 0.0214 & 19.11(27) \\
Al06 & Al & 0.027 & 7.29(13) \\
D6 & Cu & 3.7 & 3310 \\
La07 & La & 0.0243 & 14.99(27) \\
Cu07 & Cu & 0.0211 & 18.92(26) \Bstrut \\
\end{tabular}
\label{table:stack_bnl}
\end{table}

\section{Relevant Nuclear Data}
\label{nudat_appendix}
The principal gamma-ray decay data used in the activation analysis of the lanthanum foils are given in Table \ref{table:decay_data_lanthanum}. The principal gamma-ray decay data used in the activation analysis of the aluminum, copper and titanium monitor foils are given in Table \ref{table:decay_data_monitor}.


\begin{table*}[!htbp]
\scriptsize
\caption{Principle $\gamma$-ray data for lanthanum products, from ENSDF \cite{A139, A137, A136, A135, A134, A133, A132, A131, A130, A129, A128, A127, A126, A125, A123}. Uncertainties are listed in the least significant digit, that is, 137.641(20) d means 137.641 $\pm$ 0.020 d.}
\begin{tabular}{cccc}
Isotope & $\gamma$ Energy (keV) & $I_{\gamma}$ (\%) & $T_{1/2}$ \Bstrut \\
\hline
\Tstrut \ce{^{139}Ce} & 165.8575 & 80.0(80) & 137.641(20) d \\
\ce{^{137m}Ce} & 254.29 & 11.1(4) & 1.433(13) d \\
\ce{^{136}Cs} & 818.514 & 99.7(1) & 13.01(5) d \\
 & 1048.073 & 80.0(30) &  \\
\ce{^{135}Ce} & 265.56 & 41.8(14) & 17.7(3) h \\
 & 300.07 & 23.5(5) &  \\
\ce{^{135m}Ba} & 268.218 & 16.0(4) & 1.1958(83) d \\\\
\ce{^{134}Ce} & - & - & 3.14(4) d \\
\ce{^{134}La} & 604.721 & 5.04(20) & 6.45(16) min \\
\ce{^{133m}Ce} & 477.22 & 39.3(20) & 5.1(3) h \\
 & 58.39 & 19.3(4) &  \\
\ce{^{133}Ce} & 97.261 & 46.0(70) & 1.617(67) h \\
 & 76.9 & 15.9(23) &  \\
\ce{^{133}La} & 278.835 & 2.44(13) & 3.912(8) h \\
\ce{^{133m}Ba} & 275.925 & 17.69(88) & 1.6208(42) d \\
\ce{^{132}Ce} & 182.11 & 77.4(39) & 3.51(11) h \\
 & 155.37 & 10.5(5) &  \\
\ce{^{132m}La} & 135.8 & 49.0(40) & 24.3(5) min \\
 & 390.51 & 4.8(9) &  \\
\ce{^{132}La} & 464.55 & 76.0(60) & 4.8(2) h \\
 & 567.14 & 15.7(15) &  \\
\ce{^{132}Cs} & 667.714 & 97.6(49) & 6.480(6) d \\
\ce{^{131}La} & 108.081 & 25.0(8) & 59.0(20) min \\
 & 417.783 & 18.0(6) &  \\
\ce{^{131}Ba} & 496.321 & 48.0(24) & 11.50(6) d \\
 & 123.804 & 29.8(3) &  \\
\ce{^{130}La} & 357.4 & 81.0(40) & 8.7(1) min \\
 & 550.7 & 25.9(19) &  \\
\ce{^{129}Ba} & 214.3 & 13.4(7) & 2.23(11) h \\
 & 220.83 & 8.5(4) &  \\
\ce{^{129m}Ba} & 1459.1 & 50.0(2) & 2.16(2) h \\
 & 202.38 & 33.7(6) &  \\
\ce{^{129}Cs} & 371.918 & 30.6(17) & 1.3358(25) d \\
 & 411.49 & 22.3(12) &  \\
\ce{^{128}Ba} & 273.44 & 14.50(72) & 2.43(5) d \\
\ce{^{128}Cs} & 442.901 & 26.8(13) & 3.62(2) min \\
 & 526.557 & 2.41(12) &  \\
\ce{^{127}Cs} & 411.95 & 62.9(31) & 6.25(10) h \\
 & 124.7 & 11.38(22) &  \\
\ce{^{127}Xe} & 202.86 & 68.7(34) & 36.4(1) d \\
 & 172.132 & 25.7(9) &  \\
\ce{^{126}Ba} & 233.6 & 19.6(18) & 1.667(33) h \\
 & 681.8 & 4.4(6) &  \\
\ce{^{126}Cs} & 388.66 & 40.7(20) & 1.64(2) min \\
 & 491.27 & 5.0(4) &  \\
\ce{^{126}I} & 666.331 & 17.34(21) & 12.93(5) d \\
 & 388.633 & 16.84(24) &  \\
\ce{^{125}Cs} & 112.0 & 8.6(8) & 46.7(1) min \\
 & 712.0 & 3.50(34) &  \\
\ce{^{125}Xe} & 188.418 & 53.8(27) & 16.9(2) h \\
 & 243.378 & 30.0(6) &  \\
\ce{^{123}Xe} & 148.9 & 48.9(24) & 2.08(2) h \\
 & 178.1 & 14.9(8) &  \\
\ce{^{123}I} & 159.0 & 83.60(19) & 13.2230(19) h \Bstrut \\
\end{tabular}
\label{table:decay_data_lanthanum}
\end{table*}

\begin{table*}[!htbp]
\scriptsize
\caption{Principle $\gamma$-ray data for monitor foil products, from ENSDF \cite{A65, A62, A61, A60, A59, A58, A57, A56, A55, A54, A52, A51, A48, A47, A46, A44, A43, A42, A24, A22}.  Uncertainties are listed in the least significant digit, that is, 243.93(9) d means 243.93 $\pm$ 0.09 d.}
\begin{tabular}{cccc}
Isotope & $\gamma$ Energy (keV) & $I_{\gamma}$ (\%) & $T_{1/2}$ \Bstrut \\
\hline
\Tstrut \ce{^{65}Zn} & 1115.539 & 50.04(10) & 243.93(9) d \\
\ce{^{63}Zn} & 669.62 & 8.2(3) & 38.47(5) min \\
 & 962.06 & 6.5(4) &  \\
\ce{^{62}Zn} & 596.56 & 26.0(20) & 9.26(2) h \\
 & 548.35 & 15.3(14) &  \\
\ce{^{61}Cu} & 282.956 & 12.7(20) & 3.336(10) h \\
 & 656.008 & 10.4(16) &  \\
\ce{^{60}Cu} & 1332.5 & 88.0(10) & 23.7(4) min \\
 & 1791.6 & 45.4(23) &  \\
\ce{^{60}Co} & 1332.492 & 99.9826(6) & 5.27113(38) y \\
 & 1173.228 & 99.85(3) &  \\
\ce{^{59}Fe} & 1099.245 & 56.5(9) & 44.490(9) d \\
 & 1291.59 & 43.2(9) &  \\
\ce{^{58}Co} & 810.7593 & 99.45(1) & 70.86(6) d \\
\ce{^{57}Ni} & 1377.63 & 81.7(24) & 1.4833(25) d \\
 & 127.164 & 16.7(5) &  \\
\ce{^{57}Co} & 122.06065 & 85.60(17) & 271.74(6) d \\
 & 136.47356 & 10.68(8) &  \\
\ce{^{56}Ni} & 158.38 & 98.8(10) & 6.075(10) d \\
 & 811.85 & 86.0(9) &  \\
\ce{^{56}Co} & 846.77 & 99.9399(23) & 77.236(26) d \\
 & 1238.288 & 66.46(12) &  \\
\ce{^{56}Mn} & 846.7638 & 98.8(15) & 2.5789(1) h \\
 & 1810.726 & 26.9(4) &  \\
\ce{^{55}Co} & 931.1 & 75.0(35) & 17.53(3) h \\
 & 477.2 & 20.2(17) &  \\
\ce{^{54}Mn} & 834.848 & 99.976(1) & 312.2(2) d \\
\ce{^{52}Mn} & 1434.092 & 100.0(14) & 5.591(3) d \\
 & 935.544 & 94.5(13) &  \\
\ce{^{52m}Mn} & 1434.06 & 98.2(5) & 21.1(2) min \\
\ce{^{51}Cr} & 320.0824 & 9.91(1) & 27.704(3) d \\
\ce{^{48}V} & 983.525 & 99.98(4) & 15.974(3) d \\
 & 1312.106 & 98.2(3) &  \\
\ce{^{48}Sc} & 1037.522 & 97.5(20) & 1.8213(37) d \\
 & 175.361 & 7.47(18) &  \\
\ce{^{47}Sc} & 159.381 & 68.3(4) & 3.3492(6) d \\
\ce{^{46}Sc} & 1120.545 & 99.987(1) & 83.79(4) d \\
 & 889.277 & 99.984(1) &  \\
\ce{^{44m}Sc} & 271.241 & 86.72(1) & 2.4421(42) d \\
\ce{^{44}Sc} & 1157.022 & 99.8867(30) & 4.0420(25) h \\
\ce{^{43}Sc} & 372.9 & 22.5(11) & 3.891(12) h \\
\ce{^{43}K} & 372.76 & 86.8(2) & 22.3(1) h \\
 & 617.49 & 79.2(6) &  \\
\ce{^{42}K} & 1524.6 & 18.08(9) & 12.355(7) h \\
\ce{^{24}Na} & 1368.625 & 99.9940(20) & 14.956(3) h \\
 & 2754.008 & 99.867(10) &  \\
\ce{^{22}Na} & 1274.537 & 99.940(14) & 2.6018(22) y \Bstrut \\
\end{tabular}
\label{table:decay_data_monitor}
\end{table*}

\end{appendices}

\bibliography{references}

@article{Fox,
  title = {{Investigating high-energy proton-induced reactions on spherical nuclei: Implications for the preequilibrium exciton model}},
  author = {Fox, Morgan B. and Voyles, Andrew S. and Morrell, Jonathan T. and Bernstein, Lee A. and Lewis, Amanda M. and Koning, Arjan J. and Batchelder, Jon C. and Birnbaum, Eva R. and Cutler, Cathy S. and Medvedev, Dmitri G. and Nortier, Francois M. and O'Brien, Ellen M. and Vermeulen, Christiaan},
  journal = {Phys. Rev. C},
  volume = {103},
  issue = {3},
  pages = {034601},
  numpages = {34},
  year = {2021},
  month = {Mar},
  publisher = {American Physical Society},
  doi = {10.1103/PhysRevC.103.034601},
  url = {https://link.aps.org/doi/10.1103/PhysRevC.103.034601}
}

@Misc{AML_study,
	title = "{Venetoclax and Lintuzumab-Ac225 in AML Patients}",
	author = "National Library of Medicine (US)",
	howpublished = "\url{https://clinicaltrials.gov/study/NCT03867682}",
	year = "2023",
	note = "{Identifier NCT03867682}"
}

@Misc{CRMPC_study,
	title = "{A Study of JNJ-69086420, an Actinium-225-Labeled Antibody Targeting Human Kallikrein-2 (hK2) for Advanced Prostate Cancer}",
	author = "National Library of Medicine (US)",
	howpublished = "\url{https://clinicaltrials.gov/study/NCT04644770}",
	year = "2024",
	note = "{Identifier NCT04644770}"
}

@article{actinium_meta,
author = {Ma, Jiao and Li, Lanying and Liao, Taiping and Gong, Weidong and Zhang, Chunyin},
year = {2022},
month = {02},
pages = {},
title = {{Efficacy and Safety of 225Ac-PSMA-617-Targeted Alpha Therapy in Metastatic Castration-Resistant Prostate Cancer: A Systematic Review and Meta-Analysis}},
volume = {12},
journal = {Frontiers in Oncology},
doi = {10.3389/fonc.2022.796657}
}

@ARTICLE{alpha_review,
AUTHOR={Pallares, Roger M. and Abergel, Rebecca J.},   
TITLE={{Development of radiopharmaceuticals for targeted alpha therapy: Where do we stand?}},      
JOURNAL={Frontiers in Medicine},      
VOLUME={9},           
YEAR={2022},      
URL={https://www.frontiersin.org/articles/10.3389/fmed.2022.1020188},       
DOI={10.3389/fmed.2022.1020188},      
ISSN={2296-858X}
}

@article {Bobba265355,
	author = {Kondapa Naidu Bobba and Anil P. Bidkar and Niranjan Meher and Cyril Fong and Anju Wadhwa and Suchi Dhrona and Alex Sorlin and Scott Bidlingmaier and Becka Shuere and Jiang He and David M. Wilson and Bin Liu and Youngho Seo and Henry F. VanBrocklin and Robert R. Flavell},
	title = {{Evaluation of 134Ce/134La as a PET Imaging Theranostic Pair for 225Ac alpha-Radiotherapeutics}},
	year = {2023},
	doi = {10.2967/jnumed.122.265355},
	publisher = {Society of Nuclear Medicine},
	issn = {0161-5505},
	URL = {https://jnm.snmjournals.org/content/early/2023/05/18/jnumed.122.265355},
	journal = {Journal of Nuclear Medicine}
}

@article{Qaim_data,
author = {Qaim, Syed},
year = {2024},
month = {01},
pages = {},
title = {{New directions in nuclear data research for accelerator-based production of medical radionuclides}},
journal = {Journal of Radioanalytical and Nuclear Chemistry},
doi = {10.1007/s10967-023-09285-6}
}

@article{BAILEY202228,
title = {{Evaluation of 134Ce as a PET imaging surrogate for antibody drug conjugates incorporating 225Ac}},
journal = {Nuclear Medicine and Biology},
volume = {110-111},
pages = {28-36},
year = {2022},
issn = {0969-8051},
doi = {https://doi.org/10.1016/j.nucmedbio.2022.04.007},
url = {https://www.sciencedirect.com/science/article/pii/S0969805122000373},
author = {Tyler A. Bailey and Jennifer N. Wacker and Dahlia D. An and Korey P. Carter and Ryan C. Davis and Veronika Mocko and John Larrabee and Katherine M. Shield and Mila Nhu Lam and Corwin H. Booth and Rebecca J. Abergel}
}

@article{Bailey2021,
author = {Bailey, Tyler and Mocko, Veronika and Shield, Katherine and An, Dahlia and Akin, Andrew and Birnbaum, Eva and Brugh, Mark and Cooley, Jason and Engle, Jonathan and Fassbender, Michael and Gauny, Stacey and Lakes, Andrew and Nortier, Francois and O'Brien, Ellen and Thieman, Sara and White, Frankie and Vermeulen, Christiaan and Kozimor, Stosh and Abergel, Rebecca},
year = {2021},
month = {03},
pages = {},
title = {{Developing the 134Ce and 134La pair as companion positron emission tomography diagnostic isotopes for 225Ac and 227Th radiotherapeutics}},
volume = {13},
journal = {Nature Chemistry},
doi = {10.1038/s41557-020-00598-7}
}

@article{Morrell_La,
	author = {{Morrell, Jonathan T.} and {Voyles, Andrew S.} and {Basunia, M. S.} and {Batchelder, Jon C.} and {Matthews, Eric F.} and {Bernstein, Lee A.}},
	title = {{Measurement of 139La(p,x) cross sections from 35–60 MeV by stacked-target activation}},
	DOI= "10.1140/epja/s10050-019-00010-0",
	url= "https://doi.org/10.1140/epja/s10050-019-00010-0",
	journal = {Eur. Phys. J. A},
	year = 2020,
	volume = 56,
	number = 1,
	pages = "13",
}

@article{BECKER202081,
title = {{Cross section measurements for proton induced reactions on natural La}},
journal = {Nuclear Instruments and Methods in Physics Research Section B: Beam Interactions with Materials and Atoms},
volume = {468},
pages = {81-88},
year = {2020},
issn = {0168-583X},
doi = {https://doi.org/10.1016/j.nimb.2020.02.024},
url = {https://www.sciencedirect.com/science/article/pii/S0168583X20300756},
author = {K.V. Becker and E. Vermeulen and C.J. Kutyreff and E.M. O’Brien and J.T. Morrell and E.R. Birnbaum and L.A. Bernstein and F.M. Nortier and J.W. Engle}
}

@article{Tarkanyi2017,
author = {T{\'{a}}rk{\'{a}}nyi, F and Hermanne, A and Ditr{\'{o}}i, F and Tak{\'{a}}cs, S},
doi = {10.1007/s10967-017-5253-7},
issn = {0236-5731},
journal = {Journal of Radioanalytical and Nuclear Chemistry},
month = {jun},
number = {3},
pages = {691--704},
title = {{Activation cross section data of proton induced nuclear reactions on lanthanum in the 34–65 MeV energy range and application for production of medical radionuclides}},
url = {http://link.springer.com/10.1007/s10967-017-5253-7},
volume = {312},
year = {2017}
}

@article{Huang_2021,
doi = {10.1088/1674-1137/abddb0},
url = {https://dx.doi.org/10.1088/1674-1137/abddb0},
year = {2021},
month = {mar},
publisher = {Chinese Physical Society and the Institute of High Energy Physics of the Chinese Academy of Sciences and the Institute of Modern Physics of the Chinese Academy of Sciences and IOP Publishing Ltd},
volume = {45},
number = {3},
pages = {030002},
author = {W.J. Huang and Meng Wang and F.G. Kondev and G. Audi and S. Naimi},
title = {{The AME 2020 atomic mass evaluation (I). Evaluation of input data, and adjustment procedures*}},
journal = {Chinese Physics C}
}

@article{talys_level_density,
  title = {{Improved microscopic nuclear level densities within the Hartree-Fock-Bogoliubov plus combinatorial method}},
  author = {Goriely, S. and Hilaire, S. and Koning, A. J.},
  journal = {Phys. Rev. C},
  volume = {78},
  issue = {6},
  pages = {064307},
  numpages = {14},
  year = {2008},
  month = {Dec},
  publisher = {American Physical Society},
  doi = {10.1103/PhysRevC.78.064307},
  url = {https://link.aps.org/doi/10.1103/PhysRevC.78.064307}
}

@article{talys_strength,
  title = {{Gogny-HFB+QRPA dipole strength function and its application to radiative nucleon capture cross section}},
  author = {Goriely, S. and Hilaire, S. and P\'eru, S. and Sieja, K.},
  journal = {Phys. Rev. C},
  volume = {98},
  issue = {1},
  pages = {014327},
  numpages = {12},
  year = {2018},
  month = {Jul},
  publisher = {American Physical Society},
  doi = {10.1103/PhysRevC.98.014327},
  url = {https://link.aps.org/doi/10.1103/PhysRevC.98.014327}
}

@article{EXFOR,
title = "{{Towards a More Complete and Accurate Experimental Nuclear Reaction Data Library (EXFOR): International Collaboration Between Nuclear Reaction Data Centres (NRDC)}}",
journal = "Nuclear Data Sheets",
volume = "120",
pages = "272 - 276",
year = "2014",
issn = "0090-3752",
doi = "10.1016/j.nds.2014.07.065",
url = "http://www.sciencedirect.com/science/article/pii/S0090375214005171",
author = "N. Otuka and E. Dupont and V. Semkova and B. Pritychenko and A.I. Blokhin and M. Aikawa and S. Babykina and M. Bossant and G. Chen and S. Dunaeva and R.A. Forrest and T. Fukahori and N. Furutachi and S. Ganesan and Z. Ge and O.O. Gritzay and M. Herman and S. Hlava{\v c} and K. Kat{\~o} and B. Lalremruata and Y.O. Lee and A. Makinaga and K. Matsumoto and M. Mikhaylyukova and G. Pikulina and V.G. Pronyaev and A. Saxena and O. Schwerer and S.P. Simakov and N. Soppera and R. Suzuki and S. Takács and X. Tao and S. Taova and F. T{\'a}rk{\'a}nyi and V.V. Varlamov and J. Wang and S.C. Yang and V. Zerkin and Y. Zhuang"
}

@article{TALYS,
	author = {{Koning, Arjan} and {Hilaire, Stephane} and {Goriely, Stephane}},
	title = {{TALYS: modeling of nuclear reactions}},
	DOI= "10.1140/epja/s10050-023-01034-3",
	url= "https://doi.org/10.1140/epja/s10050-023-01034-3",
	journal = {Eur. Phys. J. A},
	year = 2023,
	volume = 59,
	number = 6,
	pages = "131",
}

@inproceedings{rochman2017tendl,
  title={{The TENDL library: Hope, reality and future}},
  author={Rochman, Dimitri and Koning, Arjan J and Sublet, J Ch and Fleming, Michael and Bauge, Eric and Hilaire, St{\'e}phane and Romain, Pascal and Morillon, Benjamin and Duarte, Helder and Goriely, St{\'e}phane and others},
  booktitle={EPJ web of conferences},
  volume={146},
  pages={02006},
  year={2017},
  organization={EDP Sciences}
}

@article{FLUKA,
title = {{Overview of the FLUKA code}},
journal = {Annals of Nuclear Energy},
volume = {82},
pages = {10-18},
year = {2015},
note = {Joint International Conference on Supercomputing in Nuclear Applications and Monte Carlo 2013, SNA + MC 2013. Pluri- and Trans-disciplinarity, Towards New Modeling and Numerical Simulation Paradigms},
issn = {0306-4549},
doi = {https://doi.org/10.1016/j.anucene.2014.11.007},
url = {https://www.sciencedirect.com/science/article/pii/S0306454914005878},
author = {Giuseppe Battistoni and Till Boehlen and Francesco Cerutti and Pik Wai Chin and Luigi Salvatore Esposito and Alberto Fassò and Alfredo Ferrari and Anton Lechner and Anton Empl and Andrea Mairani and Alessio Mereghetti and Pablo Garcia Ortega and Johannes Ranft and Stefan Roesler and Paola R. Sala and Vasilis Vlachoudis and George Smirnov}
}

@article{HERMAN20072655,
title = "{{EMPIRE: Nuclear Reaction Model Code System for Data Evaluation}}",
journal = "Nuclear Data Sheets",
volume = "108",
number = "12",
pages = "2655 - 2715",
year = "2007",
note = "Special Issue on Evaluations of Neutron Cross Sections",
issn = "0090-3752",
doi = "10.1016/j.nds.2007.11.003",
url = "http://www.sciencedirect.com/science/article/pii/S0090375207000981",
author = "M. Herman and R. Capote and B.V. Carlson and P. Oblo{\v z}insk{\'y} and M. Sin and A. Trkov and H. Wienke and V. Zerkin"
}

@InProceedings{CoH,
author="Kawano, Toshihiko",
editor="Escher, Jutta
and Alhassid, Yoram
and Bernstein, Lee A.
and Brown, David
and Fr{\"o}hlich, Carla
and Talou, Patrick
and Younes, Walid",
title="{{CoH3: The Coupled-Channels and Hauser-Feshbach Code}}",
booktitle="Compound-Nuclear Reactions ",
year="2021",
publisher="Springer International Publishing",
address="Cham",
pages="27--34",
doi = "10.1007/978-3-030-58082-7_3",
isbn="978-3-030-58082-7"
}

@article{ALICE,
title = {{Code ALICE/LIVERMORE 82}},
author = {Blann, M. and Bisplinghoff, J.},
abstractNote = {A revision of the OVERLAID ALICE code is described, with emphasis on new options and new physics. The function of all subroutines is described. A source code listing, sample input, and sample output is appended.},
doi = {},
place = {United States},
year = {1982},
month = {11}
}

@article{GRAVES201644,
title = "{{Nuclear excitation functions of proton-induced reactions (Ep=35--90MeV) from Fe, Cu, and Al}}",
journal = "Nuclear Instruments and Methods in Physics Research Section B: Beam Interactions with Materials and Atoms",
volume = "386",
pages = "44 - 53",
year = "2016",
issn = "0168-583X",
doi = "10.1016/j.nimb.2016.09.018",
url = "http://www.sciencedirect.com/science/article/pii/S0168583X16303998",
author = "Stephen A. Graves and Paul A. Ellison and Todd E. Barnhart and Hector F. Valdovinos and Eva R. Birnbaum and Francois M. Nortier and Robert J. Nickles and Jonathan W. Engle"
}

@article{Niobium_voyles,
title = "{{Excitation functions for (p,x) reactions of niobium in the energy range of Ep = 40--90 MeV}}",
journal = "Nuclear Instruments and Methods in Physics Research Section B: Beam Interactions with Materials and Atoms",
volume = "429",
pages = "53 - 74",
year = "2018",
issn = "0168-583X",
doi = "10.1016/j.nimb.2018.05.028",
url = "http://www.sciencedirect.com/science/article/pii/S0168583X18303458",
author = "Andrew S. Voyles and Lee A. Bernstein and Eva R. Birnbaum and Jonathan W. Engle and Stephen A. Graves and Toshihiko Kawano and Amanda M. Lewis and Francois M. Nortier"
}

@article{IAEACPR,
author = {Hermanne, A and Ignatyuk, A V and Capote, R and Carlson, B V and Engle, J W and Kellett, M A and Kib{\'{e}}di, T and Kim, G and Kondev, F G and Hussain, M and Lebeda, O and Luca, A and Nagai, Y and Naik, H and Nichols, A L and Nortier, F M and Suryanarayana, S V and Tak{\'{a}}cs, S and T{\'{a}}rk{\'{a}}nyi, F T and Verpelli, M},
doi = {10.1016/j.nds.2018.02.009},
issn = {00903752},
journal = {Nuclear Data Sheets},
month = {feb},
pages = {338--382},
title = {{Reference Cross Sections for Charged-particle Monitor Reactions}},
url = {http://www.sciencedirect.com/science/article/pii/S0090375218300280 http://linkinghub.elsevier.com/retrieve/pii/S0090375218300280 https://linkinghub.elsevier.com/retrieve/pii/S0090375218300280},
volume = {148},
year = {2018}
}

@article{ZIEGLER20101818,
title = "{{SRIM – The stopping and range of ions in matter (2010)}}",
journal = "Nuclear Instruments and Methods in Physics Research Section B: Beam Interactions with Materials and Atoms",
volume = "268",
number = "11",
pages = "1818 - 1823",
year = "2010",
note = "19th International Conference on Ion Beam Analysis",
issn = "0168-583X",
doi = "10.1016/j.nimb.2010.02.091",
url = "http://www.sciencedirect.com/science/article/pii/S0168583X10001862",
author = "James F. Ziegler and M.D. Ziegler and J.P. Biersack"
}

@article{berger2011xcom,
address = {Gaithersburg, MD},
author = {Berger, M J and Hubbell, J H and Seltzer, S M and Chang, J and Coursey, J S and Sukumar, R and Zucker, D S and Olsen, K},
institution = {National Institute of Standards and Technology},
title = {{XCOM: Photon cross section database (version 1.5)}},
year = {2010}
}

@techreport{hubbell1995tables,
  title={{Tables of X-ray mass attenuation coefficients and mass energy-absorption coefficients 1 keV to 20 MeV for elements Z= 1 to 92 and 48 additional substances of dosimetric interest}},
  author={Hubbell, John H and Seltzer, Stephen M},
  year={1995},
  institution={National Inst. of Standards and Technology-PL, Gaithersburg, MD (United States}
}

@article{VIDMAR2001533,
title = {{A semi-empirical model of the efficiency curve for extended sources in gamma-ray spectrometry}},
journal = {Nuclear Instruments and Methods in Physics Research Section A: Accelerators, Spectrometers, Detectors and Associated Equipment},
volume = {470},
number = {3},
pages = {533-547},
year = {2001},
issn = {0168-9002},
doi = {https://doi.org/10.1016/S0168-9002(01)00799-9},
url = {https://www.sciencedirect.com/science/article/pii/S0168900201007999},
author = {T. Vidmar and M. Korun and A. Likar and R. Martinčič},
}

@article{KONING200415,
title = {{A global pre-equilibrium analysis from 7 to 200 MeV based on the optical model potential}},
journal = {Nuclear Physics A},
volume = {744},
pages = {15-76},
year = {2004},
issn = {0375-9474},
doi = {https://doi.org/10.1016/j.nuclphysa.2004.08.013},
url = {https://www.sciencedirect.com/science/article/pii/S037594740400870X},
author = {A.J. Koning and M.C. Duijvestijn}
}

@article{Arsenic_Fox,
  title = {{Measurement and modeling of proton-induced reactions on arsenic from 35 to 200 MeV}},
  author = {Fox, Morgan B. and Voyles, Andrew S. and Morrell, Jonathan T. and Bernstein, Lee A. and Batchelder, Jon C. and Birnbaum, Eva R. and Cutler, Cathy S. and Koning, Arjan J. and Lewis, Amanda M. and Medvedev, Dmitri G. and Nortier, Francois M. and O'Brien, Ellen M. and Vermeulen, Christiaan},
  journal = {Phys. Rev. C},
  volume = {104},
  issue = {6},
  pages = {064615},
  numpages = {26},
  year = {2021},
  month = {Dec},
  publisher = {American Physical Society},
  doi = {10.1103/PhysRevC.104.064615},
  url = {https://link.aps.org/doi/10.1103/PhysRevC.104.064615}
}

@article{Bateman,
title = "{The solution of a system of differential equations occurring in the theory of radioactive transformations}",
journal = "In Proc. Campridge Philos. Soc.",
volume = "15",
pages = "423 - 427",
year = "1910",
issn = "V",
doi = "10.12691/ijp-4-2-3",
url = "https://archive.org/details/cbarchive_122715_solutionofasystemofdifferentia1843",
author = "H. Bateman"
}

@article{tin660,
author = {Alexandryan, V. and Ivazyan, G. and Balabekyan, A. and Danagulyan, Alita and Kalinnikov, V. and Stegailov, V. and Frána, J.},
year = {1996},
month = {04},
pages = {560-564},
title = {{Study of isomeric ratios of proton-nucleus cross sections on tin isotopes}},
volume = {59},
journal = {Physics of Atomic Nuclei - PHYS ATOM NUCL-ENGL TR}
}

@article{tin365,
author = {Balabekyan, A. and Danagulyan, Alita and Drnoyan, J. and Demekhina, Nina and Adam, Jindrich and Kalinnikov, V. and Krivopustov, M. and Pronskikh, V. and Stegailov, V. and Solnyshkin, A. and Caloun, Pavel and Tsoupko-Sitnikov, Vsevolod},
year = {2005},
month = {02},
pages = {171-176},
title = {{Investigation of spallation reactions on 120 Sn and ( d, xn ), ( d, pxn ), ( p, xn ), and ( p, pxn ) reactions on enriched tin isotopes}},
volume = {68},
journal = {Physics of Atomic Nuclei - PHYS ATOM NUCL-ENGL TR},
doi = {10.1134/1.1866372}
}

@article{KONING2003231,
author = {Koning, A J and Delaroche, J P},
doi = {10.1016/S0375-9474(02)01321-0},
issn = {0375-9474},
journal = {Nuclear Physics A},
number = {3},
pages = {231--310},
title = {{Local and global nucleon optical models from 1 keV to 200 MeV}},
url = {http://www.sciencedirect.com/science/article/pii/S0375947402013210},
volume = {713},
year = {2003}
}

@article {Kratochwil1941,
	author = {Kratochwil, Clemens and Bruchertseifer, Frank and Giesel, Frederik L. and Weis, Mirjam and Verburg, Frederik A. and Mottaghy, Felix and Kopka, Klaus and Apostolidis, Christos and Haberkorn, Uwe and Morgenstern, Alfred},
	title = {{225Ac-PSMA-617 for PSMA-Targeted $\alpha$-Radiation Therapy of Metastatic Castration-Resistant Prostate Cancer}},
	volume = {57},
	number = {12},
	pages = {1941--1944},
	year = {2016},
	doi = {10.2967/jnumed.116.178673},
	publisher = {Society of Nuclear Medicine},
	issn = {0161-5505},
	URL = {https://jnm.snmjournals.org/content/57/12/1941},
	eprint = {https://jnm.snmjournals.org/content/57/12/1941.full.pdf},
	journal = {Journal of Nuclear Medicine}
}

@Misc{curie,
  author =    {Jonathan T. Morrell},
  title =     {{Curie}: {A python toolkit to aid in the analysis of experimental nuclear data}},
  year =      {2019--},
  url = "https://jtmorrell.github.io/curie/build/html/index.html",
  note = {[Online; accessed January 24, 2024.]}
}

@article{A225,
title = "{{Nuclear Data Sheets for A = 225}}",
journal = "Nuclear Data Sheets",
volume = "110",
number = "6",
pages = "1409 - 1472",
year = "2009",
issn = "0090-3752",
doi = "10.1016/j.nds.2009.04.003",
url = "http://www.sciencedirect.com/science/article/pii/S0090375209000386",
author = "A.K. Jain and R. Raut and J.K. Tuli"
}

@article{A221,
title = "{{Nuclear Data Sheets for A = 221}}",
journal = "Nuclear Data Sheets",
volume = "108",
number = "4",
pages = "883 - 922",
year = "2007",
issn = "0090-3752",
doi = "10.1016/j.nds.2007.03.002",
url = "http://www.sciencedirect.com/science/article/pii/S0090375207000300",
author = "Ashok Kumar Jain and Sukhjeet Singh and Suresh Kumar and Jagdish K. Tuli"
}

@article{A217,
title = "{{Nuclear Data Sheets for A=217}}",
journal = "Nuclear Data Sheets",
volume = "147",
pages = "382 - 458",
year = "2018",
issn = "0090-3752",
doi = "10.1016/j.nds.2018.01.002",
url = "http://www.sciencedirect.com/science/article/pii/S0090375218300024",
author = "F.G. Kondev and E.A. McCutchan and B. Singh and K. Banerjee and S. Bhattacharya and A. Chakraborty and S. Garg and N. Jovancevic and S. Kumar and S.K. Rathi and T. Roy and J. Lee and R. Shearman"
}

@article{A213,
title = "{{Nuclear Data Sheets for A = 213}}",
journal = "Nuclear Data Sheets",
volume = "108",
number = "3",
pages = "633 - 680",
year = "2007",
issn = "0090-3752",
doi = "10.1016/j.nds.2007.02.002",
url = "http://www.sciencedirect.com/science/article/pii/S009037520700021X",
author = "M.S. Basunia"
}

@article{A209,
title = "{{Nuclear Data Sheets for A = 209}}",
journal = "Nuclear Data Sheets",
volume = "126",
pages = "373 - 546",
year = "2015",
issn = "0090-3752",
doi = "10.1016/j.nds.2015.05.003",
url = "http://www.sciencedirect.com/science/article/pii/S0090375215000149",
author = "J. Chen and F.G. Kondev"
}

@article{A152,
title = {{Nuclear Data Sheets for A = 152}},
journal = {Nuclear Data Sheets},
volume = {114},
number = {11},
pages = {1497-1847},
year = {2013},
issn = {0090-3752},
doi = {https://doi.org/10.1016/j.nds.2013.11.001},
url = {https://www.sciencedirect.com/science/article/pii/S0090375213000744},
author = {M.J. Martin}
}

@article{A140,
title = {{Nuclear Data Sheets for A=140}},
journal = {Nuclear Data Sheets},
volume = {154},
pages = {1-403},
year = {2018},
issn = {0090-3752},
doi = {https://doi.org/10.1016/j.nds.2018.11.002},
url = {https://www.sciencedirect.com/science/article/pii/S0090375218300863},
author = {N. Nica}
}

@article{A139,
title = "{{Nuclear Data Sheets for A = 139}}",
journal = "Nuclear Data Sheets",
volume = "138",
pages = "1 - 292",
year = "2016",
issn = "0090-3752",
doi = "10.1016/j.nds.2016.11.001",
url = "http://www.sciencedirect.com/science/article/pii/S0090375216300424",
author = "Paresh K. Joshi and Balraj Singh and Sukhjeet Singh and Ashok K. Jain"
}

@article{A137,
title = "{{Nuclear Data Sheets for A = 137}}",
journal = "Nuclear Data Sheets",
volume = "108",
number = "10",
pages = "2173 - 2318",
year = "2007",
issn = "0090-3752",
doi = "10.1016/j.nds.2007.09.002",
url = "http://www.sciencedirect.com/science/article/pii/S0090375207000804",
author = "E. Browne and J.K. Tuli"
}

@article{A136,
title = "{{Nuclear Data Sheets for A=136}}",
journal = {Nuclear Data Sheets},
volume = {152},
pages = {331-667},
year = {2018},
issn = {0090-3752},
doi = {https://doi.org/10.1016/j.nds.2018.10.002},
url = {https://www.sciencedirect.com/science/article/pii/S0090375218300711},
author = {E.A. McCutchan}
}

@article{A135,
title = "{{Nuclear Data Sheets for A = 135}}",
journal = "Nuclear Data Sheets",
volume = "109",
number = "3",
pages = "517 - 698",
year = "2008",
issn = "0090-3752",
doi = "10.1016/j.nds.2008.02.001",
url = "http://www.sciencedirect.com/science/article/pii/S0090375208000094",
author = "Balraj Singh and Alexander A. Rodionov and Yuri L. Khazov"
}

@article{A134,
title = "{{Nuclear Data Sheets for A = 134}}",
journal = "Nuclear Data Sheets",
volume = "103",
number = "1",
pages = "1 - 182",
year = "2004",
issn = "0090-3752",
doi = "10.1016/j.nds.2004.11.001",
url = "http://www.sciencedirect.com/science/article/pii/S0090375204000717",
author = "A.A. Sonzogni"
}

@article{A133,
title = "{{Nuclear Data Sheets for A = 133}}",
journal = "Nuclear Data Sheets",
volume = "112",
number = "4",
pages = "855 - 1113",
year = "2011",
issn = "0090-3752",
doi = "10.1016/j.nds.2011.03.001",
url = "http://www.sciencedirect.com/science/article/pii/S0090375211000202",
author = "Yu. Khazov and A. Rodionov and F.G. Kondev"
}

@article{A132,
title = "{{Nuclear Data Sheets for A=132}}",
journal = "Nuclear Data Sheets",
volume = "104",
number = "3",
pages = "497 - 790",
year = "2005",
issn = "0090-3752",
doi = "10.1016/j.nds.2005.03.001",
url = "http://www.sciencedirect.com/science/article/pii/S0090375205000128",
author = "Yu. Khazov and A.A. Rodionov and S. Sakharov and Balraj Singh"
}

@article{A131,
title = "{{Nuclear Data Sheets for A = 131}}",
journal = {Nuclear Data Sheets},
volume = {107},
number = {11},
pages = {2715-2930},
year = {2006},
issn = {0090-3752},
doi = {https://doi.org/10.1016/j.nds.2006.10.001},
url = {https://www.sciencedirect.com/science/article/pii/S0090375206000809},
author = {Yu. Khazov and I. Mitropolsky and A. Rodionov}
}

@article{A130,
title = "{{Nuclear Data Sheets for A = 130}}",
journal = {Nuclear Data Sheets},
volume = {93},
number = {1},
pages = {33-242},
year = {2001},
issn = {0090-3752},
doi = {https://doi.org/10.1006/ndsh.2001.0012},
url = {https://www.sciencedirect.com/science/article/pii/S0090375201900122},
author = {Balraj Singh}
}

@article{A129,
title = "{{Nuclear Data Sheets for A = 129}}",
journal = {Nuclear Data Sheets},
volume = {121},
pages = {143-394},
year = {2014},
issn = {0090-3752},
doi = {https://doi.org/10.1016/j.nds.2014.09.002},
url = {https://www.sciencedirect.com/science/article/pii/S0090375214006565},
author = {Janos Timar and Zoltan Elekes and Balraj Singh}
}

@article{A128,
title = "{{Nuclear Data Sheets for A = 128}}",
journal = {Nuclear Data Sheets},
volume = {129},
pages = {191-436},
year = {2015},
issn = {0090-3752},
doi = {https://doi.org/10.1016/j.nds.2015.09.002},
url = {https://www.sciencedirect.com/science/article/pii/S0090375215000472},
author = {Zoltan Elekes and Janos Timar}
}

@article{A127,
title = "{{Nuclear Data Sheets for A = 127}}",
journal = {Nuclear Data Sheets},
volume = {112},
number = {7},
pages = {1647-1831},
year = {2011},
issn = {0090-3752},
doi = {https://doi.org/10.1016/j.nds.2011.06.001},
url = {https://www.sciencedirect.com/science/article/pii/S009037521100055X},
author = {A. Hashizume}
}

@article{A126,
title = "{{Nuclear Data Sheets for A=126}}",
journal = {Nuclear Data Sheets},
volume = {180},
pages = {1-413},
year = {2022},
issn = {0090-3752},
doi = {https://doi.org/10.1016/j.nds.2022.02.001},
url = {https://www.sciencedirect.com/science/article/pii/S0090375222000011},
author = {H. Iimura and J. Katakura and S. Ohya}
}

@article{A125,
title = "{{Nuclear Data Sheets for A = 125}}",
journal = {Nuclear Data Sheets},
volume = {112},
number = {3},
pages = {495-705},
year = {2011},
issn = {0090-3752},
doi = {https://doi.org/10.1016/j.nds.2011.02.001},
url = {https://www.sciencedirect.com/science/article/pii/S009037521100010X},
author = {J. Katakura}
}

@article{A123,
title = "{{Nuclear Data Sheets for A=123}}",
journal = {Nuclear Data Sheets},
volume = {174},
pages = {1-463},
year = {2021},
issn = {0090-3752},
doi = {https://doi.org/10.1016/j.nds.2021.05.001},
url = {https://www.sciencedirect.com/science/article/pii/S0090375221000260},
author = {Jun Chen}
}

@article{A65,
title = "{{Nuclear Data Sheets for A = 65}}",
journal = {Nuclear Data Sheets},
volume = {111},
number = {9},
pages = {2425-2553},
year = {2010},
issn = {0090-3752},
doi = {https://doi.org/10.1016/j.nds.2010.09.002},
url = {https://www.sciencedirect.com/science/article/pii/S0090375210000864},
author = {E. Browne and J.K. Tuli}
}

@article{A62,
title = "{{Nuclear Data Sheets for A = 62}}",
journal = "Nuclear Data Sheets",
volume = "113",
number = "4",
pages = "973 - 1114",
year = "2012",
issn = "0090-3752",
doi = "10.1016/j.nds.2012.04.002",
url = "http://www.sciencedirect.com/science/article/pii/S0090375212000312",
author = "Alan L. Nichols and Balraj Singh and Jagdish K. Tuli"
}

@article{A61,
title = "{{Nuclear Data Sheets for A = 61}}",
journal = "Nuclear Data Sheets",
volume = "125",
pages = "1 - 200",
year = "2015",
issn = "0090-3752",
doi = "10.1016/j.nds.2015.02.001",
url = "http://www.sciencedirect.com/science/article/pii/S0090375215000022",
author = "Kazimierz Zuber and Balraj Singh"
}

@article{A60,
title = "{{Nuclear Data Sheets for A = 60}}",
journal = {Nuclear Data Sheets},
volume = {114},
number = {12},
pages = {1849-2022},
year = {2013},
issn = {0090-3752},
doi = {https://doi.org/10.1016/j.nds.2013.11.002},
url = {https://www.sciencedirect.com/science/article/pii/S0090375213000823},
author = {E. Browne and J.K. Tuli}
}

@article{A59,
title = "{{Nuclear Data Sheets for A=59}}",
journal = {Nuclear Data Sheets},
volume = {151},
pages = {1-333},
year = {2018},
issn = {0090-3752},
doi = {https://doi.org/10.1016/j.nds.2018.08.001},
url = {https://www.sciencedirect.com/science/article/pii/S0090375218300590},
author = {M. Shamsuzzoha Basunia}
}

@article{A58,
title = "{{Nuclear Data Sheets for A = 58}}",
journal = "Nuclear Data Sheets",
volume = "111",
number = "4",
pages = "897 - 1092",
year = "2010",
issn = "0090-3752",
doi = "10.1016/j.nds.2010.03.003",
url = "http://www.sciencedirect.com/science/article/pii/S0090375210000359",
author = "Caroline D. Nesaraja and Scott D. Geraedts and Balraj Singh"
}

@article{A57,
title = "{{Nuclear data sheets update for A = 57}}",
journal = {Nuclear Data Sheets},
volume = {67},
number = {2},
pages = {195-270},
year = {1992},
issn = {0090-3752},
doi = {https://doi.org/10.1016/0090-3752(92)80020-K},
url = {https://www.sciencedirect.com/science/article/pii/009037529280020K},
author = {M.R. Bhat}
}

@article{A56,
title = "{{Nuclear Data Sheets for A = 56}}",
journal = {Nuclear Data Sheets},
volume = {112},
number = {6},
pages = {1513-1645},
year = {2011},
issn = {0090-3752},
doi = {https://doi.org/10.1016/j.nds.2011.04.004},
url = {https://www.sciencedirect.com/science/article/pii/S0090375211000457},
author = {Huo Junde and Huo Su and Yang Dong}
}

@article{A55,
title = "{{Nuclear Data Sheets for A = 55}}",
journal = {Nuclear Data Sheets},
volume = {109},
number = {4},
pages = {787-942},
year = {2008},
issn = {0090-3752},
doi = {https://doi.org/10.1016/j.nds.2008.03.001},
url = {https://www.sciencedirect.com/science/article/pii/S0090375208000197},
author = {Huo Junde}
}

@article{A54,
title = "{{Nuclear Data Sheets for A = 54}}",
journal = {Nuclear Data Sheets},
volume = {107},
number = {6},
pages = {1393-1530},
year = {2006},
issn = {0090-3752},
doi = {https://doi.org/10.1016/j.nds.2006.05.003},
url = {https://www.sciencedirect.com/science/article/pii/S009037520600038X},
author = {Huo Junde and Huo Su}
}

@article{A52,
title = "{{Nuclear Data Sheets for A = 52}}",
journal = {Nuclear Data Sheets},
volume = {108},
number = {4},
pages = {773-882},
year = {2007},
issn = {0090-3752},
doi = {https://doi.org/10.1016/j.nds.2007.03.001},
url = {https://www.sciencedirect.com/science/article/pii/S0090375207000294},
author = {Junde Huo and Su Huo and Chunhui Ma}
}

@article{A51,
title = "{{Nuclear Data Sheets for A=51}}",
journal = {Nuclear Data Sheets},
volume = {144},
pages = {1-296},
year = {2017},
issn = {0090-3752},
doi = {https://doi.org/10.1016/j.nds.2017.08.002},
url = {https://www.sciencedirect.com/science/article/pii/S0090375217300546},
author = {Jimin Wang and Xiaolong Huang}
}

@article{A48,
title = "{{Nuclear Data Sheets for A=48}}",
journal = {Nuclear Data Sheets},
volume = {179},
pages = {1-382},
year = {2022},
issn = {0090-3752},
doi = {https://doi.org/10.1016/j.nds.2021.12.001},
url = {https://www.sciencedirect.com/science/article/pii/S0090375221000697},
author = {Jun Chen}
}

@article{A47,
title = "{{Nuclear Data Sheets for A = 47}}",
journal = {Nuclear Data Sheets},
volume = {108},
number = {5},
pages = {923-1056},
year = {2007},
issn = {0090-3752},
doi = {https://doi.org/10.1016/j.nds.2007.04.002},
url = {https://www.sciencedirect.com/science/article/pii/S0090375207000403},
author = {T.W. Burrows}
}

@article{A46,
title = "{{Nuclear Data Sheets for A = 46}}",
journal = {Nuclear Data Sheets},
volume = {91},
number = {1},
pages = {1-116},
year = {2000},
issn = {0090-3752},
doi = {https://doi.org/10.1006/ndsh.2000.0014},
url = {https://www.sciencedirect.com/science/article/pii/S0090375200900140},
author = {S.-C. Wu}
}

@article{A44,
title = "{{Nuclear Structure and Decay Data for A=44 Isobars}}",
journal = {Nuclear Data Sheets},
volume = {190},
pages = {1-318},
year = {2023},
issn = {0090-3752},
doi = {https://doi.org/10.1016/j.nds.2023.06.001},
url = {https://www.sciencedirect.com/science/article/pii/S009037522300042X},
author = {Jun Chen and Balraj Singh}
}

@article{A43,
title = "{{Nuclear Data Sheets for A = 43}}",
journal = {Nuclear Data Sheets},
volume = {126},
pages = {1-150},
year = {2015},
issn = {0090-3752},
doi = {https://doi.org/10.1016/j.nds.2015.05.001},
url = {https://www.sciencedirect.com/science/article/pii/S0090375215000125},
author = {Balraj Singh and Jun Chen}
}

@article{A42,
title = "{{Nuclear Data Sheets for A = 42}}",
journal = {Nuclear Data Sheets},
volume = {135},
pages = {1-192},
year = {2016},
issn = {0090-3752},
doi = {https://doi.org/10.1016/j.nds.2016.06.001},
url = {https://www.sciencedirect.com/science/article/pii/S0090375216300126},
author = {Jun Chen and Balraj Singh}
}

@article{A22,
title = "{{Nuclear Data Sheets for A = 22}}",
journal = "Nuclear Data Sheets",
volume = "127",
pages = "69 - 190",
year = "2015",
issn = "0090-3752",
doi = "10.1016/j.nds.2015.07.002",
url = "http://www.sciencedirect.com/science/article/pii/S0090375215000253",
author = "M. Shamsuzzoha Basunia"
}

@article{A24,
title = "{{Nuclear Data Sheets for A = 24}}",
journal = "Nuclear Data Sheets",
volume = "108",
number = "11",
pages = "2319 - 2392",
year = "2007",
issn = "0090-3752",
doi = "10.1016/j.nds.2007.10.001",
url = "http://www.sciencedirect.com/science/article/pii/S0090375207000877",
author = "R.B. Firestone"
}

\end{document}